\documentclass[12pt,a4paper]{article}
\usepackage[hmargin=2.5cm,vmargin=3cm]{geometry}
\usepackage[english]{babel}
\usepackage[utf8]{inputenc}
\usepackage{amsmath,amssymb,amsfonts}
\usepackage{mathrsfs}
\usepackage{caption}
\usepackage{subcaption}
\usepackage{slashed}
\usepackage{mathtools}
\usepackage{csquotes}
\usepackage[T1]{fontenc}
\usepackage{multirow}
\usepackage{multicol}

\usepackage[bbgreekl]{mathbbol}
\DeclareSymbolFontAlphabet{\mathbb}{AMSb}
\DeclareSymbolFontAlphabet{\mathbbl}{bbold}

\numberwithin{equation}{section}
\usepackage[bottom]{footmisc}

\usepackage{comment}

\usepackage{xcolor}
\usepackage[
bookmarksnumbered,
linktocpage=true,
colorlinks=true,
citecolor={green!50!black}
]{hyperref}

\usepackage{tikz}
\usetikzlibrary{arrows,calc}
\tikzset{
	dot/.style={circle, minimum size=3pt, inner sep=0, fill=black},
	every label/.append style={font=\footnotesize, label distance=-5pt}
}
\usetikzlibrary{decorations.pathreplacing}

\usepackage{array}

\newcolumntype{C}[1]{>{\centering\let\newline\\\arraybackslash\hspace{0pt}}m{#1}}

\newcommand{\ie}{\textit{i.e.}}
\newcommand{\eg}{\textit{e.g.}}
\newcommand{\cf}{\textit{cf.}}

\DeclareMathOperator{\sign}{sign}

\newcommand{\QQ}{\mathcal{Q}(q)}

\newcommand{\dQQp}{\mathcal{Q'}(q_+)}
\newcommand{\PP}{\mathcal{P}(p)}

\newcommand{\dPP}{\mathcal{P}'(p)}
\newcommand{\dPPp}{\mathcal{P}'(p_+)}
\newcommand{\dPPm}{\mathcal{P}'(p_-)}
\newcommand{\dPPpm}{\mathcal{P}'(p_\pm)}

\newcommand{\Pm}{\mathcal{P}_-(p)}
\newcommand{\Pp}{\mathcal{P}_+(p)}
\newcommand{\Qm}{\mathcal{Q}_-(q)}
\newcommand{\Qp}{\mathcal{Q}_+(q)}
\newcommand{\lens}{t}

\newcommand{\labell}{v}
\newcommand{\singp}{m_+}
\newcommand{\singm}{m_-}
\newcommand{\singpm}{m_\pm}
\newcommand{\loci}{\mathcal{L}}
\newcommand{\newsignpm}{\rho_{\pm}}
\newcommand{\newsignp}{\rho_{+}}
\newcommand{\newsignm}{\rho_{-}}

\newcommand{\parqppm}{}
\newcommand{\parqp}{}

\newcommand{\parpp}{}
\newcommand{\parpm}{}
\newcommand{\parppm}{}
\newcommand{\parp}{}

\newcommand{\angleone}{\theta_1}
\newcommand{\angletwo}{\theta_2}
\newcommand{\angles}{\theta_i}

\newcommand{\periodicone}{\varphi_1}
\newcommand{\periodictwo}{\varphi_2}
\newcommand{\periodicangles}{\varphi_i}
\newcommand{\toricone}{\phi_1}
\newcommand{\torictwo}{\phi_2}

\newcommand{\signpp}{\eta}
\newcommand{\signpm}{\delta}
\newcommand{\signqp}{\lambda}
\newcommand{\twist}{\sigma}
\newcommand{\branch}{\kappa}
\newcommand{\freetwist}{\tilde{P}}
\newcommand{\freeanti}{\overline{P}}
\newcommand{\free}{\tilde{q}_+}

\newcommand{\contform}{\eta}

\newcommand{\NN}{\mathbb{N}}
\newcommand{\ZZ}{\mathbb{Z}}

\newcommand{\RR}{\mathbb{R}}
\newcommand{\CC}{\mathbb{C}}
\newcommand{\spindle}{\mathbbl{\Sigma}}

\newcommand{\spindlesing}{\spindle_{[\singm,\singp]}}
\newcommand{\spindlesinginf}{\spindle^{\infty}_{[\singm,\singp]}}
\newcommand{\spindlesingbolt}{\spindle^{q_{+}}_{[\singm,\singp]}}
\newcommand{\bundlep}{\mathcal{O}(\singm+\singp)}

\newcommand{\bundletwist}{\mathcal{O}(\singm+\twist\singp)}

\newcommand{\AdS}{\mathrm{AdS}}

\newcommand{\dd}{\mathrm{d}}
\newcommand{\ee}{\mathrm{e}}
\newcommand{\ii}{\mathrm{i}}

\newcommand{\singpfl}{m_+}
\newcommand{\singmfl}{m_-}
\newcommand{\singpmfl}{m_\pm}
\newcommand{\Gcofl}{\eta}
\newcommand{\Fcofl}{\xi}
\newcommand{\angafl}{\phi}
\newcommand{\angbfl}{\psi}

\newcommand{\newangafl}{\theta_1}
\newcommand{\newangbfl}{\theta_2}
\newcommand{\newangtilafl}{\varphi_1}
\newcommand{\newangtilbfl}{\varphi_2}
\newcommand{\torcoafl}{\phi_1}
\newcommand{\torcobfl}{\phi_2}
\newcommand{\linparfl}{N}
\newcommand{\labellfl}{v}
\newcommand{\lensfl}{t}

\newcommand{\parflm}{}
\newcommand{\parflpm}{}

\newcommand{\newqplus}{\mathbf{q}_+}
\newcommand{\newpplus}{\mathbf{p}_+}
\newcommand{\newpminus}{\mathbf{p}_-}
\newcommand{\newpplusminus}{\mathbf{p}_\pm}
\newcommand{\newlabell}{\labell}
\newcommand{\acc}{\mathbf{A}}
\newcommand{\Pfunc}{\mathbf{P}(p)}
\newcommand{\Qfunc}{\mathbf{Q}(q)}
\newcommand{\newlens}{\lens}

\newcommand{\newsingp}{\singp}
\newcommand{\newsingm}{\singm}
\newcommand{\newsingpm}{\singpm}

\begin{document}

	\begin{titlepage}
		\vskip 2cm
		
		\begin{center}
			
			\vspace*{2.5cm}
		
			{\Large \bf NUTs,  Bolts, and Spindles}
			
			\vskip 1.5cm
			
			{Matteo Kevin Crisafio, Alessio Fontanarossa and Dario Martelli}
			
			\vskip 1cm
			
			\textit{Dipartimento di Matematica \enquote{Giuseppe Peano}, Universit\`a di Torino,\\
				Via Carlo Alberto 10, 10123 Torino, Italy}
			
			\vskip 0.4cm
			
			\textit{INFN, Sezione di Torino,\\Via Pietro Giuria 1, 10125 Torino, Italy}
			
		\end{center}

		\vskip 3.8cm

		\begin{abstract}
\noindent
We construct  new infinite classes of Euclidean supersymmetric  solutions of four dimensional minimal gauged supergravity  comprising a $U (1) \times  U (1)$-invariant asymptotically locally hyperbolic metric on the total space of orbifold line bundles over a spindle (bolt).~The conformal boundary is generically a squashed, branched, lens space and the graviphoton gauge field can have either twist or anti-twist  through the spindle bolt.~Correspondingly, the boundary geometry inherits  two types of rigid Killing spinors, that we refer to as twist and anti-twist for the three-dimensional Seifert orbifolds, as well as some specific flat connections for the background gauge field, determined by the data of the spindle bolt.~For all our solutions we compute the holographically renormalized on-shell action and compare it to the expression obtained via equivariant localization, uncovering a markedly distinct behaviour in the cases of twist and anti-twist.~Our results provide precise predictions for the large $N$ limit of the corresponding localized partition functions of three-dimensional $\mathcal{N}=2$ superconformal field theories placed on Seifert orbifolds.

		 		 		\end{abstract}
		
	\end{titlepage}
	
	\tableofcontents

	\section{Introduction}

In the past few years supergravity solutions featuring spindles  
have significantly  enriched the landscape of holographic dualities, indicating  that quantum theories of gravity should be addressed in a context broader than that of  smooth manifolds.  In almost all cases in the literature so far, the spindle $\spindle $ appears as a factor of an 
AdS$\times \spindle$ space-time, suggesting that it should be interpreted as the near-horizon geometry  of a brane
wrapped on a spindle~\cite{Ferrero:2020laf,Ferrero:2021wvk,Faedo:2021nub}.
One of the main novelties of the spindle
is that it leads to new possibilities for  preserving  supersymmetry, with  direct implications for  quantum field theories with rigid supersymmetry. In particular, a supersymmetric field theory coupled to a background R-symmetry gauge field can be compactified on a spindle in two ways~\cite{Ferrero:2021etw}, referred to as twist and anti-twist, leading to different types of supersymmetric field theories in two lower dimensions in the IR.
In the context of holography the expectation that the low energy limit of the wrapped brane theory is a SCFT is fully vindicated in the case of the supersymmetric accelerating black hole~\cite{Ferrero:2020twa} (and its complex non-extremal deformation~\cite{Cassani:2021dwa}). In that context  one can take the spindle appearing in the near-horizon region all the way to infinity, where (in Euclidean) the geometry at the conformal boundary is $\spindle\times S^1$.  In turn, this led to the study of three-dimensional supersymmetric field theories on such backgrounds~\cite{Inglese:2023tyc} and the computation of the corresponding localized partition function, namely the spindle index~\cite{Inglese:2023wky}. The agreement of the large $N$ limit  of the spindle index with the entropy of the accelerating black hole was  achieved in~\cite{Colombo:2024mts}.

In this paper we will construct  various classes of explicit supersymmetric solutions of four-dimensional minimal gauged supergravity, where generically the spindle appears  in the bulk of an asymptotically locally hyperbolic space. More precisely, our solutions comprise global 
metrics on the total space of certain orbifold line bundles over the spindle, 
together with a graviphoton field and Killing spinors. The spindle corresponds to the zero section of these line bundles, meaning that it appears as a \emph{bolt}~\cite{Gibbons:1979xm} in the solutions. Generically, the three-dimensional 
conformal boundaries are squashed lens spaces, possibly with orbifold singularities, which we refer to as squashed, branched, lens spaces~\cite{Inglese:2023tyc}. 
Interestingly, we will show that the topological data of the bulk are encoded in the boundary geometry, through some subtle flat connections for the boundary gauge field. 
	In particular, we will see that the twist/anti-twist for the graviphoton through the spindle bolt is in one-to-one correspondence with 
	two types of rigid supersymmetry in the three-dimensional Seifert orbifolds discussed in \cite{Inglese:2023tyc},  that we shall therefore again refer to as \emph{twist and anti-twist}. This information is encoded 
	in some flat connections for the three dimensional gauge field which, surprisingly,  turn out to encapsulate the complete data of the spindle in the bulk.
The results of this paper were first announced in~\cite{conference:Dario} and presented in~\cite{conference:Alessio}.

Our analysis includes in special limits several previously known supersymmetric solutions of Euclidean minimal gauged supergravity~\cite{Martelli:2011fu,Martelli:2011fw,Martelli:2012sz,Martelli:2013aqa,Cassani:2021dwa} and sheds new light on them.  For example, we will clarify that the  1/4-BPS and 1/2-BPS families of  spherical bolt 
solutions in~\cite{Martelli:2012sz} arise as limits of spindle bolt solutions in the Carter-Pleba\'nski family~\cite{Carter:1968ks,PLEBANSKI1975196}, with twist and anti-twist for the graviphoton field, respectively.
It is perhaps surprising that the Carter-Pleba\'nski family leads to spindles, since its Lorentzian version contains
only the Kerr-Newman-NUT-AdS, Reissner-Nordstr\"om-NUT-AdS and  the  AdS-C-metric~\cite{Alonso-Alberca:2000zeh,Klemm:2013eca}\footnote{Note  that in~\cite{Alonso-Alberca:2000zeh} it is incorrectly stated that the Carter-Pleba\'nski  solution is the most general Pleba\'nski-Demianski solution.},
whereas the conical  singularities of the spindle are associated to a non-zero acceleration, which is present only in the more general Pleba\'nski-Demianski family~\cite{PLEBANSKI197698}\footnote{See also~\cite{Griffiths:2005qp,Podolsky:2006px,Podolsky:2022xxd,Astorino:2024bfl}.}.
Furthermore, whilst in the accelerating black hole~\cite{Ferrero:2021etw} only the anti-twist is realized, the solutions that we find here generically allow for both twist and anti-twist.

With rare exceptions,  given a set of boundary conditions and/or topological data,  the existence of suitably regular supergravity solutions remains a difficult open problem. For example, building on the 
results of~\cite{Martelli:2005tp,Martelli:2006yb}, it has been proved 
that for toric Sasakian manifolds, a necessary and sufficient condition for the existence of Sasaki-Einstein metrics is that the Sasakian volume is extremized with respect to the allowed Reeb vector fields~\cite{Futaki:2006cc}. 
More generally, whenever one can formulate an \emph{extremization problem} for a given class of solutions, one hopes that, under  suitable conditions, the solution of the extremization problem will be also sufficient 
for the existence of the supergravity solutions. This paradigm has been reinforced by the recent advances in the techniques for computing on-shell actions using localization methods. 
In particular, in the context of four dimensional supergravity this problem was initially tackled in~\cite{Farquet:2014kma} (for supersymmetric self-dual solutions of the minimal theory) and in~\cite{BenettiGenolini:2019jdz} (for generic supersymmetric solutions of the minimal theory) and then fully solved in~\cite{BenettiGenolini:2023kxp} using equivariant 
localization\footnote{Extensions to matter-coupled  four-dimensional gauged supergravities have been discussed in~\cite{BenettiGenolini:2024xeo}.}. 
Interestingly, our findings strongly suggest that for the spindle bolt solutions  the cases of twist and anti-twist behave differently: in the first case, solutions seem to exist only upon 
extremizing the localized action with respect to the choices of supersymmetric Killing vector, while in the second case it appears that for any given supersymmetric Killing vector 
supersymmetric solutions should exist.  We will comment more on this aspect of our new  solutions in the final section.
While this paper was being completed, the paper~\cite{BenettiGenolini:2024hyd} appeared where general properties of Euclidean supersymmetric solutions of four-dimensional matter-coupled gauged supergravity
with $U(1)^2$ isometry are  discussed.

 The outline of the rest of the paper is as follows. In section~\ref{sect: The Plebanski-Demianski family of  solutions} we recall the local form of the (Euclidean) Pleba\'nski-Demianski family of  solutions, review their  supersymmetry conditions and summarize some notable old solutions. In section~\ref{sect:Non-accelerating solutions}  we specialize to the Carter-Pleba\'nski
class, that we refer to as \enquote{non-accelerating}. We begin presenting the explicit local Killing spinors for this class of solutions and then we move to the global  analysis of the boundary and the bulk in section~\ref{subsect:Regularity}. In particular, 
we show that with a specific choice of gauge for the graviphoton, the Killing spinor is regular both on the bolt and at the boundary. This analysis will reveal that all the topological data of the bulk are already contained in the boundary, as discussed in section \ref{subsect: new Seifert}. 
The methods developed in~\cite{Faedo:2022rqx,Faedo:2024upq}  allow us to uncover  the underlying toric orbifolds, which may be characterised in terms of two-dimensional labelled polyhedral cones.
We show that there exist solutions with both twist and anti-twist in section~\ref{subsect:Quantization} and discuss how these two sub-classes reduce to the old \enquote{spherical} bolt solutions of~\cite{Martelli:2012sz}, in appropriate limits. 
In section~\ref{subsect:On-shell action} we discuss the holographically renormalized 
on-shell action of the non-accelerating class of solutions, comparing the results obtained evaluating the integrals using the explicit solutions with the localization results, pointing out the different role of extremization 
for the cases of twist and anti-twist.  In section~\ref{sect:Accelerating solutions} we return to the accelerating solutions. We begin presenting
the complete holographically renormalized on-shell action for the general class of non-supersymmetric 
Pleba\'nski-Demianski solutions. In order to make analytic progress in the global analysis
we make a judicious choice of the parameters, such that the supersymmetry conditions take a form similar to the non-accelerating case. Despite this, we are able to construct 
solutions of generic type, where again both twist and anti-twist are  realized.  In section~\ref{sect:Discussion} we discuss our findings and make comments on the implications that these have for holography. Four appendices complete this paper. In appendix~\ref{appendix:Old-bolt-solutions} we recall some relevant features of the old $1/4$-BPS and $1/2$-BPS spherical bolt solutions of~\cite{Martelli:2012sz}.
In appendix~\ref{appendix:A-spindle-Calabi-Yau-metric} we construct a Ricci-flat version of the solutions described in the main body, with nevertheless a spindle (bolt) in the bulk. 
In appendix~\ref{appendix: Values} we present some values of the parameters for which the solutions are correctly defined. Finally, appendix \ref{appendix:Uplift} is devoted to global aspects of the uplift of the solutions to eleven dimensions.

	\section{The Pleba\'nski-Demianski solutions}{\label{sect: The Plebanski-Demianski family of  solutions}}
		In this paper will consider asymptotically locally (Euclidean) AdS solutions of the Maxwell-Einstein-$\Lambda$ action. Alternatively, this can be seen as the bosonic sector of the minimal  $d=4$, $\mathcal{N}=2$ gauged supergravity~\cite{Freedman:1976aw}, whose Euclidean action reads
		\begin{equation}\label{4daction}
			S_E = -\frac{1}{16\pi G_4}\int \dd^4 x \sqrt{g}\,\Bigl(R-F_{\mu\nu}F^{\mu\nu}+\frac{6}{\ell^2}\Bigr)\, .
		\end{equation}
		Here the cosmological constant $\Lambda$ is related to the length parameter $\ell$ via $\Lambda=-3/\ell^2$, $R$ is the Ricci scalar of the four dimensional metric $g$ and $F=\dd A$ is the field strength of the abelian graviphoton $A$. The equations of motion stemming from~\eqref{4daction} are
		\begin{equation} \label{minimalSUGRAeom}
			R_{\mu\nu}+\frac{3}{\ell^2}g_{\mu\nu}=2\Bigl(F_{\mu\rho}F_{\nu}^{\,\,\rho}-\frac{F^2}{4}g_{\mu\nu}\Bigr)\, ,\quad \dd\star F=0\, ,
		\end{equation}
		which may also be rewritten in the equivalent form 
		\begin{equation} \label{minimalSUGRAeomMATH}
			\begin{aligned}
				\dd F =0\, , \qquad \quad \dd\star F=0\, ,\qquad   \left[\mathrm{Ric} + 2 F \circ  F \right]_0 = 0	\, , \qquad R = -\frac{12}{\ell^2}\, , 
			\end{aligned}	
		\end{equation}
		where $\mathrm{Ric}$ is the Ricci tensor, 
		$(F \circ  F )_{\mu\nu} \equiv - F_{\mu\rho}F_{\nu}^{\,\,\rho}$	and $\left[~\right]_0$ denotes the trace-free part with respect to the metric $g$.
		Notice that, in particular, all solutions must have constant scalar curvature. The form of the equations~\eqref{minimalSUGRAeomMATH} appears usually
		in the mathematical literature\footnote{The factor of 2 difference with~\cite{LeBrun:2008kh} in the third equation is purely conventional and can be reabsorbed by rescaling $F$.}, where typically it is assumed that the underlying manifold (or orbifold) is compact, and then the constant scalar curvature condition can be shown to be implied by the first three equations in~\eqref{minimalSUGRAeomMATH}~\cite{LeBrun:2008kh}. In the present paper we shall assume that the space is non-compact and in
		particular asymptotically locally hyperbolic, but in principle one might also consider the compact case and we shall make some comments about this possibility in the 
		concluding section.

		We will be particularly interested in the sub-class of solutions to~\eqref{minimalSUGRAeomMATH} that are \emph{supersymmetric}, namely they admit
		at least one non-identically zero Dirac spinor $\varepsilon$ satisfying the Killing spinor equation 
		\begin{equation}\label{KSE}	\hat{D}_{\mu}\varepsilon\equiv\bigl[D_{\mu}-\frac{\ii}{\ell}A_\mu+\frac{1}{2\ell}\Gamma_{\mu}+\frac{\ii}{4}F_{\nu\rho}\Gamma^{\nu\rho}\Gamma_{\mu}\bigr]\varepsilon=0\, ,
		\end{equation}
		where $D_{\mu}\varepsilon=\partial_{\mu}\varepsilon+ 1/4 \omega_{\mu a b} \Gamma^{ab}\varepsilon$ is the standard covariant derivative.~Here $\Gamma_a$, $a=1,\dots,4$, generate the Clifford algebra $\text{Cliff}(4,0)$, so that their curved counterparts satisfy $\{\Gamma_{\mu},\Gamma_{\nu}\}=2 g_{\mu\nu} \text{Id}_{4\times4}$. 
		The supersymmetry condition~\eqref{KSE} can be re-cast equivalently as  coupled PDEs for a set of bosonic tensors constructed as bilinears in the Killing 
		spinors, in terms of which the metric and gauge field take a 
		canonical form. This was done in Lorentzian signature in~\cite{Caldarelli:1998hg}  and in Euclidean signature, that is relevant for the present paper, in~\cite{BenettiGenolini:2019jdz}. However, here we will take advantage of the fact that 
		the supersymmetry conditions for the class of local Pleba\'nski–Demianski  solutions can be neatly written in terms of some constraints among the parameters of the solutions~\cite{Klemm:2013eca}, as we shall review below.

		Using the formulas of~\cite{Gauntlett:2007ma, Gauntlett:2009zw} one can uplift locally  any supersymmetric
		solution of the four-dimensional theory to a supersymmetric solution of M-theory. Global aspects of the uplift to eleven dimensions are discussed in appendix~\ref{appendix:Uplift}.
		 As it happens in the context of black holes~\cite{Cabo-Bizet:2018ehj,Cassani:2019mms,Cassani:2021dwa},  in our solutions the gauge field and the metric can   take complex values. 
		Specifically, we will see that for all the solutions with a twist for the graviphoton  the metric will take real values, while for the solutions with anti-twist generically the metric can take complex values.
		In principle one could investigate systematically for which values of the parameters the metric becomes real, as was done in~\cite{Martelli:2012sz}. However, from the point of view of holography such requirement does not appear to be motivated and therefore we will not dwell on this presently. From now on we set $\ell=1$ without loss of generality.

		\subsection{Non-accelerating vs accelerating solutions}\label{subsect:Non-accelerating vs accelerating solutions}

		Our starting point is the local form of the  solutions 
		found by Pleba\'nski-Demianski (PD)~\cite{PLEBANSKI197698} almost fifty years ago. This generalized an earlier solution found independently by  Carter~\cite{Carter:1968ks} and Pleba\'nski~\cite{PLEBANSKI1975196} (that we will refer to as Carter-Pleba\'nski (CP) solution) by the addition of a parameter that, 
		in Lorentzian signature, may be interpreted as acceleration. Although we will be interested in the class of Euclidean (possibly complex) solutions, we will denote such parameter as $\acc$ and continue to refer to this 
		throughout the paper as \enquote{acceleration} parameter. 
		We will write the solutions directly in Euclidean signature, using coordinates essentially as in~\cite{Klemm:2013eca}, but we introduce the additional free 
		parameter\footnote{The form given in~\cite{Klemm:2013eca} is recovered setting  $\acc=1$. The parameter $k$ appearing there should be identified as $k=P^2-\alpha$. Several authors have studied slightly different parametrization of the PD solutions, see \eg~\cite{Griffiths:2005qp,Podolsky:2006px,Cassani:2021dwa,Podolsky:2022xxd,Astorino:2024bfl}} $\acc$ from the outset.
		We will keep track of $\acc$, distinguishing between the CP solutions ($\acc=0$) that we will refer to as \enquote{non-accelerating},  and the PD solutions 
		($\acc\neq 0$) that, by contrast, we will refer to as \enquote{accelerating}. We start presenting the more general PD solution, but soon we will set $\acc=0$, 
		where we will be able to perform all the relevant computations analytically in full generality, constructing also the Killing spinor $\varepsilon$ solving~\eqref{KSE}. Having studied in  detail the CP case, we will come back to the PD solutions later on, focussing on a sub-case, where we can make analytic progress.
		
		We start with the general solution solving the equations of motion while we will impose the supersymmetry conditions later. The metric  reads
		\begin{equation}\label{accelerating:metric}
			\begin{aligned}
				{\dd s_4 ^2}=\frac{1}{(1-\acc p q)^2}\biggl\{&(q^2-\omega^2p^2)\biggl[\frac{\dd q^2}{\Qfunc}+\frac{\dd p^2}{-\Pfunc}\biggr]
				\\
				&+\frac{1}{q^2 -\omega^2p^2}\biggl[\Qfunc(\dd\tau+\omega p^2\dd\sigma)^2-\Pfunc(\omega \dd\tau+q^2\dd\sigma)^2\biggr]\biggr\}\, ,
			\end{aligned}
		\end{equation}
		while the graviphoton is 
		\begin{equation}\label{accelerating_graviphoton}
			\begin{aligned}
				{A}=\frac{\omega p P-q Q}{q^2-\omega^2p^2}\dd\tau + p q \frac{q P-\omega p Q}{q^2-\omega^2 p^2}\dd\sigma\,,
			\end{aligned}
		\end{equation}
		where $\omega$ is a scaling parameter usually called \enquote{twist}~\cite{Podolsky:2022xxd}, which can be set to one. The metric functions are given by
		\begin{equation}{\label{accelerating:metricfunc}}
			\begin{aligned}
				\Pfunc&=\PP-2\acc M p^3+\acc^2\bigl[-Q^2+\alpha \omega^2 -P^2 (\omega^2-1)\bigr]p^4\,,
				\\
				\Qfunc&=\QQ-2\frac{\acc N}{\omega} q^3+\acc^2(-P^2+\alpha)q^4\,,
			\end{aligned}
		\end{equation}
		where $\PP$ and $\QQ$ are the \enquote{non-accelerating} functions (\cf\ the metric~\eqref{solution:metric})
		\begin{equation}\label{non-acc:metricfunc}
			\begin{aligned}
				\PP&= \omega^2 p^4+ E p^2-2 \frac{N}{\omega} p - P^2+\alpha\, ,
				\\
				\QQ&=q^4+E q^2-2 Mq -Q^2+\alpha\omega^2-P^2(\omega^2-1)\, .
			\end{aligned}
		\end{equation}
		Explicitly, we have
		\begin{equation}
			\begin{aligned}
				\Pfunc&=-P^2+\alpha -2 \frac{N}{\omega}p + E p^2 -2 \acc M p^3+\Bigl[\acc^2\bigl[\omega^2(-P^2+\alpha)-Q^2 + P^2\bigr]+\omega^2\Bigr]p^4\,,
				\\
				\Qfunc&=\bigl[\omega^2(-P^2+\alpha)-Q^2 + P^2\bigr]-2 M q + E q^2-2 \frac{\acc N}{\omega}q^3+ \bigl[\acc^2(-P^2+\alpha)+1\bigr]q^4\,.
			\end{aligned}
		\end{equation}
		The CP solution, as considered for example in~\cite{Alonso-Alberca:2000zeh,Martelli:2013aqa}, is obtained by setting $\acc=0$ and $\omega=1$. With these two conditions, the solution becomes highly symmetric in $(p,q)$, thus allowing for a general analysis.
		The parameters $(\acc,N,M,Q,P,\omega)$ can be loosely identified as the Euclidean counterpart of acceleration, NUT parameter, mass, electric and magnetic charge and rotation, respectively. However, the correct interpretation is 
		more subtle and in the Lorentzian set-up we refer for example  to~\cite{Podolsky:2022xxd,Astorino:2023ifg} for a more complete classification, where some interesting limits are also discussed. The interpretation of the various parameters in Euclidean signature is again complicated and case-dependent and will be a matter of discussion throughout the paper.

		We recall that the  solution has a self-dual Weyl tensor if and only if  $P=Q$ and $N=M$, and possesses a scaling symmetry
		\begin{equation}\label{scaling}
		\begin{aligned}
(q,p)\rightarrow& \lambda \,(q,p)\,,\quad (Q,P,E)\rightarrow \lambda ^2 \,(Q,P,E)\,,\quad (N,M)\rightarrow \lambda^3\, (N,M)\,,
				\\
				&\alpha\rightarrow \lambda^4\, \alpha\,,\quad \tau\rightarrow\lambda^{-1} \,\tau\,,\quad\acc\rightarrow\lambda^{-2}\,\acc\,,\quad \sigma\rightarrow\lambda^{-3}\,\sigma\,.
		\end{aligned}
		\end{equation}
		with $\lambda\in \RR \setminus \{0\}$.

		In this paper we will be interested in supersymmetric solutions with a holographic interpretation, namely we will require that they are asymptotically locally  hyperbolic, while 
		in the interior, we will allow local conical singularities.  As we shall see, this will lead to solutions  with the topology of a complex line bundle over a spindle and with  a lens space  $L(\lens,1)$ as conformal boundary\footnote{The boundary may be also an orbifold, \ie\ a branched lens space~\cite{Inglese:2023tyc} and in some cases we can also have $\lens=0$, corresponding to  $\spindle\times S^1$ topology.}. For these reasons, we will need a \enquote{radial} coordinate (that we take to be $q$) and an angular one (that will therefore be $p$). For this set-up we can take the ranges of definition 
		\begin{equation} \label{coordinate-range}
			\newqplus\le q\le \frac{1}{\acc p}\,,\quad \newpminus\le p\le \newpplus\,,
		\end{equation}
		where $q=1/(\acc p)$ is the location of the conformal boundary, $\newqplus$ is the largest real root of $\Qfunc$ and similarly $\newpplusminus$ are the largest real roots of $\Pfunc$. Notice that when $\acc=0$ the conformal boundary moves to infinity, so that the range of the coordinate $q$ is modified to $\newqplus\le q \le +\infty$. Moreover,
		without loss of generality we may take the following conditions in order to ensure the correct signature of the metric 
		\begin{equation}
			q^2-\omega^2p^2>0\,,\quad \Pfunc<0\,,\quad\Qfunc>0\,.
		\end{equation}
		We emphasise that while later we shall allow solutions to take complex values, the procedure we adopt is analogous to that used to study the thermodynamics of supersymmetric black holes, put forward in~\cite{Cabo-Bizet:2018ehj}. 
		Namely, we assume that in the general non-supersymmetric solution the metric and gauge field take real values, so that 
		we can require that for example the metric has correct Euclidean signature. After imposing supersymmetry we will find that in general it may not be possible to keep the metric real. However, we will require that in this more general set-up the Killing spinors have correct global properties and the bosonic fields are regular 
		 in the neighbourhood of fixed points\footnote{This is the Euclidean counterpart of requiring regularity near the tip of the cigar in the complexified black hole geometries~\cite{Cabo-Bizet:2018ehj}. }.
		For simplicity, and for future comparison with~\cite{Martelli:2013aqa}, we (partially) use the scaling symmetry~\eqref{scaling} to require $\newpminus+\newpplus\ge0$. We thus introduce the ordering
		\begin{equation}
			\newpminus \le p \le  \newpplus \le \newqplus\le q \,, \quad \newqplus ^2 -\newpplus ^2 \ge 0\,,\quad \newpplus^2-\newpminus^2\ge0\,.
		\end{equation} 
		From the signature conditions, we conclude that $\mathbf{Q}'(\newqplus)>0, \mathbf{P}'(\newpplus)>0$ and $\mathbf{P}'(\newpminus)<0$.
		
		We now switch gears and study supersymmetry. We do not attempt to solve~\eqref{KSE} in the general case $\acc\neq 0$, even if we will construct explicitly the Killing spinor $\varepsilon$ for $\acc=0$ in section~\ref{subsect:Killing spinors}, showing that the solution is $1/4$-BPS.
		The integrability condition $M_{\mu\nu}\varepsilon\equiv[\hat{D}_\mu,\hat{D}_\nu]\varepsilon=0$ of~\eqref{KSE} reduces, after some work, to
		\begin{equation}\label{intergrability}
			\begin{aligned}
				M_{\mu\nu}\varepsilon\equiv\bigg[\frac{1}{4} R_{\mu\nu}^{\,\,\,\,\,\,\,ab}\Gamma_{ab} + \frac{1}{2} \Gamma_{\mu\nu}^{} -& \ii F_{\mu\nu}^{} + \frac{\ii}{2}\Gamma_{ab}\nabla_{[\mu}F^{ab}\Gamma_{\nu]}^{} + \frac{\ii}{4}F_{ab}\Gamma_{[\mu}\Gamma^{ab}\Gamma_{\nu]}^{} 
				\\
				&- \frac{1}{16}F_{ab}F_{cd}[\Gamma^{ab}\Gamma_{\mu}, \Gamma^{cd}\Gamma_{\nu}]^{} + \frac{\ii}{4}F_{ab}\Gamma^{ab}\Gamma_{\mu\nu}^{}\bigg]\varepsilon=0\,.
			\end{aligned}
		\end{equation}
		We choose the coordinates to be ordered as $(p,\tau,\sigma,q)$, the vierbein to be
		\begin{equation}\label{frame}
			\begin{aligned}
				e^1 &=\frac{1}{(1-\acc p q)}\sqrt{\frac{q^2-\omega^2p^2}{-\Pfunc}}\dd p\,,\quad &&e^2=\frac{1}{(1-\acc p q)}\sqrt{\frac{-\Pfunc}{q^2-\omega^2p^2}}(\omega \dd\tau+q^2\dd\sigma)\,,
				\\
				e^3&=\frac{1}{(1-\acc p q)}\sqrt{\frac{\Qfunc}{q^2-\omega^2p^2}}(\dd\tau+\omega p^2\dd\sigma)\,,\quad &&e^4=\frac{1}{(1-\acc p q)}\sqrt{\frac{q^2-\omega^2p^2}{\Qfunc}}\dd q\,,
			\end{aligned}
		\end{equation}
		and adopt the following explicit representation of the Clifford algebra
		\begin{equation}\label{gammas}
			\Gamma_{a}=\sigma_1\otimes\sigma_a\,, a=1,\ldots,3\,,\quad \Gamma_{4}=-\sigma_2\otimes \text{Id}_2\,,\quad\Gamma_{\star}=\Gamma_{1}\Gamma_{2}\Gamma_{3}\Gamma_{4}\,,
		\end{equation}
		with $\sigma_a$ the standard Pauli matrices.
		It is then straightforward to show that $\det(M_{1,4})=0$ is equivalent to
		\begin{equation}\label{acceleration-susy}
			\begin{aligned}
				2(q^2+\omega^2 p^2)(N \Sigma_1-&\omega M \Sigma_2)+4 \omega (P^2-Q^2)( \acc q p+1 )(p \Sigma_1+q \Sigma_2)
				\\
				&-2 p q \omega (2 M \Sigma_1-2 \omega N \Sigma_2+\acc \omega \Sigma_3)-\omega^2 (\acc^2 p^2 q^2+1)\Sigma_3=0\,,
			\end{aligned}
		\end{equation}
		where
		\begin{equation}
			\begin{aligned}
				-\Sigma_1&= 2 Q \Omega \omega^2+\acc M \Pi\omega+2 \acc^2 N(P^2-Q^2)\Phi\,,
				\\
				-\Sigma_2&= 2 P \Omega \omega+\acc N \Pi+2 \acc^2 M(P^2-Q^2)(-P^2+\alpha)\omega\,,
				\\
				\Sigma_3&=\Pi^2-4(P^2-Q^2)^2\bigl[(-P^2+\alpha)\omega^2+P^2+\acc^2 (-P^2+\alpha)\Phi\bigr]\,,
			\end{aligned}
		\end{equation}
		and we defined
		\begin{equation}\label{Omega-Pi-Phi}
			\begin{aligned}
				\Omega= MP-NQ\,, \,\,\,\,\Pi=M^2-N^2-E(P^2-Q^2)\,,\,\,\,\, \Phi=\omega^2(-P^2+\alpha)-Q^2+P^2\,.
			\end{aligned}
		\end{equation}
		Moreover it can be verified that $\Sigma_3 =\sum_{i,j=1}^{2} a_{ij}\Sigma_{i}\Sigma_{j}$ for some coefficients $a_{ij}$, thus we only require $\Sigma_{1,2}\equiv 0$. 
		This system is entirely solved by the self-duality condition
		\begin{equation}
			M=N\,,\quad P=Q\,,
		\end{equation}
		or, in general, by fixing
		\begin{equation}\label{accelerating-susy-easy}
			\begin{aligned}
				\alpha&=\frac{-\omega^2(M P-N Q)^2 +\acc^2(P^2-Q^2)\bigl[N^2(P^2-Q^2)+(M^2-N^2)P^2\omega^2\bigr]}{\acc^2\omega^2(P^2-Q^2)(M^2-N^2)}\,,
				\\
				E&=\frac{-2(M P- N Q)(N P- M Q)\omega^2+\acc \omega (M^2-N^2)^2+2 \acc^2 M N (P^2-Q^2)^2}{\acc\omega(P^2-Q^2)(M^2-N^2)}\,.
			\end{aligned}
		\end{equation}
		If useful, we can also adopt the following (more complicated) expressions that are perfectly smooth in the limit $\acc\rightarrow 0$\footnote{Notice that in general the parameters in $\Sigma_i$ can be complex, and we have selected a branch for the square root.}
		\begin{equation}\label{accelerating-susy-complicated}
			\begin{aligned}
				M&=\frac{\omega NPQ+  \acc N(P^2-Q^2)\sqrt{P^2-       (P^2-\alpha)\Bigl[\omega^2-\acc^2\bigl[-P^2+Q^2+\omega^2(P^2-\alpha)\bigr]\Bigr]}}{\omega\bigl[P^2-\acc^2(P^2-Q^2)(P^2-\alpha)\bigr]}\,,
				\\
				E&=\frac{M^2-N^2}{P^2-Q^2}-\frac{2\acc \bigl[N P (-P^2+Q^2)+\omega^2(N P- M Q)(P^2-\alpha)\bigr]}{\omega(MP-N Q)}\,.
			\end{aligned}
		\end{equation}
		These are compatible with the results in~\cite{Klemm:2013eca} (with $\acc=1$) and we will refer to~\eqref{accelerating-susy-easy} or~\eqref{accelerating-susy-complicated} as the supersymmetry conditions. Even if we will not construct explicitly the spinor for the accelerating case, we believe that these conditions are (necessary and) sufficient for supersymmetry, analogously the CP case. In particular when $\acc=0$, $\Sigma_1\propto\Sigma_2$ and then $\Sigma_3$ becomes independent, and as such it is a genuine constraint. Then the supersymmetry conditions for the non-accelerating case become  \cite{Alonso-Alberca:2000zeh,Martelli:2011fu}

		\begin{equation}\label{non-acc:susy}
			\acc=0:\quad	M=\frac{NQ}{P}\,,\quad E=-\frac{N^2}{P^2}+2  \sqrt{\alpha \omega^2+P^2 (1-\omega^2)}\,,
		\end{equation}
		and they further simplify for $\omega=1$.

\subsection{Ambitoric structure} 
\label{subsect:Hermitian and symplectic properties of PD}

Let us briefly discuss some additional aspects of the local geometry associated to the (Euclidean) Pleba\'nski-Demianski solutions, which were first 
pointed out in~\cite{Apostolov:2013oza}. These features will not play a prominent role in our analysis, but it is nevertheless worth mentioning them,
 to establish a bridge with the mathematical literature.  

Firstly, one can show that there exist two integrable commuting almost complex structures $(J_\pm)_{\mu}{}^{\nu}$ compatible with the metric \eqref{accelerating:metric}, meaning that 
these are \emph{ambiHermitian}. Specifically, we introduce two almost symplectic structures
\begin{equation}
\omega_{\pm}=e^1\wedge e^2\pm e^3\wedge e^4\,,
\end{equation}
inducing two opposite orientations, namely such that  $\pm\frac{1}{2}\omega_{\pm}\wedge\omega_{\pm}=\mathrm{vol}_4$. 
It can be verified that $(J_\pm)_{\mu}{}^{\nu}=(\omega_{\pm})_{\mu\rho}g^{\rho\nu}$ define two (commuting) almost complex structures, since $(J_\pm)_{\mu}{}^{\rho}(J_\pm)_{\rho}{}^{\nu}=-\delta_{\mu}^{\nu}$. Moreover, the associated Nijenhuis tensors  vanish, namely
\begin{equation}
(N_{\pm})_{\mu\nu}{}^{\rho}\equiv -2 (J_{\pm})_{[\mu|}{}^{\sigma}\partial_{\sigma}(J_{\pm})_{|\nu]}{}^\rho+2 (J_{\pm})_{\sigma}{}^{\rho}\partial_{[\mu}(J_{\pm})_{\nu]}{}^{\sigma}=0\,,
\end{equation}
implying that $(J_{\pm},g)$ are Hermitian structures.  
Equivalently, it can be seen that  
the holomorphic (2,0)-forms
\begin{equation}
\Omega_{\pm}=(e^1+\ii e^2)\wedge (e^3\pm\ii e^4)\,,
\end{equation}
satisfy $\dd \Omega_{\pm}=\ii P_{\pm}\wedge\Omega_{\pm}$, where $P_\pm$ are connections one-forms on the anti-canonical bundles, from which the associated Ricci curvature two-forms are given by $\rho_\pm=\dd P_\pm$. Explicitly we find
\begin{equation}\label{Ricci potentials}
	\begin{aligned}
		 P_\pm =&2\left[\acc\frac{ \omega q\Pfunc\pm p\Qfunc}{1-\acc p q}+\frac{\omega\mathbf{P}'(p)\pm\mathbf{Q}'(q)}{4}\right]\frac{\dd\tau}{q^2-\omega^2 p^2} \\
		&+2\left[\acc\frac{ q^3\Pfunc\pm  \omega p^3\Qfunc}{1-\acc p q}+\frac{ q^2\mathbf{P}'(p)\pm\omega p^2\mathbf{Q}'(q)}{4}\right]\frac{\dd\sigma}{q^2-\omega^2 p^2}\,.
	\end{aligned}
\end{equation}

Although the triples $(J_{\pm},\omega_{\pm}, g)$ do not define Kähler structures, since $\omega_{\pm}$ are not closed,
 it can be shown that $\dd\omega_{\pm}=\eta_{\pm}\wedge\omega_{\pm}$, where $\eta_{\pm}$ are closed one-forms, implying that 
there exist two (oppositely oriented) conformally related K\"ahler structures, with K\"ahler two-forms given by  
\begin{equation}
\omega'_{\pm}=\left(\frac{1-\acc p q}{q\mp p\omega}\right)^2\omega_{\pm}\equiv \Xi_{\pm}^2\omega_{\pm}\,,\quad \dd\omega'_{\pm}=0\,,
\end{equation}
The rescaled triples  $(g'_{\pm},J'_{\pm},\omega'_{\pm})\equiv (\Xi_{\pm}^2g,J_{\pm},\Xi_{\pm}^2 \omega_{\pm})$ define an \emph{ambiK\"ahler structure}. Indeed, one can verify that the standard relation between the Ricci tensor and the Ricci form\footnote{Notice that this relation holds only for the rescaled metrics $g'_\pm$, which are K\"ahler, but not for the Hermitian one $g$. However, both $(P_\pm,P'_\pm)$ are connections on the anti-canonical bundles and their curvatures $(\rho_\pm,\rho'_\pm)$ are representative of the first Chern class, \ie\ $c_1 (T\mathcal{M}_4)=[\rho]/2\pi$.} $(\dd P' _\pm)_{\mu\nu}=(\rho'_\pm)_{\mu\nu} =(J'_\pm)^{\sigma}{}_{\mu} R'_{\sigma\nu}$, where $R'_{\mu\nu}$ is computed from the rescaled metrics $g'_{\pm}$.

 Since  both K\"ahler metrics are toric, with common torus action (as will be discussed in detail in section~\ref{subsect:Toric data}), these define an \emph{ambitoric structure}. This property of the local (Euclidean) Pleba\'nski-Demianski solutions was proven in~\cite{Apostolov:2013oza}. We note that 
the follow-up 
paper~\cite{2013arXiv1302.6979A} constructed global ambitoric structures on \emph{compact orbifolds}, focussing on 
extremal K\"ahler  and conformally Einstein metrics. As we shall discuss extensively, the examples that we will construct here concern non-compact orbifolds, solving (the supersymmetric sector of)  the  Maxwell-Einstein-$\Lambda$ equations, thus in particular the metrics  have constant scalar curvature. 

As follows from the discussion  around \eqref{coordinate-range}, the conformal factors $\Xi_{\pm}^2$ are positive semi-definite, vanishing precisely at the conformal boundary 
of the orbifolds. This implies that the rescaled K\"ahler metrics  $(g'_{\pm},J'_{\pm},\omega'_{\pm})$ are defined on the \enquote{conformal compactifications}, which are four-dimensional compact orbifolds with boundaries.

	\subsection{Notable old solutions}\label{subsect:A list of notable solutions}
	
	 In this section we summarize a number of  known asymptotically locally hyperbolic solutions sharing some features with 
the solutions that we will discuss presently. In fact, as we will show later, 
all of these old solutions can be recovered as special cases or limits  from our more general analysis. See figure~\ref{newoldsolutions:fig}.
\begin{figure}[h]
	\centering
	\begin{tikzpicture}

		\node (one) [rectangle, draw, rounded corners,align=center,thick,inner sep=7pt] {\small $1/4$-BPS NUTs};
		
		\node (three) [rectangle, draw, rounded corners, xshift=10cm, align=center,thick,inner sep=7pt]  {\small $1/2$-BPS NUTs};
		
		\node (four) [rectangle, draw, rounded corners,below of=one, yshift=-1cm, ,inner sep=7pt, thick,align=center] {\small $1/4$-BPS Bolts};
		
		\node (five) [rectangle, draw, rounded corners, xshift=5cm,inner sep=7pt, yshift=-1cm,thick,align=center]  {\small\small\textbf{NUTs \&  Bolts} \\
			\small $SU(2)\times U(1)$~\cite{Martelli:2012sz}};
		
		\node (six) [rectangle, draw,rounded corners, below of=three, yshift=-1cm ,inner sep=7pt,thick,align=center] {\small $1/2$-BPS Bolts};

		\node (seven) [rectangle,draw,rounded corners,below of=one, thick,yshift=-3cm,inner sep=7pt,align=center] {\small $1/4$-BPS Twist};
		
		\node (eight) [rectangle,draw,rounded corners,below of=five, yshift=-2cm,thick,inner sep=7pt,align=center] {\small\textbf{Spindle Bolts}\\\small $U(1)\times U(1)$ $[\text{\textcolor{green!50!black}{here}}]$};
		
		\node (nine) [rectangle,draw,rounded corners,below of=three, yshift=-3cm,inner sep=7pt, thick,align=center] {\small $1/4$-BPS Anti-Twist};

		\node (ten) [rectangle,draw,rounded corners,below of=one, yshift=-5cm,inner sep=7pt, thick,align=center] {\small $1/4$-BPS Type I};
		
		\node (eleven) [rectangle,draw,rounded corners,below of=five, yshift=-4cm,inner sep=7pt,thick,align=center] {\small\textbf{NUTs}\\\small $U(1)\times U(1)$~\cite{Martelli:2013aqa}};
		
		\node (twelve) [rectangle,draw,rounded corners,inner sep=7pt,below of=three, yshift=-5cm, thick,align=center] {\small  $1/4$-BPS Type II};

		\draw[-stealth,line width=1pt] (eight.north) -- (five.south);
		\draw[-stealth,line width=1pt] (eight.south) -- (eleven.north);
		\draw[-stealth,line width=01pt] (eight.west) -- (seven.east);
		\draw[-stealth,line width=01pt] (eight.east) -- (nine.west);
		
		\draw[-stealth,line width=01pt] (five.east) -- (three.west);
		\draw[-stealth,line width=01pt] (five.east) -- (six.west);
		
		\draw[-stealth,line width=01pt] (five.west) -- (one.east);
		\draw[-stealth,line width=01pt] (five.west) -- (four.east);
		
		\draw[-stealth,line width=01pt] (four.north) -- (one.south);
		\draw[-stealth,line width=01pt] (six.north) -- (three.south);
		
		\draw[-stealth,line width=01pt] (seven.north) -- (four.south);
		\draw[-stealth,line width=01pt] (nine.north) -- (six.south);
		
		\draw[-stealth,line width=01pt] (seven.south) -- (ten.north);
		\draw[-stealth,line width=01pt] (nine.south) -- (twelve.north);
		
		\draw[-stealth,line width=01pt] (eleven.east) -- (twelve.west);
		\draw[-stealth,line width=01pt] (eleven.west) -- (ten.east);
		
		\draw[-stealth, line width=1pt, bend left=45] (ten.west) to (one.west);
		\draw[-stealth, line width=1pt, bend right=45] (twelve.east) to (three.east);
	\end{tikzpicture}
	\caption{Summary of supersymmetric euclidean $\AdS_4$ solutions to $d=4$, $\mathcal{N}=2$ minimal gauged supergravity containing nuts and bolts. In the central column the symmetry and the bulk content are sketched, while the lateral columns show when a certain supersymmetric solution is contained in another one by some limiting procedure which will be explained in detail in section~\ref{subsect:Limits to the old solutions}.}
	\label{newoldsolutions:fig}
\end{figure}

Let us start by considering the $U(1)\times U(1)$-invariant two-parameter Euclidean solutions of~\cite{Martelli:2013aqa}. These are $1/4$-BPS solutions in  the CP family  ($\acc=0$) where the conformal boundary is a  squashed $S^3$ and have the topology of the ball, namely 
they contain a single nut  in the bulk.  They belong to the class of self-dual solutions~\cite{Farquet:2014kma}, but this aspect will not be particularly relevant for our discussion. 
According to the values of the parameters, there is a splitting into two distinct  families, referred to as Type I and Type II, which are neatly distinguished 
by the form of their  on-shell action, which read\footnote{The parameter $\beta$ is related to the parameter $Q$ that appears in~\eqref{non-acc:metricfunc} as $2Q= (\beta^2-1)/(\beta^2+1)$.} 
\begin{equation}\label{Type I and Type II actions}
\begin{aligned}
\text{Type I}:\quad S=\frac{\pi}{2 G_4}\,, \quad \text{Type II}:\quad S=\frac{\pi}{8 G_4}\biggl[\beta+\frac{1}{\beta}\biggr]^2 \, ,
\end{aligned}
\end{equation}
where $\beta$ is a combination of the two parameters of the solutions and it can also take  complex values. It is interesting to notice that in both families there are two free parameters, which however do not appear in the Type I action, while the Type II action depends on one combination ($\beta$). Accordingly, the supersymmetric Killing vectors, up to an irrelevant normalization constant, read\footnote{Here the vectors $\vec{\epsilon}$ are written in the basis $(-\partial_{\varphi_{1}},\partial_{\varphi_{2}})$ defined in~\cite{Martelli:2013aqa}.}
\begin{equation}
	\begin{aligned}
		\text{Type I}&:\quad \vec{\epsilon}\propto(1,-1)\,,\quad 
		\text{Type II}&:\quad \vec{\epsilon}\propto (\beta^2,1) \, .
	\end{aligned}
\end{equation}
Notice that these  are consistent with the general results of~\cite{Farquet:2014kma,BenettiGenolini:2016tsn}, which showed that for supersymmetric 
 self-dual solutions with the topology of the ball the only values of $\epsilon_2/\epsilon_1$ that correspond to non-singular solutions  are either $\epsilon_2/\epsilon_1=-1$ or $\epsilon_2/\epsilon_1>0$.

A similar behaviour is present in the one-parameter $SU(2)\times U(1)$-invariant solutions of~\cite{Martelli:2012sz}, where the boundary is  a  squashed lens space $L(\hat{\lens},1)=S^3/\ZZ_{\hat{\lens}}$, while the bulk may be either 
a nut or a bolt, which  is always a round $S^2$. For each type of bulk topology, there are again two distinct families, referred to as  $1/4$-BPS and $1/2$-BPS.
The bolt solutions are characterised by their  on-shell actions as well as the flux of the graviphoton field through the $S^2$ bolt, that read
\begin{equation}\label{bolt actions}
	\begin{aligned}
\frac{1}{4}\text{-BPS}&: \quad &&\frac{1}{2\pi}\int_{S^2} F=-1 +\branch\frac{\hat{\lens}}{2}\,,\quad &&S_{\text{bolt}}=\frac{\pi}{2 G_4}\biggl[1-\branch\frac{\hat{\lens}}{4}\biggr]\,,\quad\hat{t}\ge 2\,,
\\
\frac{1}{2}\text{-BPS}&:\quad &&\frac{1}{2\pi}\int_{S^2} F=-\branch\frac{\hat{\lens}}{2}\,,\quad &&S_{\text{bolt}}=\frac{\pi}{2 G_4}\biggl[1-\branch\frac{2\sqrt{4s^2-1}}{s \hat{\lens}}\Bigl[s^2-\frac{\hat{\lens}^2}{16}\Bigr]\biggr]\,,
\end{aligned}
\end{equation}
where $\hat\lens \in \mathbb{N}$, $\branch=\pm 1$ is a sign further distinguishing two  branches\footnote{These were denoted Bolt$_\pm$ in~\cite{Martelli:2012sz}.} of the solutions and $s$ is the squashing of the lens space boundary, 
which in~\cite{Martelli:2012sz} was assumed to take real values. Here $\hat{\lens}\ge 2$ for $\branch=-1$ and $\hat{\lens}\ge 3$ for $\branch=+1$~\cite{Martelli:2012sz}. Again we see that the on-shell action of the $1/4$-BPS bolt is completely fixed by the topology (\ie\ the value of $\hat{\lens}$), whilst the  free parameter $s$ appears in the $1/2$-BPS case. More details
about these solutions are given in appendix~\ref{appendix:Old-bolt-solutions}. 
We record here also the supersymmetric Killing vector $\vec{\epsilon}$ for these solutions, which is given (up to an irrelevant normalization constant) by~\cite{BenettiGenolini:2019jdz}
\begin{equation}\label{Killing vectors spherical bolts}
\begin{aligned}
\frac{1}{4}\text{-BPS}&: \quad \vec{\epsilon}\propto (1,0)\,,\quad
 \frac{1}{2}\text{-BPS}&: \quad \vec{\epsilon}\propto \Bigl(\frac{\hat{\lens}}{4s}-2s-\sqrt{4s^2-1},2s+\sqrt{4s^2-1}\Bigr)\,.
\end{aligned}
\end{equation}

For the nut solutions the on-shell actions read
	 \begin{equation}\label{nut actions}
	\begin{aligned}
		\frac{1}{4}\text{-BPS}&:\quad S_{\text{nut}}=\frac{\pi}{2 G_4}\,,\quad
		\frac{1}{2}\text{-BPS}&:\quad S_{\text{nut}}=  \frac{\pi}{2G_4}4s^2\,,
	\end{aligned}
\end{equation}
and again the free parameter $s$ appears only in $1/2$-BPS case. Now the supersymmetric Killing vectors read~\cite{Farquet:2014kma}
\begin{equation}
	\begin{aligned}
		\frac{1}{4}\text{-BPS}&: \quad \vec{\epsilon}\propto (1,-1)\,,\quad
		\frac{1}{2}\text{-BPS}&: \quad \vec{\epsilon}\propto \Bigl(\frac{1}{4s},-\frac{1}{4s}+2s+\sqrt{4s^2-1}\Bigr)\,.
	\end{aligned}
\end{equation} 
For completeness, let us also mention that the $1/4$-BPS  spherical Bolt solutions have been generalized in~\cite{Toldo:2017qsh} to allow bolts with the topology of a Riemann surface with genus $\mathrm{g}$. Correspondingly, the metrics have only $U(1)$ isometry and do not fall in the PD family. The reason why there is no Riemann surface analogue of the $1/2$-BPS spherical Bolt solutions will become clear in due course.

Finally we recall the accelerating  black hole solution, which has  boundary topology $\spindle\times S^1$~\cite{Ferrero:2020twa}. In this case the underlying metric is PD ($\acc\neq 0$) and it is the acceleration itself responsible for the conical singularities of the spindle. Allowing the parameters of the solution to take complex values one obtains
a \enquote{non-extremal} supersymmetric solution~\cite{Cassani:2021dwa} that in the bulk has a bolt with spindle topology\footnote{In the extremal limit the solution in the near horizon becomes AdS$_2\times \spindle$.}. We introduce some standard notation
\begin{equation}\label{integers of BH}
\mu=\frac{\singm+\singp}{\singm-\singp}\,,\quad G_4 Q_m =\frac{\singm-\singp}{4\singm\singp}\,,\quad \chi_{\spindle}=\frac{\singm+\singp}{\singm\singp}\,,
\end{equation}
in terms of which 
\begin{equation}\label{action of the BH}
\frac{1}{2\pi}\int_{\spindle} F=2 G_4 Q_m\,,\quad  S_{\pm}=\pm\frac{1}{2 \ii G_4 }\biggl[\frac{\varphi^2}{z}+(G_4 Q_m)^2 \,z\biggr]\,,\quad \varphi-\frac{\chi_\spindle}{4}z=\pm\ii\pi\,,
\end{equation}
where $z$ is a complicated function of the two  parameters of the solution.  This shows that the solution preserves supersymmetry via anti-twist through the spindle, while the on-shell action depends on the single complex parameter $z$. The supersymmetric Killing vector for this solution is~\cite{Cassani:2021dwa}
\begin{equation}
\vec{\epsilon}\propto (2 \pi, \ii z)\,.
\end{equation}Notice that the supersymmetric accelerating black hole reduces to the Kerr-Newman-AdS one~\cite{Klemm:2013eca} when the acceleration is turned off. This results in a spherical horizon with $\singm=\singp=1$, for which the on-shell  action has been computed in~\cite{Cassani:2021dwa} and coincides with~\eqref{action of the BH} for $Q_m=0$ and $\chi_{S^{2}}=2$.

In all the above cases, the on-shell action of the solutions computed through holographic renormalization coincides with the general expression obtained with localization~\cite{BenettiGenolini:2019jdz,BenettiGenolini:2023kxp}, see~\eqref{Genolini}.  This  has contributions from the fixed points of the supersymmetric Killing vector, which is  the  single nut for~\eqref{Type I and Type II actions} and~\eqref{nut actions}
  and the two poles of the sphere, or the spindle, for~\eqref{bolt actions} and~\eqref{action of the BH}.
  We will review the results of~\cite{BenettiGenolini:2019jdz,BenettiGenolini:2023kxp} in Section~\ref{subsubsect:Equivariant localization}.

 It is striking  that there is an asymmetry in the expressions for the various on-shell actions above. In particular, the actions for the Type I family and for the $1/4$-BPS nuts and bolts does not depend on any continuous parameter and can depend only on topological data of the underlying space (namely $\hat \lens$). Instead, for 
   the Type II family, the $1/2$-BPS nuts and bolts and  the black hole,  the action is always written in terms of a single parameter.  Throughout the paper we will recover these features  from our more general solutions containing a spindle in the bulk and a (possibly branched) lens space in the boundary. 
   In particular,  we will show  that this asymmetry in the behaviour of the on-shell actions is in 1-1 correspondence with the type  of supersymmetry on the spindle: if the twist is  realized, the action is fixed in terms of topological data of the underlying space, otherwise, if the anti-twist is realized,  the action depends on  a single free parameter.
   In turn, from the point of view of the boundary, this dichotomy between twist and anti-twist will be related to certain \enquote{flat connections} for the boundary gauge field, generalising the observations of~\cite{Martelli:2012sz}\footnote{For the $1/4$-BPS spherical Bolt solutions the role of the flat connections has been emphasised in the very recent paper \cite{Hong:2024uns}.}.  We conclude this section noticing that while  the complex deformation of the accelerating black hole is the only known solution that displays a spindle bolt,
 we will show that spindle bolts can also   arise  in the CP family with $\acc=0$.

	\section{Non-accelerating solutions }\label{sect:Non-accelerating solutions}
In this section we study the non-accelerating solution, \ie\ the Carter-Pleba\'nski case, which is obtained from the general solution~\eqref{accelerating:metric}-\eqref{non-acc:metricfunc} by setting $\acc=0$ and scaling away the twisting parameter, or $\omega=1$. We will show that it is possible to obtain, within the CP class of solutions, a bulk topology $\CC/\ZZ_{\labell}\hookrightarrow\mathcal{O}_{} (-\lens)\rightarrow\spindlesing$. For the ease of the reader we repeat here the metric
	\begin{equation}\label{solution:metric}
	\dd s_4 ^2=(q^2-p^2)\biggl[\frac{\dd q^2}{\QQ}+\frac{\dd p^2}{-\PP}\biggr]+\frac{1}{(q^2 -p^2)}\biggl[\QQ(\dd\tau+p^2\dd\sigma)^2-\PP(\dd\tau+q^2\dd\sigma)^2\biggr]\, ,
\end{equation}
and the graviphoton
\begin{equation}\label{solution:graviphoton}
	A=\frac{p P-q Q}{q^2-p^2}\dd\tau + p q \frac{q P-p Q}{q^2-p^2}\dd\sigma\,.
\end{equation}
The metric functions coming from~\eqref{non-acc:metricfunc} read
\begin{equation}\label{metricfunc}
	\begin{split}
		\PP&= p^4+ E p^2-2 N p - P^2+\alpha\, ,
		\\
		\QQ&=q^4+E q^2-2 Mq -Q^2+\alpha\, ,
	\end{split}
\end{equation}
	where the high symmetry in $(p,q)$ is now evident. In fact, it is this symmetry that will allow us to perform a general analysis. The main difference with respect to the accelerating case is that the conformal boundary is now located at infinity, so that
	\begin{equation}
q_+\le q \le +\infty\,,\quad p_-\le p \le p_+\,.
	\end{equation}
In appendix~\ref{appendix:A-spindle-Calabi-Yau-metric} we show that upon sending $\Lambda\rightarrow0$ and setting $A=0$, we can obtain a Ricci-flat version of this set-up. In this case we will show that is possible again to obtain a bulk topology $\mathcal{O}(-\lens)\rightarrow\spindle$ but with a Calabi-Yau metric.
	Firstly, we continue considering local aspects and we construct the Killing spinor $\varepsilon$ solving~\eqref{KSE}.  Global issues will be addressed later on.

	\subsection{Local Killing spinors}\label{subsect:Killing spinors}
	Recall that the supersymmetry conditions~\eqref{non-acc:susy} for the Euclidean CP solution with $\omega=1$ read\footnote{Recall that, for simplicity, we have chosen a specific branch for the square root $\sqrt{\alpha}$.}
		\begin{equation}{\label{susy}}
		MP=NQ\,,\quad E=-\frac{N^2}{P^2}+2  \sqrt{\alpha}\,.
	\end{equation}
	In the following we will assume $N\neq 0$. In this case~\eqref{susy} are the correct BPS conditions. When instead $N=0$, from the first equation in~\eqref{susy} wee see that it is possible to choose $M=0$ or $P=0$ independently, and the analysis splits into two sub-cases. The $N=0$ situation will be discussed separately, in section~\ref{subsubsect: N=0}. In the following we will construct the spinor in the general case, with all the parameters different from zero.
	
	The solution presented in~\cite{Martelli:2013aqa} corresponds to the sub-case in which $P=Q$ and $N=M$ by regularity of the metric, so that $\Omega=0$ identically (see~\eqref{Omega-Pi-Phi}). As we will discuss, we are in the general situation for which $\Omega=0$ is a constraint for having a supersymmetric solution. Using these expressions to eliminate $M$ and $E$ in favour of the other parameters, the metric functions~\eqref{metricfunc} factorize as 
	\begin{equation}\label{susy-func-dec}
		P^2 \PP = \Pm \Pp\,,\quad P^2 \QQ= \Qm \Qp\,,
	\end{equation}
	where
	\begin{equation}\label{susy-func}
		\mathcal{P}_{\pm}(p)=P p^2 \mp N p \mp P^2+ P \sqrt\alpha\,,\quad\mathcal{Q}_{\pm}(q)=P q^2\mp N q \mp PQ + P  \sqrt\alpha\,.
	\end{equation}
	Writing the four-dimensional Dirac spinor in terms of its chiral components $\varepsilon_\pm$ as
	\begin{equation}
		\varepsilon=\begin{pmatrix}
			\varepsilon_+ \\
			\varepsilon_- \\
		\end{pmatrix}
	\end{equation}
	and employing~\eqref{gammas}, the integrability condition~\eqref{intergrability} can be used to relate $\varepsilon_-$ to $\varepsilon_+$. In particular we find the relation 
	\begin{equation}
		(X^+ + Y^+ )\varepsilon_+ +(X^-+ Y^-)\varepsilon_-=0\,,
	\end{equation}
	where $X^\pm$ and $Y^\pm$ are $2\times 2$ matrices with non-vanishing elements
	\begin{equation}
		\begin{aligned}
			X^{\pm}_{1,2}=&\pm\ii\frac{2\big[\PP-\QQ\big]+(p\pm q)\Big\{2(p\mp q)\big[(q\pm p)^2-P\mp Q]-\dPP\pm\mathcal{Q}'(q)\Big\}}{4(p\pm q)^2 \sqrt{-\PP}\sqrt{\QQ}}\,,
			\\
			X^{\pm}_{2,1}=&\pm\ii\frac{2\big[\PP-\QQ\big]+(p\pm q)\Big\{2(p\mp q)\big[(q\pm p)^2+P\mp Q]-\dPP\pm\mathcal{Q}'(q)\Big\}}{4(p\pm q)^2 \sqrt{-\PP}\sqrt{\QQ}}\,,
		\end{aligned}
	\end{equation}
	and
	\begin{equation}
		\begin{aligned}
			Y^{\pm}=\frac{(p\mp q)(P\pm Q)}{2(q\pm p)^2\sqrt{q^2-p^2}}
			\begin{pmatrix}
				\mp\ii/\sqrt{\QQ}&& \pm 1/\sqrt{-\PP} \\
				\mp 1/\sqrt{-\PP}&& \pm\ii/\sqrt{\QQ} \\
			\end{pmatrix}
		\end{aligned}\,.
	\end{equation}
	Using these relations it is easy to construct the Killing spinor, that is
	\begin{equation}\label{four-spinor}
		\begin{aligned}
			\varepsilon=c\begin{pmatrix}
				-\ii	\sqrt{\Pm}\sqrt{\Qm}/\sqrt{q+p}\\
				-\ii\sqrt{\Pp}\sqrt{\Qp}/\sqrt{q+p}\\
				\langle P\rangle  \sqrt{\Pm}\sqrt{\Qp}/\sqrt{q-p}\\
				\langle P\rangle  \sqrt{\Pp}\sqrt{\Qm}/\sqrt{q-p} \\
			\end{pmatrix}\ee^{\ii \frac{N}{2P}(\tau+ \sqrt{\alpha}\sigma)}\,,
		\end{aligned}
\end{equation}
where $\langle P\rangle\equiv P/\sqrt{P^2}\in\CC\setminus \{0\}$ and $c\in \CC$ a normalization constant.
Notice that it takes remarkably the same (implicit) form as in~\cite{Martelli:2013aqa}, and is then the most general form of the Killing spinor for the Euclidean CP solution even when $M\neq N$ and $P\neq Q$.

Before tackling the regularity of the metric and the spinor, let us express the parameters of the solution~\eqref{metricfunc} in terms of the real roots $p_\pm$ and $q_+$. For $p_+$ to be a root of $\PP$, it must be a zero of one between $\mathcal{P}_{+}(p)$ or $\mathcal{P}_{-}(p)$ in~\eqref{susy-func}. We then introduce a sign $\signpp=\pm1$ such that $\mathcal{P}_{\signpp}(p_+)=0$. Similarly for $p_-$ we impose $\mathcal{P}_{\signpm}(p_-)=0$, with $\signpm=\pm1$. The combinations $ \mathcal{P}_{\signpp}(p_+)\signpp- \mathcal{P}_{\signpm}(p_-)\signpm=0$ and $ p_- \mathcal{P}_{\signpp}(p_+)\signpp- p_+ \mathcal{P}_{\signpm}(p_-)\signpm=0$ can be inverted to give\footnote{In writing the following formulas we assume $p_++p_-> 0$. The $p_+=-p_-$ case is treated at the end of section~\ref{subsubsect: Regularity in the boundary}. }
\begin{equation}\label{N-alpha}
\begin{split}
	N&
	=\signpp P\frac{(p_+ + p_-)( p_++\twist p_-)+\signpp P (\twist - 1)}{p_+ + p_-}\, ,
	\\
	\sqrt{\alpha}&
	=\twist\frac{\signpp P(p_+ - p_- )( p_+ + \twist p_-)+p_+ p_- (p_+ ^2 - p_- ^2)}{p_+ ^2 - p_- ^2}\, .
\end{split}
\end{equation}
where we introduced $\twist\equiv\signpp\signpm=\pm 1$.
Similarly, the condition $\mathcal{Q}_{\signqp}(q_+)=0$ with $\signqp=\pm1$, supplemented with~\eqref{N-alpha}, is equivalent to
\begin{equation}\label{Q}
Q
=\lambda\frac{(p_+ + p_-)\big[q_+ ^2 +\twist p_+ p_- - q_+ (p_+ + \twist p_-)\branch\big]+\signpp P\big[\twist p_+ + p_-+\branch\twist q_+ (\twist-1)\big]}{p_+ + p_-}\,,
\end{equation}
where similarly we introduced $\branch\equiv\signpp \signqp=\pm 1$. It is evident from the previous expressions that the system is distinguished by the value of $\twist$. Indeed, in the next sections we will see that $\twist=\pm 1$ correspond to the twist or anti-twist realization of the supersymmetry for the spindle~\cite{Ferrero:2021etw}. The sign $\branch$ instead leads to different branches that, as we will see later, generalize those present in~\cite{Martelli:2012sz,Martelli:2013aqa}.

For each value of the signs $\signpp$, $\signpm$ and $\signqp$, it holds 
\begin{equation}\label{constraint}
\begin{aligned}
	\dPPp &=\frac{2\,(p_+ - p_-)\bigl[N+ p_+ (p_+ + p_-)^2\bigr]}{p_+ + p_-}\, ,
\end{aligned}
\end{equation}
where recall that $p_++p_->0$ from the scaling symmetry. A necessary and sufficient condition for this to be positive (as required by the signature of the metric) is $N\in \RR_+$\footnote{$N=0$ is also possible, but we will continue in the general case $N>0$ for the moment. See section~\ref{subsubsect: N=0} for the reality analysis when $N=0$.}. When $\twist=1$, $0<N=(p_++p_-) \signpp P $ from~\eqref{N-alpha}, and we conclude that
\begin{equation}\label{constraint-twist}
\twist=1:\quad P\in \RR\,,\quad \signpp=\sign P\,.
\end{equation}
In the case $\twist=-1$, from
\begin{equation}
\twist=-1:\quad  N=\frac{\signpp P}{p_++p_-}\bigl[(p_+^2-p_-^2)-2 \signpp P\bigr]\,,
\end{equation}
we see that in order for $N$ to be real and positive it is necessary that
\begin{equation}\label{constraint-tantiwist}
\twist=-1:\quad 
\begin{aligned}
	P\in\CC\implies& \text{Re}(P)=\frac{\eta(p_+^2-p_-^2)}{4}\,, \quad\signpp=\sign\text{Re}(P)\,,
	\\
	P\in\RR\implies& 0<\signpp P<\frac{(p_+^2-p_-^2)}{2}\,, \quad\signpp=\sign P\,.
\end{aligned}
\end{equation}
When $P\in\RR$, which can happen for both $\twist=\pm1$, all the parameters of the solution are real from~\eqref{N-alpha} and~\eqref{Q}. Then also the metric and the gauge field are real. However, only for $\twist=-1$, $P$ can be complex with its real part fixed by~\eqref{constraint-tantiwist}. However, in this case one can verify that the parameter $E$ and the combination $-P^2+\alpha$ appearing in the structure function $\PP$ in~\eqref{metricfunc} are nevertheless real. Instead $M$ and $-Q^2+\alpha$ are complex, unless the real root $q_+$ satisfies the further condition $q_+=p_+$ with $\branch=1$.
Finally, $Q\in\CC$ iff $P\in\CC$.

Summarizing, for $\twist=1$ the solution is real, whilst for $\twist=-1$ the function $\PP$ is always real but the metric remains complex except if $q_+=p_+$. This sub-case corresponds to the solution presented in~\cite{Martelli:2013aqa}, where the metric was always real but the gauge field can be complex if $P$ (and therefore $Q$) is complex. Reality properties of the solution when $N=0$ will be addressed throughout section~\ref{subsubsect: N=0}.

\subsection{Global analysis}\label{subsect:Regularity}
Having established the local form of the solutions of interest, we now examine their global properties. Following closely~\cite{Martelli:2007pv,Martelli:2013aqa}, we divide the regularity analysis into two steps: firstly we require the boundary to be a lens space $L(\lens,1)$ (eventually branched~\cite{Inglese:2023tyc}) and then we consider the conditions under which the bulk topology is that of $\mathcal{M}_4=\mathcal{O}(-\lens)\rightarrow \spindle$, a (complex) line bundle over the spindle. Finally, we will show the regularity of the Killing spinor~\eqref{four-spinor}.

\subsubsection{Metric at the boundary}\label{subsubsect: Regularity in the boundary}
We now study the regularity of the boundary metric $\dd s^2_{b}$, obtained as the leading order in ${q\rightarrow +\infty}$. In~\cite{Martelli:2013aqa}, the boundary was topologically an $S^3$, while more generally here we require it to be a (branched) lens space. As it is well known~\cite{Closset:2018ghr, Ferrero:2020twa,Ferrero:2021etw, Inglese:2023tyc}, the metric on the very same $S^3$ (or orbifold thereof) can be written in different ways, more precisely as a standard Hopf fibration over $S^2$, or as a weighted $U(1)$ fibration over the spindle.
For these reasons, it is convenient to adopt coordinates in which the boundary is already written as a fibration over $\spindlesing$, differently from~\cite{Martelli:2013aqa}.
To this end, it proves useful to write the solution~\eqref{solution:metric} in coordinates adapted to the degenerate Killing vectors at the end-points of the coordinate range, namely $p_\pm$ and $q_+$. A generic Killing vector $K=a_1\partial_\tau + a_2 \partial_\sigma$, in the full four-dimensional metric~\eqref{solution:metric}, has norm
\begin{equation}\label{Killing-norm}
K_\mu K^\mu=\frac{\big[-(a_1+a_2 q^2)^2\PP+(a_1+a_2 p^2)^2\QQ\big]}{q^2-p^2}\,,
\end{equation}
and since it is the sum of positive quantities, they should be nil separately. At this point it is worth emphasizing that, in order to describe correctly global issues, we need to introduce patches on $\mathcal{M}_4$. In particular we consider two patches $U_\pm$ which include $p_\pm$ respectively, so that $\mathcal{M}_4=U_+ \cup U_-$. A generic Killing vector is more precisely written as $K=a_1^{\pm}\partial_{\tau^{\pm}} + a_2^{\pm} \partial_{\sigma^{\pm}}$, where $(\tau^{\pm},\sigma^{\pm})$ are coordinates on $U_\pm$. We then take $K_{q_{+}}=\parqppm(q_+ ^2 \partial_{\tau^{\pm}}-\partial_{\sigma^{\pm}})$ and $K_{\pm}=K_{p_{\pm}}=\parppm(-p_\pm ^2 \partial_{\tau^{\pm}}+\partial_{\sigma^{\pm}})$, which degenerate at $q_+$ and at $p_\pm$, respectively\footnote{Notice that if we take $a_1=0$, then or $a_2=0$ or $q_+=p_\pm$, which are both degenerate cases. Similarly it happens if we start with $a_2=0$, so we continue assuming $a_i\neq0$.}. These Killing vectors can be conveniently written as $K_{q_+}=\partial_{\angleone^{\pm}}$ and $K_\pm=\partial_{\angletwo^{\pm}}$, with again $(\angleone^{\pm},\angletwo^{\pm})$ on $U_\pm$, if we change coordinates as
\begin{equation}{\label{change:tau-theta}}
	U_\pm:\quad	\tau^{\pm} = \parqppm q_+ ^2\angleone^{\pm} -\parppm p_\pm^2\angletwo^{\pm}\,,\quad \sigma^{\pm}=-\parqppm \angleone^{\pm}+\parppm\angletwo^{\pm}\,.
\end{equation}
Since the Jacobian of~\eqref{change:tau-theta} is $q_+^2-p_\pm^2$, the periodicities are related simply as
\begin{equation}\label{period}
	\Delta\tau^{+} \Delta\sigma^{+}=\Delta\tau ^{-}\Delta\sigma^{-}
	=(q_+^2-p_+^2) \Delta\angleone^{+} \Delta\angletwo^{+}=(q_+^2-p_-^2) \Delta\angleone^{-} \Delta\angletwo^{-}\,.
\end{equation}
In these new coordinates $(q,p,\angleone^{\pm},\angletwo^{\pm})$, the four-dimensional metric reads
\begin{equation}\label{metric-for-spindle}
	\begin{split}
		U_{\pm}:\quad \dd s^2 _4=&\frac{q^2-p^2}{-\PP}\dd p^2-  \frac{(q^2-p_\pm^2)^2\PP}{q^2-p^2} \dd( \angletwo^{\pm})^2+\frac{q^2-p^2}{\QQ}\dd q^2
		\\
		&+(q^2-q_+ ^2)\frac{ \dd(\angleone^{\pm}) ^2 (q^2-q_+^2)-2 \parqppm\parppm( q^2-p_\pm^2)\dd\angleone^{\pm} \dd\angletwo^{\pm}}{p^2-q^2}\PP
		\\
		& -\frac{\big[\parppm(p^2-p_\pm^2)\dd\angletwo^{\pm}+\parqppm(q_+ ^2-p^2)\dd\angleone^{\pm} \big]^2}{p^2-q^2}\QQ\,.
	\end{split}
\end{equation}

We are now in the position to analyze the boundary of~\eqref{metric-for-spindle}, which is easily found to be
\begin{equation}\label{boundary-metric}
	\begin{split}
		U_\pm:\quad			\dd s^2 _b =&\frac{\dd p^2}{-\PP}-\frac{(q_+ ^2-p_\pm^2)^2 \PP}{(q_+^2-p ^2)^2-\PP}\dd(\angletwo^{\pm})^2
		\\
		&+[(q_+^2-p ^2)^2-\PP]\biggl\{\parqppm\dd\angleone^{\pm} + \frac{(q_+ ^2 - p^2)(p^2-p_\pm^2)+ \PP}{\left[(q_+^2-p ^2)^2-\PP\right]}\parppm\dd\angletwo^{\pm}\biggr\}^2\,.
	\end{split}
\end{equation}
 Notice that the boundary metric~\eqref{boundary-metric} contains only the function $\PP$ and is therefore real as a consequence of the discussion at the end of the previous section.
We now zoom in near the zeros of $\PP$ introducing a new coordinate $R_{\pm}$
\begin{equation}\label{bound-limit}
	U_\pm :\quad	p=p_{\pm}- \frac{\dPPpm}{4} R_{\pm}^2\,,
\end{equation}
for which the boundary metric, at the leading order in $R_\pm$, reads
\begin{equation}\label{boundary near poles}
	U_\pm:\quad	\dd s^2_b \underset{p\rightarrow p_\pm}{\simeq}\dd R_{\pm}^2+ \frac{\dPPpm^2}{4}R_{\pm}^2 \dd (\angletwo^{\pm}) ^2+ (q_+ ^2 - p_\pm ^2)^2\dd(\angleone^{\pm})^2\,.
\end{equation}
Provided 
\begin{equation}{\label{spindle-conditions-boundary}}
	U_{\pm}:\quad	\frac{|\dPPpm| }{2 }\Delta\angletwo^{\pm} =\frac{2\pi}{\singpm} \,,
\end{equation}
the base of this fibration in~\eqref{boundary-metric} can be made topologically a spindle $\spindlesing$, for some co-prime positive integers $\singpm$, with $\singm>\singp$.  Recall also that $\dPPm<0$, for which we take the absolute value in~\eqref{spindle-conditions-boundary}. This condition is equivalent to
\begin{equation}{\label{singm-singp}}
	\singm>\singp\implies(q_+^2-p_+^2)\dPPm+(q_+^2-p_-^2)\dPPp>0\,.
\end{equation}	
We will refer to this spindle as $\spindlesinginf$, since its shape is different for different values of $q$, but the topology will remain the same, \ie\ the volume is a function of $q$, but the conical deficits remain the same along the flow. 
We stress that in the boundary the spindle is a topologically trivial cycle, but nevertheless we will see that it plays an important role.
Indeed, with reference to the two dimensional base
\begin{equation}\label{bound-base}
	U_\pm:\quad	\dd s^2 _{\spindle_{\infty}} =\frac{\dd p^2}{-\PP}-\frac{(q_+ ^2-p_\pm^2)^2 \PP}{(q_+^2-p ^2)^2-\PP}\dd(\angletwo^{\pm})^2\,,
\end{equation}
it can be seen that
\begin{equation}
	\sqrt{g_{\spindle_{\infty}}}R_{\spindle_{\infty}}=(q_+ ^2-p_\pm^2)\partial_p\biggl\{(q_+ ^2-p^2)\Bigl[4\PP p+ (q_+ ^2-p^2)\dPP\Bigr]\Bigl[(q_+ ^2 - p^2)^2-\PP\Bigr]^{-3/2}\biggr\}\,,
\end{equation}
and as a consequence
\begin{equation}
	U_{\pm}\,:\quad\frac{1}{4\pi}\int_{\spindle_{\infty}} \dd p \,\dd \angletwo^{\pm}\, \sqrt{g_{\spindle_{\infty}}}R_{\spindle_{\infty}}=\frac{q_+^2-p_{\pm}^2}{4\pi}\biggl|  \frac{ \dPP}{q_+^2-p^2} \biggr|^{p_{+}}_{p_{-}}\Delta\angletwo^{\pm}\,.
\end{equation}
This computation must agree in the two patches $U_\pm$, and using~\eqref{period} we see that necessarily $\Delta\angleone^{+}=\Delta\angleone^{-}$.
In turn, this implies
\begin{equation}
\frac{1}{4\pi}\int_{\spindle_{\infty}} \dd p \,\dd \angletwo^{\pm}\, \sqrt{g_{\spindle_{\infty}}}R_{\spindle_{\infty}} =\frac{\singm+\singp}{\singm\singp}\equiv\chi_{\spindle}\,,
\end{equation}
where $\chi_{\spindle}$ is the (orbifold) Euler characteristic of the spindle. Then, with~\eqref{spindle-conditions-boundary} and
\begin{equation}\label{period-one}
	\Delta\angleone\equiv \Delta\angleone^{+}=\Delta\angleone^{-}\,,
\end{equation}
the metric~\eqref{bound-base} represents indeed $\spindlesinginf$. With these conditions imposed, it is also useful to introduce
\begin{equation}\label{period-two}
	\Delta\equiv (q_+^2-p_+^2) \Delta\angletwo^{+}=(q_+^2-p_-^2) \Delta\angletwo^{-}\,,
\end{equation}
in terms of which~\eqref{spindle-conditions-boundary} is rewritten as
\begin{equation}\label{quantization_cond_spindle}
 \pm\frac{\dPPpm }{2(q_+^2-p_\pm^2) }\Delta=\frac{2\pi}{\singpm}\,.
\end{equation}
Moreover we require that the fibration is well-defined in the orbifold sense~\cite{Ferrero:2021etw}, that is
\begin{equation}\label{lens-condition}
	\frac{1}{2\pi}\int_{\spindle_{\infty}} \frac{2\pi}{\Delta\angleone}\dd\biggl\{\frac{\parppm}{\parqp}\frac{(q_+ ^2 - p^2)(p^2-p_{\pm}^2)+ \PP}{\left[(q_+^2-p ^2)^2-\PP\right]}\dd\angletwo^{\pm}\biggr\}\equiv\parqp\parppm \frac{ \lens}{\singm\singp}\,,
\end{equation}
for $t\in\NN$, which is equivalent to
\begin{equation}\label{bound-lens}
	U_\pm:\quad	\lens=\singm\singp \frac{p_+^2-p_-^2}{q_+^2-p_\mp^2}\frac{\Delta\angletwo^{\pm}}{\Delta\angleone}\implies \lens=\singm\singp\frac{p_+^2-p_-^2}{(q_+^2-p_-^2)(q_+^2-p_+^2)}\frac{\Delta}{\Delta\angleone}\,.
\end{equation}
Notice that combining~\eqref{period},~\eqref{period-one},~\eqref{period-two},~\eqref{quantization_cond_spindle} and~\eqref{bound-lens} we can write 
\begin{equation}\label{period-product}
	\Delta\tau^{\pm}\Delta\sigma^{\pm}=\Delta\angleone\Delta =\frac{\singm\singp}{\lens}\Delta^2\frac{p_+^2-p_-^2}{(q_+^2-p_-^2)(p_+^2-p_-^2)}=\frac{(4\pi)^2}{\lens}\frac{p_+^2-p_-^2}{-\dPPm\dPPp}\,,
\end{equation}
where the last two expressions are valid only for $\lens\neq 0$.

To summarize the analysis so far, we have shown that by choosing suitably the parameters of the local solutions, the boundary is a circle orbifold bundle over the spindle $\spindlesing$, namely
the bundle $S^1 \hookrightarrow {\mathcal M}_3 \rightarrow  \spindlesing$,  which is a \emph{Seifert orbifold}.  It is in fact known from general results on Seifert orbifolds~\cite{2016arXiv160806844G,Closset:2018ghr} that topologically  ${\mathcal M}_3$  must be a lens space, possibly with residual orbifold 
singularities. More precisely ${\mathcal M}_3$ is a smooth lens space $L(\lens,1)=S^3/\ZZ_\lens$ if and only if $\gcd(\lens,\singpm)=1$. When $\lens$ and $\singpm$ have a common divisor there are various sub-cases that we do not discuss in detail now, but the simplest example in this context is perhaps when $\lens=\overline{\lens}\, \singp\,\singm$. This symmetric situation corresponds to an \enquote{orbifold lens space} $\mathbb{L}_{[\singm,\singp]}(\lens,1)$, also referred to as branched lens space, which consists in a three-sphere with orbifold poles of local topology $\CC/\ZZ_{\singpm}$ and with a further global $\ZZ_\lens$ quotient. We will demonstrate explicitly these facts in section~\ref{subsect: new Seifert}, but see also~\cite{Faedo:2024upq} for a similar discussion in four dimensions. For simplicity, we assume $\gcd(\lens,\singpm)=1$ for the time being.

When $\lens=0$, the boundary is generically a direct product $\spindle\times S^1$. Since $\lens\propto (p_++p_-)$, we can study separately the case where $p_+=-p_-\equiv r$. Indeed, the discussion around~\eqref{N-alpha} assumed $p_++p_->0$. We can repeat now that analysis, requiring \emph{a priori} $\lens=0$. Imposing again $ \mathcal{P}_{\signpp}(p_+)\signpp- \mathcal{P}_{\signpm}(p_-)\signpm=0$, $ p_- \mathcal{P}_{\signpp}(p_+)\signpp- p_+ \mathcal{P}_{\signpm}(p_-)\signpm=0$ and $\mathcal{Q}_{\signqp}(q_+)=0$ we find
\begin{equation}\label{t=0 implies N=0}
\begin{aligned}
\twist=+1&:\quad N=0\,,\quad  \sqrt{\alpha}=-r^2+\signpp P\,,\quad Q=\signqp(q_+^2-r^2+\signpp P)\,,
\\
\twist=-1&:\quad N=0\,,\quad P=0\,.
\end{aligned}
\end{equation}
with $\twist=\signpp\signpm=\pm 1$ as before. We see that $\lens=0\implies N=0$,  but the vice-versa is not true. The case $\lens=0$ is thus excluded from the generic analysis, and will be treated in all details in section~\ref{subsubsect: N=0}.

Let us conclude this section by working out explicitly the coordinate change between $U_+$ and $U_-$ from~\eqref{change:tau-theta}. To this end it is useful to write the Killing vectors expressed in the two patches
\begin{equation}\label{Killing-vectors-non-toric}
	\begin{aligned}
		K_{p_{+}}&=\parpp\biggl[-\parqp\frac{p_+^2-p_-^2}{q_+^2-p_-^2}\partial_{\angleone^{-}}+\parpm\frac{q_+^2-p_+^2}{q_+^2-p_-^2}\partial_{\angletwo^{-}}\biggr]=\partial_{\angletwo^{+}}\,,
		\\
		K_{q_{+}}&=\partial_{\angleone^{-}}=\partial_{\angleone^{+}}\,,
	\\
	K_{p_{-}}&=\partial_{\angletwo^{-}}=\parpm\biggl[\parqp\frac{p_+^2-p_-^2}{q_+^2-p_+^2}\partial_{\angleone^{+}}+\parpp\frac{q_+^2-p_-^2}{q_+^2-p_+^2}\partial_{\angletwo^{+}}\biggr]\,,
	\\
	\end{aligned}
\end{equation}
which follow from
\begin{equation}\label{coordinate-in-patches}
	\parqp\angleone^+=\parqp\angleone^{-}+\parpm \frac{p_+^2-p_-^2}{q_+^2-p_+^2}\angletwo^{-}\,,\quad
	\parpp\angletwo^{+}=\parpm \frac{q_+^2-p_-^2}{q_+^2-p_+^2}\angletwo^{-}\,.
\end{equation}
These can be rewritten introducing $2\pi$-periodic coordinates
\begin{equation}\label{equal-periodic}
	\periodicone^{\pm}=\frac{2\pi}{\Delta\angleone^{}}\angleone^{\pm}\,,\quad\periodictwo^{\pm}=\frac{2\pi}{\Delta\angletwo^{\pm}}\angletwo^{\pm}\,,
\end{equation}
and using~\eqref{period-one},~\eqref{period-two} and~\eqref{bound-lens} to obtain
\begin{equation}\label{change:periodic-patch}
	\parqp\periodicone^+=\parqp\periodicone^{-}+ \frac{\lens}{\singm\singp}\parpm\periodictwo^{-}\,,\quad
	\parpp\periodictwo^{+}=\parpm\periodictwo^{-}\,.
\end{equation}
Thus we can define
\begin{equation}\label{periodicangletwodef}
	\parp\,\periodictwo\equiv\parpp\periodictwo^{+}=\parpm\periodictwo^{-}\,,
\end{equation}
which is identified in the gluing
and use it independently of the patch $U_\pm$.
Notice that when $p_-=-p_+$, \ie\ when $\lens=0$, we can always choose $\angleone^+=\angleone^-\equiv\angleone$ and $\angletwo^+=\angletwo^-\equiv\angletwo$ from~\eqref{coordinate-in-patches}. This is the reason for which, for example in~\cite{Martelli:2013aqa,Cassani:2021dwa}, it is not needed the description in patches.

\subsubsection{Metric in the bulk}\label{subsubsect:Regularity in the bulk}

We now study the regularity of the bulk, with a finite value of the radial coordinate $q$. 
In principle one can try to consider the case $p_+=q_+$, as in~\cite{Martelli:2013aqa}. However, this is incompatible with a bulk topology\footnote{With a standard abuse of notation we will omit often the fibre, since it is $\CC$ or orbifold thereof.} $\mathcal{O}(-\lens)\rightarrow \spindlesing$. Indeed in that case the Killing vector $K_{q_+}$ has a zero-dimensional fixed point (called a \emph{nut}), but we want it to develop a two-dimensional fixed submanifold $\Sigma\subset \mathcal{M}_4$ (called a \emph{bolt}), using a standard terminology~\cite{Gibbons:1979xm}. In the regularity analysis of~\cite{Martelli:2013aqa}, for $p_+=q_+$, the authors obtained that $N=M$, and in turn $P=Q$. Indeed the point $p=q=p_+=q_+$ contains a curvature singularity unless $P=Q$. For example, the square root of the Ricci tensor is
\begin{equation}
	R_{\mu\nu}R^{\mu\nu}=4\frac{\bigl[9(q^2-p^2)^4+ (P^2-Q^2)^2\bigr]}{(q^2-p^2)^4}\,.
\end{equation}
From the previous discussion, we continue with $Q\neq P$ and $N\neq M$. The situation is summarized as 
\begin{equation}\label{signature}
	\begin{split}
		&\,\QQ\ge0\, , \quad \PP\le 0\, , \quad  q^2 > p^2\, ,
		\\
		&p_- \le p \le p_+ < q_+\le q\,,\quad q_+ ^2 > p_\pm ^2\,,
	\end{split}
\end{equation}
where the range of $p$ and $q$ is in general disjoint. In order to obtain a complete metric, the space must necessarily \enquote{close off} at the largest root of $\mathcal{Q}(q)$, the location of the bolt.

We can now look at~\eqref{metric-for-spindle}. Noticing that for $q=q_+$ the second line is zero, we argue that $(p,\angletwo^{\pm})$, when $q$ is fixed at $q_+$, are again the coordinates on (a different shaped) spindle $\spindlesingbolt$. Since $q_+$ is the position of the bolt, we will refer to this $\spindlesingbolt$ as a \enquote{spindle bolt}. Near the points $(p_{\pm},q_+)$ we use 
\begin{equation}\label{doule-lim}
	U_\pm:\quad q=q_+ +\frac{\dQQp}{4 (q_+ ^2 - p_\pm ^2)}R^2\,,\quad p=p_\pm- \frac{\dPPpm}{4(q_+ ^2-p_\pm ^2)} R_\pm ^2\,,
\end{equation}
on~\eqref{metric-for-spindle}, obtaining at the leading order
\begin{equation}{\label{the-sing-points}}
	U_\pm :\quad	\dd s^2 _4 \underset{q_+ \, ,\,p_\pm}{\simeq} \dd R^2+ \frac{ \dQQp^2}{4} R^2\dd (\angleone^{\pm}) ^2+ \dd R_{\pm}^2+ 
	\frac{R_{\pm}^2}{\singpm^2} \dd \periodictwo ^2\,,
\end{equation}
where we already used~\eqref{equal-periodic},~\eqref{periodicangletwodef} and~\eqref{spindle-conditions-boundary}.
Locally this metric describes $ \CC/\mathbb{Z}_{\labell}\times \CC/\ZZ_{\singpm}$, once we impose
\begin{equation}\label{spindle-cond1}
	\frac{\dQQp}{2}  \Delta\angleone=\frac{2\pi}{\labell}\,,
\end{equation}
which is the analogous of~\eqref{spindle-conditions-boundary} and recall that $\Delta\angleone=\Delta\angleone^{+}=\Delta\angleone^{-}$. Summarizing,  equations~\eqref{quantization_cond_spindle},~\eqref{bound-lens} and~\eqref{spindle-cond1} are \enquote{quantization conditions}, which we will study in detail in section~\ref{subsect:Quantization}.

Defining the (real-valued) spindle bolt metric to be the first line of~\eqref{metric-for-spindle} in $q=q_+$
\begin{equation}\label{spindle_bolt}
	U_\pm\quad\dd s_{\spindle_{q_{+}}} ^2 =\frac{q_+ ^2-p^2}{-\PP}\dd p^2-  \frac{(q_+ ^2-p_\pm^2)^2\PP}{q_+ ^2-p^2} \dd( \angletwo^{\pm})^2\,,
\end{equation}
we can observe that
\begin{equation}
	U_\pm:\quad\sqrt{g_{\spindle_{q_+}}}R_{\spindle_{q_+}} =(q_+^2-p_\pm^2)\partial_p \bigg[\frac{2p \PP + (q_+ ^2-p^2)\dPP}{(q_+^2 -p^2)^2}\bigg]\,,
\end{equation}
and the (orbifold) Euler characteristic is, again, found to be
\begin{equation}
	\chi_\spindle=\frac{1}{4\pi}\int \sqrt{g_{\spindle_{q_+}}}R_{\spindle_{q_+}}\dd p \,\dd \angletwo^{\pm}=\frac{\singm+\singp}{\singm \singp}\,.
\end{equation}
The metric of the spindle~\eqref{spindle_bolt} is different from~\eqref{bound-base}, but it describes topologically the same spindle, with conical singularities $\CC/\ZZ_{\singpm}$. The four-dimensional metric~\eqref{metric-for-spindle}, with~\eqref{quantization_cond_spindle},~\eqref{bound-lens} and~\eqref{spindle-cond1} describes the bundle $\CC/\ZZ_{\labell}\hookrightarrow \mathcal{O}(-t)\rightarrow \spindlesingbolt$, which has a normal singularity $\CC/\ZZ_{\labell}$.

We conclude this section by considering the collapse of the metric at  $q=q_0>q_+$ fixed. 
To zoom in near $p_\pm$ we moreover use $p=p_{\pm}- \frac{\dPPpm}{4(q_0 ^2 - p_\pm ^2)} R_{\pm}^2$ for which~\eqref{metric-for-spindle} becomes
\begin{equation}
	U_\pm:\quad	\dd s^2 \bigl|_{q=q_0}\,\underset{p\rightarrow p_\pm}{\simeq} \dd R_{\pm}^2+ \frac{R_\pm ^2}{\singpm^2}(\dd \periodictwo+ c_\pm (q_0)\dd\angleone^{\pm} )^2+\frac{(q_+^2-p_\pm^2)^2}{q_0 ^2 - p_\pm ^2}\mathcal{Q}(q_0)\dd(\angleone^{\pm}) ^2\,,
\end{equation}
where $c_\pm(q_0)$ is an irrelevant constant that goes to zero as $q_0$ approaches $q_+$.
Then, the regularity follows from the previous conditions imposed to the topology of the boundary.

\subsubsection{Gauge field and Killing spinors in the bulk}\label{subsubsect:Regularity of the Killing spinors}

	Let us come back to the study of the regularity of the spinor~\eqref{four-spinor}.  The analysis is divided into two parts: firstly, we pick a gauge in which the graviphoton~\eqref{solution:graviphoton} is regular on the spindle bolt, then we choose an adapted frame in which the spinor is smooth and well-defined. In the patch $U_\pm$ which contains $p_\pm$ we perform the gauge transformation
\begin{equation}\label{regular gauge}
	U_{\pm}\,:\quad A_{\pm}'
	=A+\parqp q_+ Q \dd\angleone^{\pm}-\parppm p_{\pm}P\dd\angletwo^{\pm}\,,
\end{equation}
so that $A_\pm '(q_+,p_\pm)=0$. Correspondingly, the spinor~\eqref{four-spinor} acquires a phase. Summing up the contributions, the overall phase $\Phi_\pm$ in the $2\pi$-periodic coordinates $\periodicangles^{\pm}$ can be written as
\begin{equation}\label{phase-gauge}
	\begin{aligned}
		U_{\pm}\,:\quad\Phi_\pm=&\frac{2 q_+ P Q+N (q_+^2-\sqrt{\alpha})}{2P}\frac{\parqp\Delta\angleone}{2\pi}\periodicone^{\pm}-\frac{2 p_\pm P Q+N (p_\pm^2-\sqrt{\alpha})}{2P}\frac{\parppm\Delta\angletwo^{\pm}}{2\pi}\periodictwo^{}
		\\
		=&\frac{\signqp\mathcal{Q}'(q_+)}{4}\frac{\parqp\Delta\angleone}{2\pi}\periodicone^{\pm}-\frac{\rho^\pm \mathcal{P}'(p_\pm)}{4}\frac{\parppm\Delta\angletwo^{\pm}}{2\pi}\periodictwo^{}=	\frac{\signqp\parqp}{2\labell}\periodicone^{\pm}\mp\frac{\newsignpm\parppm }{2\singpm}\periodictwo^{}\,,
	\end{aligned}
\end{equation}
where we have used~\eqref{equal-periodic},~\eqref{periodicangletwodef},~\eqref{susy-func-dec},~\eqref{susy-func} and~\eqref{spindle-conditions-boundary},~\eqref{spindle-cond1} in the last step. Moreover we have introduced the sign $\newsignpm$, defined as $\newsignp=\signpp$ and $\newsignm=\signpm$, so that the product $\newsignp\newsignm=\signpp\signpm=\twist$.  Next, we notice that the frame~\eqref{frame} is singular at the poles of the bolt. It is therefore convenient to write the frame in $\periodicangles^{\pm}$ coordinates and focus on $(q_+,p_\pm)$ by means of~\eqref{doule-lim}. We obviously obtain (see~\eqref{the-sing-points})
\begin{equation}\label{infinitesimal-vielbeins}
	U_\pm\,:\quad e^a\simeq\Bigl\{\mp\dd R_{\pm}\,,\parppm \frac{R_{\pm}}{\singpm}\dd\periodictwo^{}\,,\parqp\frac{R}{\labell}\dd\periodicone^{\pm}\,,\dd R\Bigr\}\,,
\end{equation}
where the sign in $e^1$ is due to the fact that approaching $p_+$ ($p_-$) the coordinate $p$ is increasing (decreasing), while $R_\pm$ is decreasing (increasing).
A well-defined frame on the spindle, in the patch $U_\pm$, is given by 
\begin{equation}
	U_\pm:\quad	( \hat{e}^{1})+\ii(\hat{e}^2)=\dd\biggl[\mp R_{\pm}\exp\biggl(\mp\ii\parppm\frac{\periodictwo^{}}{\singpm}\biggr)\biggr]=\exp\biggl(\mp\ii\parppm\frac{\periodictwo^{}}{\singpm}\biggr)( e^1+\ii e^2)\,,
\end{equation}
which is obtained by means of a $U(1)\simeq SO(2)$ rotation. Similarly it happens for the transverse part described by $(e^3,e^4)$, so that the whole $SO(2)\times SO(2)$ transformation is
\begin{equation}\label{SOxSO}
	U_\pm:\begin{pmatrix}
		\hat{e}^1   \\
		\hat{e}^2\\
		\hat{e}^3\\
		\hat{e}^4
	\end{pmatrix}=\begin{pmatrix}
		\cos\big[\mp\parppm\periodictwo^{}/\singpm\big] & -\sin\big[\mp\parppm\periodictwo^{}/\singpm\big] &0&0 \\
		\sin\big[\mp\parppm\periodictwo^{}/\singpm\big]& \cos\big[\mp\parppm\periodictwo^{}/\singpm\big] &0&0\\
		0 & 0 &\cos\big[\parqp\periodicone^{\pm}/\labell\big]&\sin\big[\parqp\periodicone^{\pm}/\labell\big]\\
		0 & 0 &-\sin\big[\parqp\periodicone^{\pm}/\labell\big]&\cos\big[\parqp\periodicone^{\pm}/\labell\big]
	\end{pmatrix}\begin{pmatrix}
		{e}^1   \\
		{e}^2\\
		{e}^3\\
		{e}^4
	\end{pmatrix}\,,
\end{equation}
Since under a transformation of the vielbein $\hat{e}^a=\exp(\lambda^{a}{}_{b}) \,e^{b}$ a four-dimensional Dirac spinor $\psi$ transforms in the spin representation as $\psi'=\exp(1/2 \lambda^{ab}\Gamma_{ab})\psi$, with the $\Gamma_a$ given in \eqref{gammas}, we have that the four dimensional spinor~\eqref{four-spinor} becomes
\begin{equation}{\label{vielbein-transf}}
	U_\pm:\quad	\varepsilon '=\exp\biggl[\frac{\ii}{2}\text{diag}\Bigl\{\pm\frac{\parppm\periodictwo^{}}{\singpm}-\frac{\periodicone^\pm}{\labell}\,,\mp\frac{\parppm\periodictwo^{}}{\singpm}+\frac{\periodicone^\pm}{\labell}\,,\pm\frac{\parppm\periodictwo^{}}{\singpm}+\frac{\periodicone^\pm}{\labell}\,,\mp\frac{\parppm\periodictwo^{}}{\singpm}-\frac{\periodicone^\pm}{\labell}\Bigr\}\biggr]\varepsilon\,.
\end{equation}
Explicitly, summing the contributions from~\eqref{phase-gauge} and~\eqref{vielbein-transf}, the spinor~\eqref{four-spinor} in the regular gauge and frame reads
\begin{equation}{\label{regular spinor}}
	U_\pm:\quad \varepsilon'(q,p,\periodicone^\pm,\periodictwo)=c\begin{pmatrix}
		-\ii\frac{\sqrt{\mathcal{P}_- (p)}\sqrt{\mathcal{Q}_- (q)}}{\sqrt{q+p}} \ee^{\frac{\ii}{2}[\frac{\signqp-1}{\labell}\periodicone^{\pm}\mp \frac{\rho^\pm -1}{\singpm}\periodictwo]}
		\\
		-\ii	\frac{\sqrt{\mathcal{P}_+ (p)}\sqrt{\mathcal{Q}_+ (q)}}{\sqrt{q+p}}\ee^{\frac{\ii}{2}[\frac{\signqp+1}{\labell}\periodicone^{\pm}\mp \frac{\rho^\pm +1}{\singpm}\periodictwo]}
		\\
		\langle P\rangle\frac{\sqrt{\mathcal{P}_- (p)}\sqrt{\mathcal{Q}_+ (q)}}{\sqrt{q-p}}\ee^{\frac{\ii}{2}[\frac{\signqp+1}{\labell}\periodicone^{\pm}\mp \frac{\rho^\pm -1}{\singpm}\periodictwo]}
		\\
		\langle P\rangle\frac{	\sqrt{\mathcal{P}_+ (p)}\sqrt{\mathcal{Q}_- (q)}}{\sqrt{q-p}}\ee^{\frac{\ii}{2}[\frac{\signqp-1}{\labell}\periodicone^{\pm}\mp \frac{\rho^\pm +1}{\singpm}\periodictwo]}
	\end{pmatrix}\,.
\end{equation}
We can now see that the spinor in~\eqref{regular spinor} is smooth everywhere by noticing that, at the fixed points, it can be written as 
\begin{equation}\label{four-spinor-fixed-twist}
	\twist=1:\,\, \varepsilon'(q_+,p_\pm)=\frac{c}{4}\begin{pmatrix}
		-\ii\frac{\sqrt{\mathcal{P}_- (p_\pm)}\sqrt{\mathcal{Q}_- (q_+)}}{\sqrt{q_++p_\pm}}     (1+\signpp)(1+\signqp)\ee^{\frac{\ii}{2}[\frac{\signqp-1}{\labell}\periodicone^{\pm}\mp \frac{\signpp -1}{\singpm}\periodictwo]}
		\\
		-\ii	\frac{\sqrt{\mathcal{P}_+ (p_\pm)}\sqrt{\mathcal{Q}_+ (q_+)}}{\sqrt{q_++p_\pm}}(1-\signpp)(1-\signqp)\ee^{\frac{\ii}{2}[\frac{\signqp+1}{\labell}\periodicone^{\pm}\mp \frac{\signpp +1}{\singpm}\periodictwo]}
		\\
		\langle P\rangle\frac{\sqrt{\mathcal{P}_- (p_\pm)}\sqrt{\mathcal{Q}_+ (q_+)}}{\sqrt{q_+-p_\pm}}(1+\signpp)(1-\signqp)\ee^{\frac{\ii}{2}[\frac{\signqp+1}{\labell}\periodicone^{\pm}\mp \frac{\signpp -1}{\singpm}\periodictwo]}
		\\
		\langle P\rangle\frac{	\sqrt{\mathcal{P}_+ (p_\pm)}\sqrt{\mathcal{Q}_- (q_+)}}{\sqrt{q_+-p_\pm}}(1-\signpp)(1+\signqp)\ee^{\frac{\ii}{2}[\frac{\signqp-1}{\labell}\periodicone^{\pm}\mp \frac{\signpp+1}{\singpm}\periodictwo]}
	\end{pmatrix}\,,
\end{equation}
or 
\begin{equation}\label{four-spinor-fixed-antitwist}
	\twist=-1:\,\, \varepsilon'(q_+,p_\pm)=\frac{c}{4}\begin{pmatrix}
		-\ii\frac{\sqrt{\mathcal{P}_- (p_\pm)}\sqrt{\mathcal{Q}_- (q_+)}}{\sqrt{q_++p_\pm}}     (1\pm\signpp)(1+\signqp)\ee^{\frac{\ii}{2}[\frac{\signqp-1}{\labell}\periodicone^{\pm}\mp \frac{\pm\signpp -1}{\singpm}\periodictwo]}
		\\
		-\ii	\frac{\sqrt{\mathcal{P}_+ (p_\pm)}\sqrt{\mathcal{Q}_+ (q_+)}}{\sqrt{q_++p_\pm}}(1\mp\signpp)(1-\signqp)\ee^{\frac{\ii}{2}[\frac{\signqp+1}{\labell}\periodicone^{\pm}\mp \frac{\pm\signpp +1}{\singpm}\periodictwo]}
		\\
		\langle P\rangle\frac{\sqrt{\mathcal{P}_- (p_\pm)}\sqrt{\mathcal{Q}_+ (q_+)}}{\sqrt{q_+-p_\pm}}(1\pm\signpp)(1-\signqp)\ee^{\frac{\ii}{2}[\frac{\signqp+1}{\labell}\periodicone^{\pm}\mp \frac{\pm\signpp -1}{\singpm}\periodictwo]}
		\\
		\langle P\rangle\frac{	\sqrt{\mathcal{P}_+ (p_\pm)}\sqrt{\mathcal{Q}_- (q_+)}}{\sqrt{q_+-p_\pm}}(1\mp\signpp)(1+\signqp)\ee^{\frac{\ii}{2}[\frac{\signqp-1}{\labell}\periodicone^{\pm}\mp \frac{\pm\signpp+1}{\singpm}\periodictwo]}
	\end{pmatrix}\,.
\end{equation}
It is evident from the above expressions that, for a fixed choice of the signs $(\signpp,\signqp)$, the single component of the spinor which is non-zero does not contain any phase. Choosing for example the roots $q_+$ and $p_+$ such that $\mathcal{P}_+ (p_+)=\mathcal{Q}_+(q_+)=0$, which is realized by fixing $\signpp=\signqp=1$,  we see from~\eqref{four-spinor-fixed-twist} that all the components are zero but the first, which however does not depend any more on $(\periodicone^+,\periodictwo)$. In the twist case this example is equivalent to
\begin{equation}
 \varepsilon'(q_+,p_+)=c\begin{pmatrix}
		-\ii\frac{\sqrt{\mathcal{P}_- (p_+)}\sqrt{\mathcal{Q}_- (q_+)}}{\sqrt{q_++p}} 
		\\
	0
		\\
	0
		\\
	0
	\end{pmatrix}\,.
\end{equation}
 Notice also that these spinors are regular at the location of the bolt $q=q_+$, without moving to its poles. Indeed the $\periodicone^\pm$ are cancelled in the non-vanishing components by requiring $\mathcal{Q}_\pm (q_+)=0$. Moreover equations \eqref{four-spinor-fixed-twist} and \eqref{four-spinor-fixed-antitwist} show that the spinor has the same chirality at the poles of the spindle in the $\twist=1$ case and opposite chiralities when $\twist=-1$. In turn, this explicitly shows that ($\twist=-1$) $\twist=1$ corresponds to the (anti-) twist realization of the supersymmetry on the spindle bolt~\cite{Ferrero:2021etw}. 

We end this section by writing the spinor transition function coming from~\eqref{vielbein-transf}, obtained by the composition of the transformation on $U_+$ and the (inverse) transformation on $U_-$
\begin{equation}
	\begin{aligned}
	\exp\biggl[	\parp\frac{\ii\,\periodictwo}{2}	\text{diag}\Bigl\{\chi_\spindle-\frac{\lens/\labell}{\singm\singp}\,,-\chi_\spindle+\frac{\lens/\labell}{\singm\singp}\,,\chi_\spindle+\frac{\lens/\labell}{\singm\singp}\,,-\chi_\spindle-\frac{\lens/\labell}{\singm\singp}\Bigr\}\biggr]\,,
	\end{aligned}
\end{equation}
where we have used~\eqref{change:periodic-patch} and~\eqref{periodicangletwodef}. Notice that these transition function is valid for both the twist and the anti-twist set-up, as for example in~\cite{Ferrero:2020twa,Faedo:2021nub}, even if only $\chi_\spindle$ appears.

\subsubsection{Gauge field and Killing spinors at  the boundary}
\label{boundary gauge spinor}.

 We conclude the global analysis of our solutions by discussing  the gauge field and Killing spinors at the boundary. Here we will employ 
the local $2\pi$-periodic coordinates $(\periodicone^\pm,\periodictwo)$  inherited from the study of regularity of the metric near to the bolt, see~\eqref{equal-periodic} and~\eqref{change:tau-theta}. We will come back to 
the boundary geometry later, after introducing an alternative set of coordinates in the next subsection.

The  gauge field $A_{(3)}$ and a Killing spinor $\chi_{(3)}$ induced at the boundary are  obtained from  the $q\rightarrow+\infty $ expansion of~\eqref{solution:graviphoton} and~\eqref{four-spinor}, respectively. In particular,  in the gauge~\eqref{regular gauge} the  boundary gauge field $A_{(3)}=pP\dd\sigma$ reads 
	\begin{equation}\label{boundary-gauge}
			U_{\pm}\,:\quad \bigl(A_{(3)}\bigr)_{\pm}'
		=\frac{\Delta\angleone}{2\pi}(q_+ Q-pP)\dd\periodicone^\pm+\frac{\Delta}{2\pi}\frac{P(p-p_\pm)}{(q_+^2-p_\pm^2)}\dd\periodictwo\, .
	\end{equation}
Using the three-dimensional  vierbein induced from the bulk~\eqref{frame}, namely 
	\begin{equation}\label{boundary frame}
	 e_{(3)}^a = \Bigl\{ \frac{\dd p}{\sqrt{-\PP}}\,, \sqrt{-\PP}\dd\sigma\,, (\dd\tau+p^2\dd\sigma)\Bigr\}\underset{p\rightarrow p_\pm}{\simeq} \Bigl\{ \mp\dd R_\pm \,, \frac{R_\pm}{\singpm}\dd\periodictwo\,,(q_+^2-p_\pm ^2) \frac{\Delta\angleone}{2\pi}\dd\periodicone^\pm\Bigr\}\,,
	\end{equation}
	the boundary spinor is~\cite{Martelli:2013aqa}
	\begin{equation}\label{boundary-spinor}
		\begin{aligned}
		U_\pm:\quad	\chi_{(3)}=\begin{pmatrix}
				\sqrt{\Pm}\\
				\sqrt{\Pp}\\
			\end{pmatrix}\ee^{\frac{\ii}{2}[\frac{\signqp\parqp}{\labell}\periodicone^{\pm}\mp\frac{\newsignpm\parppm }{\singpm}\periodictwo^{}]}=\begin{pmatrix}
			\sqrt{\Pm}\\
			\sqrt{\Pp}\\
		\end{pmatrix}\ee^{\frac{\ii\signpp}{2}[\frac{\branch\parqp}{\labell}\periodicone^{\pm}\mp\frac{\twist\pm\parppm }{\singpm}\periodictwo^{}]}\,.
		\end{aligned}
	\end{equation}
Here we have used the definitions $\rho_+ \rho_-=\twist$, $\branch=\signpp\signqp$ and introduced the symbol $\twist_\pm=\signpp\newsignpm$ which is $\twist_+=1$ and $\twist_-=\twist$. 
We emphasise that since we used the \emph{same regular gauge} for the bulk and boundary gauge field,  the phase of the spinor is the same as in~\eqref{phase-gauge}.

Following the steps outlined in~\eqref{phase-gauge}, we can easily show that the transition functions for the gauge field~\eqref{boundary-gauge} across the patches $U_+$ and $U_-$ are
\begin{equation}\label{anticipating-flux}
(A_{(3)}\bigr)_- -(A_{(3)}\bigr)_+=\frac{1}{2}\biggl[\frac{\rho^+ \singm+\rho^- \singp}{\singm\singp}-\signqp \frac{\lens/\labell}{\singm\singp}\biggr]\dd\periodictwo=\frac{\signpp}{2}\biggl[\frac{\singm+\twist \singp}{\singm\singp}-\branch \frac{\lens/\labell}{\singm\singp}\biggr]\dd\periodictwo\,.
\end{equation}
This formula shows that  the choice of the sign $\twist$, that will 
determine the type of twist for the graviphoton field in the bulk (see~\eqref{twist-anti-twist} below), is captured by the boundary geometry.
Notice that this formula is perfectly valid also when $\lens=0$, for which the boundary is $\spindle\times S^1$ and contains the spindle as a non-trivial two-cycle. In section~\ref{subsect:Quantization} we will reproduce the same expression  integrating $F=\dd A$ over the spindle bolt defined in~\eqref{spindle_bolt}.

Let us now show that the spinor~\eqref{boundary-spinor} is well defined, in each patch. The boundary frame~\eqref{boundary frame} suffers from singularities at the poles $p=p_\pm$ exactly as the one in~\eqref{infinitesimal-vielbeins}. We then perform a frame rotation analogous to that  in~\eqref{SOxSO} on $e_{(3)}^{1,2}$,
\begin{equation}
U_\pm:\quad	\begin{pmatrix}
	\hat{e}^1_{(3)}\\
		\hat{e}^2_{(3)}\\
			\hat{e}^3_{(3)}
	\end{pmatrix}=
\begin{pmatrix}
				\cos\big[\mp\parppm\periodictwo^{}/\singpm\big] & -\sin\big[\mp\parppm\periodictwo^{}/\singpm\big]&0 \\
				\sin\big[\mp\parppm\periodictwo^{}/\singpm\big] & \cos\big[\mp\parppm\periodictwo^{}/\singpm\big]&0\\
				0&0&1
			\end{pmatrix}	\begin{pmatrix}
			{e}^1_{(3)}\\
			{e}^2_{(3)}\\
			{e}^3_{(3)}
		\end{pmatrix}\,,		
\end{equation}
with the corresponding action on the spinor\footnote{We are using the standard Pauli matrices 
$\sigma_a$, $a=1,2,3$  as a basis of the three-dimensional Clifford algebra since they are inherited~from \eqref{gammas}.}
given by 
\begin{equation}{\label{vielbein trasf bound}}
	\begin{aligned}
U_\pm:\quad	\chi_{(3)}'=	\begin{pmatrix}
			\ee^{\pm\frac{\ii}{2} \frac{\periodictwo}{\singpm}} & 0\\
			0 & \ee^{\mp\frac{\ii}{2} \frac{\periodictwo}{\singpm}} \\
		\end{pmatrix}\chi_{(3)}\,,
	\end{aligned}
\end{equation}
	so that at the two poles we have
\begin{equation}
		\begin{aligned}
			\twist=1:\quad 	\chi'_{(3)}(p_\pm)&=\frac{1}{2}\begin{pmatrix}
				\sqrt{\Pm}(1+\signpp)\ee^{\mp\frac{\ii}{2}\frac{\signpp\parppm-1 }{\singpm}\periodictwo^{}}
				\\
				\sqrt{\Pp}(1-\signpp)\ee^{\mp\frac{\ii}{2}\frac{\signpp\parppm+1 }{\singpm}\periodictwo^{}}\\
			\end{pmatrix}e^{\frac{\ii}{2}\frac{\signqp\parqp}{\labell}\periodicone^{\pm}}\,.
		\\
			\twist=-1:\quad 	\chi'_{(3)}(p_\pm)&=\frac{1}{2}\begin{pmatrix}
			\sqrt{\Pm}(1\pm\signpp)e^{\mp\frac{\ii}{2}\frac{\pm\signpp\parppm-1 }{\singpm}\periodictwo^{}}
			\\
			\sqrt{\Pp}(1\mp\signpp)e^{\mp\frac{\ii}{2}\frac{\pm\signpp\parppm+1 }{\singpm}\periodictwo^{}}\\
		\end{pmatrix}e^{\frac{\ii}{2}\frac{\signqp\parqp}{\labell}\periodicone^{\pm}}\,
		\end{aligned}
	\end{equation}
	which are always well defined. 
We then conclude that the transition function for the boundary spinor coming from~\eqref{vielbein trasf bound} are given by
\begin{equation}\label{trans spinor}
	\begin{aligned}
		\begin{pmatrix}
			\ee^{\frac{\ii}{2}\chi_{\spindle} \periodictwo} & 0\\
			0 & \ee^{-\frac{\ii}{2}\chi_{\spindle} \periodictwo} \\
		\end{pmatrix}\,,
	\end{aligned}
\end{equation}
which is  the same as that for the spinors defined on a spindle~\cite{Ferrero:2021etw}! 
This might be surprising, but it is a direct consequence of the regularity analysis at the bolt. We will make further comments about the boundary geometry and the relationship with the analysis of~\cite{Inglese:2023tyc} in section~\ref{subsect: new Seifert}.

\subsection{Toric data}\label{subsect:Toric data}

To conclude the presentation of our solution~\eqref{solution:metric}, we will provide a description of the orbifold $\CC/\ZZ_{\labell}\hookrightarrow\mathcal{O}_{}(-\lens)\rightarrow\spindlesing$ in the framework of toric geometry, albeit the metric does not admit a compatible symplectic structure\footnote{\label{symnotoric}As we commented in section ~\ref{subsect:Hermitian and symplectic properties of PD} there exist two conformally related, oppositely oriented, K\"ahler structures.
 From these it is straightforward to compute the associated moment maps for their common torus action. However, the resulting spaces are compact orbifolds with boundaries, and therefore cannot be described in terms of standard toric geometry. We will therefore not pursue this approach.}. Indeed the space it describes is toric with a $U(1)^2$ action, corresponding to the $U(1)^2$ isometry of~\eqref{solution:metric}. Since in the following sections we will be primarily interested in the topological properties of our solution, we now construct a~\textit{labelled polytope} \cite{Lerman:1995aaa}  in the spirit of~\cite{Faedo:2022rqx,Faedo:2024upq}.
We start defining the loci 
\begin{equation}
	\loci_1=\{p=p_+\}\,,\qquad\loci_2=\{q=q_+\}\,,\qquad \loci_3=\{p=p_-\}\,,
\end{equation}
where $\loci_{1,3}$ are non compact whilst $\loci_2$ defines the bolt, and it is indeed compact. These meet at the two fixed points
\begin{equation}\label{fixed-points}
	p_1=\{p_+,q_+\}\,,\quad p_2=\{p_-,q_+\}\,,
\end{equation}
where by convention we take $\loci_a$ to be the one that joins $p_{a-1}$ and $p_a$.
We can extract the toric data, following the prescription of~\cite{Faedo:2022rqx} (see also~\cite{Martelli:2004wu}). Indeed, given a generic base $(e_1,e_2)$ of the torus action (not necessarily effective), and a set of Killing vectors $\xi_a$ degenerating on $\loci_a$, we use the relation
\begin{equation}\label{relation}
	\xi_a =\vec{v}_a \cdot (e_1,e_2)\,, \quad\frac{\partial_{\mu}  |\xi_a|^2 \partial^{\mu}|\xi_a|^2}{4 |\xi_a|^2}\Big|_{\loci_a}=1\,,
\end{equation}
to extract the vectors of the fan $\vec{v}_a\in\ZZ^2$. These in general will not be primitive, and the greatest common divisor of the entries of each vector will represent a label $m_a$.  Such a number is the order of the transverse singularity at \textit{every point} of the toric divisor $\loci_a$, meaning that the structure group $\ZZ_{m_a}$ is acting on the complex direction $\CC$ normal to $\loci_a$.
Notice that we normalized $\xi_a$ to have unit surface gravity. In the following, without loss of generality, we choose the patch $U_-$ in which the  normalized Killing vectors read (see~\eqref{Killing-vectors-non-toric} and~\eqref{change:periodic-patch})
\begin{equation}\label{normalized Killing vectors}
	U_-:\quad
	\begin{aligned}
			\xi_1 &=\frac{2}{-\dPPp}(p_+ ^2 \partial_{\tau^{-}}-\partial_{\sigma^{-}})=\singp\biggl[\parpm \partial_{\periodictwo^{-}}-\parqp\frac{\lens}{\singp\singm} \partial_{\periodicone^{-}}\biggr]\,,
		\\
		\xi_2&=\frac{2 }{\dQQp}(q_+ ^2 \partial_{\tau^{-}}-\partial_{\sigma^{-}})=\labell \parqp \partial_{\periodicone^{-}}\,,
	\\
		\xi_3&=\frac{2 }{\dPPm}(p_- ^2 \partial_{\tau^{-}}-\partial_{\sigma^{-}})=\singm\parpm \partial_{\periodictwo^{-}}\,.
		\end{aligned}
\end{equation}
Taking inspiration from~\cite{Faedo:2022rqx}, we take as effective toric basis 
\begin{equation}\label{effective-coordinates}
	\begin{aligned}
	U_-&:\quad	E_1^- \equiv \partial_{\toricone^-}=\parqp \partial_{\periodicone^{-}}\,,\quad E_{2}^-\equiv \partial_{\torictwo^-}=\parpm  \partial_{\periodictwo^{-}}-\parqp\frac{r_+}{\singm}  \partial_{\periodicone^{-}}\,,
	\\
	U_+&:\quad E_1^+\equiv \partial_{\toricone^+}=\partial_{\periodicone^{+}}\,,\quad  E_{2}^+\equiv \partial_{\torictwo^+}=\parpp  \partial_{\periodictwo^{+}}+\parqp\frac{r_-}{\singp}  \partial_{\periodicone^{+}}
	\end{aligned}
\end{equation}
where $r_\pm$ are integers satisfying the B\'ezout's lemma
\begin{align}\label{Bezout}
	r_- \singm + r_+ \singp = \lens \,,
\end{align}
which always exist for coprime $\singpm$. Notice that the coordinates $(\periodicone^\pm,\periodictwo)$ are related to the $2\pi$-periodic effective toric ones $(\toricone^\pm,\torictwo^\pm)$ as 
\begin{equation}\label{true new toric coords}
	\begin{aligned}
		U_-:\quad \periodicone^- =\toricone^- -\frac{r_+}{m_-}\torictwo^-\,, \quad \periodictwo=\torictwo^-\,,
		\\
		U_+:\quad \periodicone^+ =\toricone^+ +\frac{r_-}{m_+}\torictwo^+\,, \quad \periodictwo=\torictwo^+\,,
	\end{aligned}
\end{equation}
and employing the transition functions~\eqref{change:periodic-patch} we see easily that
\begin{equation} \label{trivial-effective-toric-transition-functions}
	\toricone^+=\toricone^-\equiv \toricone\,,\quad \torictwo^+=\torictwo^-\equiv \torictwo\,.
\end{equation}
Then, we will continue without indicating the $\pm$ for $(\toricone,\torictwo)$ and also for $E_i^-=E_i^+\equiv E_i$.
In this way, we have 
\begin{equation}{\label{Killing in toric basis}}
	\begin{aligned}
				\xi_1&=\singp E_2 + E_1 (\singp r_+ - \lens)/\singm=\singp E_2 - r_- E_1\,,
				\\
		\xi_2&=\labell E_1\,,
		\\
		\xi_3&=\singm E_2+ r_+ E_1\,,
	\end{aligned}
\end{equation}
From the relation~\eqref{relation} we extract the following non-primitive vectors
\begin{align}{\label{toric-fan}}
 \vec{v}_1 =(r_- , -\singp)	\,,\quad \vec{v}_2=(\labell,0)\,,\quad\vec{v}_3=(r_+,\singm)\,,
\end{align}
which generate the non-compact polytope in figure~\ref{fig:non-compact polytope non-zerot}. This polytope is the same described in~\cite{Martelli:2023oqk} as the \enquote{blow-up} of $\CC/\ZZ_{\lens}$, and from a supergravity point of view it is the non compact version of the solution presented in~\cite{Faedo:2022rqx}, with the first facet therein is sent to infinity. As anticipated, $\labell$ is a label for $\loci_2$, whilst $\loci_{1,3}$ are standard toric divisors.  As pointed out in~\cite{Faedo:2024upq}, one can create singularities along $\loci_{1,3}$ as well by tuning the value of the parameter $\lens=r_-\singm+r_+\singp$. Indeed it is easy to see from the vectors~\eqref{toric-fan} that, for the \enquote{symmetric} case $\lens=\overline{\lens}\,m_+\,m_-$, the labels are $m_a=(m_-,\labell,m_+)$. We will demonstrate this statement explicitly in the next section. The signs of the vectors $\vec{v}_a$ are chosen requiring them to be ordered counter-clockwise, in our conventions.  Whilst here we do not have a symplectic closed two-form $\omega_{(2)}$, in appendix~\ref{appendix:A-spindle-Calabi-Yau-metric} we show that for a Ricci-flat version of this construction it is possible to construct the labelled polytope in a standard way, and the result is consistent with the prescription \eqref{relation}.

For future reference in figure~\ref{fig:non-compact polytope} we have drawn also the weights of the toric action, \ie\ the outward-pointing vectors $\vec{\mu}^{\,(A)}_{i}\equiv\vec{\mu}^{\,(A)}_{A_{i}}$ from each vertex $A=1,2$ of the polytope. Here $i=1,2$ indicates the number of facets that meet at $p_A$. Given two vectors $\vec{v}_{1,2}$, the respective weights can be extracted as 
\begin{equation}{\label{generic-weights}}
	\begin{aligned}
		\vec{\mu}_1 &=\pm\frac{(\vec{v}_1 \cdot \vec{v}_2)\, \vec{v}_1 - |\vec{v}_1|^2\,\vec{v}_2}{d_{12}}\,,
		\\
		\vec{\mu}_2 &=\mp\frac{|\vec{v}_2|^2\,\vec{v}_1-(\vec{v}_1 \cdot \vec{v}_2)\, \vec{v}_2 }{d_{12}}\,,
	\end{aligned}
\end{equation}
where the sign ambiguities can be treated as before, and for~\eqref{toric-fan} they read
\begin{equation}\label{weights}
	\begin{aligned}
		\vec{\mu}_{1}^{\,(1)}&=(-\singp,  -r_-)\;,\quad &&\vec{\mu}_{2}^{\,(1)}=(0,-\labell)\,,
		\\
		\vec{\mu}_{1}^{\,(2)}&=(0, \labell)\;,\quad &&\vec{\mu}_{2}^{\,(2)}=(-\singm,r_+)\,,
	\end{aligned}
\end{equation} 
We have normalized them such that
\begin{align}\label{normalization1}
	|\vec{\mu}^{\,(A+1)}_{1}|^2=|\vec{\mu}^{\,(A)}_{2}|^2=|\vec{v}_{A+1}|^2\,,\quad \vec{\mu}^{(A)}_1 \cdot \mu^{(A)}_2 =  \vec{v}_A \cdot \vec{v}_{A+1}\,,
\end{align}
and
\begin{align}\label{normalization2}
	\det(\vec{\mu}^{\,(A)}_1 ,\vec{\mu}^{\,(A)}_{2})=\det(\vec{v}_{A},\vec{v}_{A+1})=d_A\,.
\end{align}
where $d_A$ is the order of the singularity at the $A$-th fixed-point.
\begin{figure*}[h!]
	\centering
	\begin{subfigure}[t]{0.5\textwidth}
		\centering
	\begin{tikzpicture}[rotate=0]
\def\frazione{0.8}
\def\modulo{0.75}
\def\normallenght{1}

\def\n{2}
\def\m{-4}
\def\l{3.5}
\def\p{-4}
\def\q{-0.7}

\coordinate (A) at (0,\n); 
\node[above] at (A) {$p_2$};
\coordinate (B) at (\m,\l);
\coordinate (C) at ($ (A)!\frazione!(B) $);
\draw[black] (A) -- (C);
\draw[black, dashed] (C) -- (B);
\coordinate (D) at (0,0);
\node[below] at (D) {$p_1$};
\coordinate (E) at (\p,\q);
\coordinate (F) at ($ (D)!\frazione!(E) $);
\draw[black] (D) -- (A);
\draw[black] (D) -- (F);
\draw[ black, dashed] (F) -- (E);

\pgfmathsetmacro{\objone}{-sqrt(\m*\m*\modulo*\modulo)/sqrt(\m*\m+(\l-\n)*(\l-\n))}
\pgfmathsetmacro{\objtwo}{(\l*\modulo-\n*\modulo+\n*sqrt(\m*\m+(\l-\n)*(\l-\n)))/sqrt(\m*\m+(\l-\n)*(\l-\n))}
\coordinate (G) at (\objone,\objtwo);
\draw[thick,red,-latex] (A) -- (G);
\node[above] at (G) {$\vec{\mu}^{\,(2)}_2$};
\pgfmathsetmacro{\objthree}{(\modulo*\p)/(sqrt(\p*\p+\q*\q))}
\pgfmathsetmacro{\objfour}{(\modulo*\q)/(sqrt(\p*\p+\q*\q))}
\coordinate (H) at (\objthree,\objfour);
\draw[thick,red,-latex] (D) -- (H);
	\node[above ] at (H) {$\vec{\mu}^{\,(1)}_1$};
\draw[thick,red,-latex] (D) -- (0,\modulo);	
\node[below right] at (0,\modulo) {$\vec{\mu}^{\,(2)}_1$};
\draw[thick,red,-latex] (A) -- (0,\n-\modulo);	
\node[above left] at (0,\n-\modulo) {$\vec{\mu}^{\,(1)}_2$};
\node[left] at (0,\n/2) {$\labell$};
\draw[blue,-latex] (0,\n/2)-- (\normallenght,\n/2);	
\node[above left] at (\normallenght,\n/2) {$\vec{v}_2$};
\pgfmathsetmacro{\objfive}{(\l+\n)/2}
\coordinate (I) at (\m/2,\objfive);
\coordinate (J) at (\p/2,\q/2);
\pgfmathsetmacro{\objsix}{(-\m)/(\l-\n)}
\pgfmathsetmacro{\objseven}{(\l*\l+\m*\m-\n*\n)/(2*\l-2*\n)}
\pgfmathsetmacro{\dx}{\normallenght / sqrt(1 + \objsix*\objsix)} 
\pgfmathsetmacro{\dy}{\objsix * \dx} 
\pgfmathsetmacro{\xone}{\m/2 + \dx}      
\pgfmathsetmacro{\yone}{\objfive + \dy} 
   \draw[ blue,-latex] (\m/2, \objfive) -- (\xone, \yone);
   \node[ left] at (\xone, \yone) {$\vec{v}_3$};
  \pgfmathsetmacro{\objeight}{-\p/\q}
  \pgfmathsetmacro{\dxone}{\normallenght / sqrt(1 + \objeight*\objeight)} 
  \pgfmathsetmacro{\dyone}{\objeight * \dxone} 
  \pgfmathsetmacro{\xoneone}{\p/2 + \dxone}      
  \pgfmathsetmacro{\yoneone}{\q/2 + \dyone} 
   \draw[ blue,-latex] (\p/2, \q/2) -- (\xoneone, \yoneone);
   \node[above right] at (\xoneone, \yoneone) {$\vec{v}_1$};
	\end{tikzpicture}
		\caption{$\lens\neq 0$}
			\label{fig:non-compact polytope non-zerot}
	\end{subfigure}%
	~ 
	\begin{subfigure}[t]{0.5\textwidth}
		\centering
\begin{tikzpicture}

\def\frazione{0.8}
\def\modulo{0.75}
\def\normallenght{1}

\def\n{2}
\def\m{-4}
\def\l{2}
\def\p{-4}
\def\q{0}

\coordinate (A) at (0,\n); 
\node[above] at (A) {$p_2$};
\coordinate (B) at (\m,\l);
\coordinate (C) at ($ (A)!\frazione!(B) $);
\draw[black] (A) -- (C);
\draw[black, dashed] (C) -- (B);
\coordinate (D) at (0,0);
\node[below ] at (D) {$p_1$};
\coordinate (E) at (\p,\q);
\coordinate (F) at ($ (D)!\frazione!(E) $);
\draw[black] (D) -- (A);
\draw[black] (D) -- (F);
\draw[black, dashed] (F) -- (E);

\pgfmathsetmacro{\objone}{-sqrt(\m*\m*\modulo*\modulo)/sqrt(\m*\m+(\l-\n)*(\l-\n))}
\pgfmathsetmacro{\objtwo}{(\l*\modulo-\n*\modulo+\n*sqrt(\m*\m+(\l-\n)*(\l-\n)))/sqrt(\m*\m+(\l-\n)*(\l-\n))}
\coordinate (G) at (\objone,\objtwo);
\draw[thick,red,-latex] (A) -- (G);
\node[above ] at (G) {};
\pgfmathsetmacro{\objthree}{(\modulo*\p)/(sqrt(\p*\p+\q*\q))}
\pgfmathsetmacro{\objfour}{(\modulo*\q)/(sqrt(\p*\p+\q*\q))}
\coordinate (H) at (\objthree,\objfour);
\draw[thick,red,-latex] (D) -- (H);
\node[above] at (H) {};
\draw[thick,red,-latex] (D) -- (0,\modulo);	
\node[below left] at (0,\modulo) {};
\draw[thick,red,-latex] (A) -- (0,\n-\modulo);	
\node[above right] at (0,\n-\modulo) {};
\node[left] at (0,\n/2) {$\labell$};
\draw[blue,-latex] (0,\n/2)-- (\normallenght,\n/2);	
\node[above left ] at (\normallenght,\n/2) {$\vec{v}_2$};
\pgfmathsetmacro{\objfive}{(\l+\n)/2}
\coordinate (I) at (\m/2,\objfive);
\coordinate (J) at (\p/2,\q/2);
\draw[blue,-latex] (I)-- (\m/2,\objfive+\normallenght);	
\draw[blue,-latex] (J)-- (\p/2,\q/2-\normallenght);	
\node[left] at (\m/2,\objfive+\normallenght) {$\vec{v}_3$};
\node[above right] at  (\p/2,\q/2-\normallenght) {$\vec{v}_1$};
\draw[white,->] (J) -- (0,-1.35); 

\end{tikzpicture}
		\caption{$\lens=0$}
			\label{fig:non-compact polytope zerot}
	\end{subfigure}
	\caption{Labelled polytope associated to normal vectors~\eqref{toric-fan}. The normal singularity $\CC/\ZZ_{\labell}$ shows up as a label in the polytope. figure \ref{fig:non-compact polytope non-zerot} is valid for $\lens\neq 0$, whilst figure \ref{fig:non-compact polytope zerot} is for $\lens=0$.}
	\label{fig:non-compact polytope}
\end{figure*}

We conclude this section by considering the special case $\lens=0$. From~\eqref{normalized Killing vectors} we see immediately that the polytope is an open rectangle as in figure~\ref{fig:non-compact polytope zerot}. Notice that when $\lens=0$ the integers $r_\pm$ in~\eqref{Bezout} lose their meaning. In particular, all the previous formulas are applicable by simply setting $r_\pm=0$.

\subsection{Seifert orbifolds with twist and anti-twist}

\label{subsect: new Seifert}

In this section we analyze in more details the Seifert orbifolds and spinors obtained in the boundary in section~\ref{boundary gauge spinor}. Up to this point, we have used the local coordinates $(\periodicone^\pm,\periodictwo)$ to study the regularity of the metric, spinor and gauge field. However, as seen in the previous section, the coordinates $(\toricone,\torictwo)$ describe correctly the effective action of the torus.
These are the appropriate coordinates to study the singularities of the three-dimensional Seifert orbifold~\cite{Closset:2018ghr,Ferrero:2021etw}.  More precisely, the metric near to the poles~\eqref{boundary near poles} takes the form
\begin{equation}\label{bound metric near poles toric}
	U_\pm:\quad	\dd s^2_b \underset{p\rightarrow p_\pm}{\simeq}\dd R_{\pm}^2+ \frac{R_\pm^2}{\singpm^2} \dd \torictwo^2+ \left(\frac{(q_+ ^2 - p_\pm ^2)\Delta\theta_1}{2\pi}\right)^2\Bigl[\dd\toricone\pm\frac{r_\mp}{\singpm}\dd\torictwo\Bigr]^2\,.
\end{equation}
It is now straightforward to introduce complex coordinates adapted to the poles, namely
\begin{equation}
	U_\pm:\quad z_{\pm}=R_\pm \ee^{\ii \frac{\torictwo}{\singpm}}\,,\quad f_\pm=\frac{(q_+ ^2 - p_\pm ^2)\Delta\theta_1}{2\pi}\ee^{\ii (\toricone\pm \frac{r_\mp}{\singpm}\torictwo)}\,,
\end{equation}
where $z_\pm$ are local coordinates on the base  and $f_\pm$ on the fibre. Performing a $2\pi$ rotation along both $\toricone$ and $\torictwo$ we have the following identifications
\begin{equation}{\label{complex action}}
	U_\pm:\quad m\cdot(z_\pm,f_\pm)= \left(\ee^{\frac{2\pi\ii\,m}{\singpm}}z_\pm, \ee^{\pm\frac{2\pi\ii \,m\,r_\mp}{\singpm}}f_\pm\right)\,,
\end{equation}
with $m\in\ZZ_{\singpm}$. The analysis now splits into two cases. If $\gcd(r_\mp,m_\pm)=1$, the action is free and  the three-dimensional space is smooth, and in particular it is a lens space $L(\singm r_- +\singp r_+=\lens,1)$. Notice that we can rephrase this statement by saying that the boundary space is smooth if and only if $\gcd(\lens, \singpm)=1$. Indeed given that the B\'ezout's lemma holds and $\gcd(\singm,\singp)=1$, we have that $\gcd (\lens,\singpm)=\gcd(r_\mp,\singpm)$. If instead $\gcd(r_\mp,\singpm)\neq 1$, we can parameterize $r_\mp= k_\pm \overline{r}_\mp$ and $\singpm=k_\pm \overline{m}_\pm$ (with $\gcd(k_+,k_-)=1$) so that $\gcd (\lens,\singpm)=\gcd(r_\mp,\singpm)=k_\pm>0$. Then we can act $m=\overline{m}_\pm$ times on~\eqref{complex action}, and the result is
	\begin{equation}{\label{complex action-1}}
		U_\pm:\quad \overline{m}_\pm\cdot(z_\pm,f_\pm)= \left(\ee^{\frac{2\pi\ii}{k_{\pm}}}z_\pm, f_\pm\right)\,.
	\end{equation}
	This explicitly shows that there is a $\ZZ_{k_\pm}$ sub-action which leaves the fibre fixed and, in turn, that the three-dimensional space is an orbifold, with conical singularities near the poles $p_\pm$ of order $\CC/\ZZ_{k_\pm}$, respectively. In terms of the lens space parameter $\lens$, we can state that the space has a singularity $\gcd(\lens,\singpm)$ near $p_\pm$. Of course one can also have a singularity at a single pole if one among $k_\pm$ is one, but in the symmetric case with both $k_\pm\neq 1$ the validity of the B\'ezout's lemma implies $\lens=k_- k_+ \overline{\lens}$.
This configuration corresponds to the branched lens space $\mathbb{L}_{[k_-,k_+]}(k_-k_+ \overline{\lens},1)$.

In these coordinates, the boundary gauge field~\eqref{boundary-gauge} and the three-dimensional Killing spinor~\eqref{boundary-spinor} read
\begin{equation}
	\begin{aligned}
	U_{\pm}\,:\quad \bigl(A_{(3)}\bigr)_{\pm}'
=&\frac{\Delta\angleone}{2\pi}\biggl[(q_+ Q-pP)\dd\toricone
\\
&+\frac{1}{\singm\singp}\left(\frac{\lens (q_+^2-p_{\mp}^2)(p-p_\pm)P}{p_+^2-p_-^2}\pm (q_+Q-pP)(\lens -r_{\pm}\singpm)\right)\dd\torictwo\biggr]\, ,
\end{aligned}
\end{equation}
where we have used~\eqref{bound-lens}, and
\begin{equation}\label{boundary spinor in real toric}
		U_\pm:\quad	\chi_{(3)}=\begin{pmatrix}
			\sqrt{\Pm}\\
			\sqrt{\Pp}\\
		\end{pmatrix}\ee^{\frac{\ii\signpp}{2}[\frac{\branch\parqp}{\labell}\toricone^{}\pm \frac{\branch r_{\mp}-\labell \twist_{\pm}}{\labell\singpm}\torictwo^{}]}\,.
\end{equation}
Notice that the transition functions for the gauge field~\eqref{anticipating-flux} and for the spinor~\eqref{trans spinor} are essentially unchanged, since they depend only on $\periodictwo=\torictwo$, namely
\begin{equation}\label{transition functions gauge again}
	\begin{aligned}
	(A_{(3)}\bigr)_- -(A_{(3)}\bigr)_+=\frac{\signpp}{2}\biggl[\frac{\singm+\twist \singp}{\singm\singp}-\branch \frac{\lens/\labell}{\singm\singp}\biggr]\dd\torictwo\,,
	\end{aligned}
\end{equation}
and 
\begin{equation}\label{transition functions spinor again}
	\begin{pmatrix}
	\ee^{\frac{\ii}{2}\chi_{\spindle} \torictwo} & 0\\
	0 & \ee^{-\frac{\ii}{2}\chi_{\spindle} \torictwo} \\
\end{pmatrix}\,.
\end{equation}

Notice that from~\eqref{boundary spinor in real toric} one can see  that for $\twist=1$ the spinor has the same \enquote{transverse chirality}\footnote{The boundary spinor~\eqref{boundary-spinor} is already decomposed in a product of a two dimensional spinor on the base (with coordinates $p,\periodictwo$) and a phase on the Seifert fibre $\periodicone^\pm$. Accordingly, the two dimensional gamma matrices can be taken to be just $\sigma_1$ and $\sigma_2$, with chirality matrix $\sigma_\star=-\ii \sigma_1\sigma_2=\sigma_3$. Under $\sigma_\star$ the two-dimensional spinor~\eqref{boundary-spinor} is chiral or anti-chiral at the poles of the Seifert orbifold.} at the poles $p=p_\pm$ of the Seifert orbifold, 
whilst for  $\twist=-1$ it has opposite chiralities. 
We therefore refer to the choices $\sigma=\pm 1$ as to \emph{twist and anti-twist for the Seifert orbifold}. 
We emphasize that the mechanism by which the bulk \enquote{informs} the boundary is through the regularity of the gauge field at the bolt, which then determines a preferred gauge at the boundary\footnote{In particular, 
the phase $\ee^{\frac{\ii}{2}\frac{\signqp\parqp}{\labell}\periodicone^{\pm}}$ in the Killing spinor~\eqref{boundary-spinor} is precisely the phase $\ee^{\frac{\ii}{2}n\psi}$ that appears in the spindle index~\cite{Inglese:2023wky}, with $n=\pm 1$ for the accelerating black holes. Indeed as remarked 
the $\spindle \times S^1$ geometry is a special case $\lens=0$ of the general set-up discussed above. See also footnote~\eqref{footnp}.}. 
This is the same idea that has proved crucial in the context of supersymmetric black holes,  put forward in~\cite{Cabo-Bizet:2018ehj}. In a context analogous to the present one, similar observations were made in~\cite{Martelli:2012sz} and~\cite{Closset:2018ghr}.

We can now make a comparison with with the rigid supersymmetry results for a large class of three-dimensional Seifert orbifolds analyzed in~\cite{Inglese:2023tyc}. In particular, let us consider the gauge field $A^C$ (\cf\ (2.103) in~\cite{Inglese:2023tyc})\footnote{The relation between our variables and the ones in section 2.6 of~\cite{Inglese:2023tyc} is $n_\pm |_{\text{there}}=\singpm|_{\text{here}}$, $n|_{\text{there}}=\lens|_{\text{here}}$, $n|_{\text{there}}=\lens|_{\text{here}}$, $t_\pm |_{\text{there}}=\mp \frac{r_\mp}{\lens}|_{\text{here}}$, $\varphi|_{\text{there}}=-\torictwo|_{\text{here}}$, $\psi|_{\text{there}}=\toricone|_{\text{here}}$ and finally we have mapped $b_1 \rightarrow-b_1$ and $b_2\rightarrow \twist b_2$.}
\begin{equation}
A^C=\frac{1}{2}\frac{b_1 \singm+\twist b_2 \singp}{\lens f(\hat{\theta})}\dd\toricone+\frac{1}{2}\frac{\twist b_2 r_- - b_1 r_+}{\lens f(\hat{\theta})}\dd \torictwo\,,
\end{equation}
and its transition functions. Here $b_1$ and $b_2$ represent the squashing of the three-sphere from which the lens space is obtained and $f(\hat{\theta}=0)=-b_2$ and $f(\hat{\theta}=\pi)=-b_1$. To ensure that  this gauge field is well-defined, we perform a gauge transformation in each patch
\begin{equation}
	\begin{aligned}
		U_-:\quad A_{(0)}^{C}&=\frac{1}{2}\left(\frac{b_1 \singm+\twist b_2 \singp}{\lens f(\hat{\theta})}+\alpha_3^{(0)}\right)\dd\toricone+\frac{1}{2}\left(\frac{\twist b_2 r_- - b_1 r_+}{\lens f(\hat{\theta})}+\alpha_2^{(0)}\right)\dd \torictwo\,,
		\\
		U_+:\quad A_{(\pi)}^{C}&=\frac{1}{2}\left(\frac{b_1 \singm+\twist b_2 \singp}{\lens f(\hat{\theta})}+\alpha_3^{(\pi)}\right)\dd\toricone+\frac{1}{2}\left(\frac{\twist b_2 r_- - b_1 r_+}{\lens f(\hat{\theta})}+\alpha_2^{(\pi)}\right)\dd \torictwo\,,
	\end{aligned}
\end{equation}
 and require
 \begin{equation}
\alpha_2^{(0)}=\frac{\twist -r_+ \alpha_3^{(0)}}{\singm}\,,\quad \alpha_2^{(\pi)}=\frac{r_- \alpha_3^{(\pi)}-1}{\singp}\,.
 \end{equation}
Notice that $A^C$ is well-defined in the patches $U_\pm$ near $\hat{\theta}=0,\pi$ if  it depends on the angular coordinates 
only through the well-defined (in each patch) combinations $\dd\toricone\pm ({r_\mp}{/\singpm})\dd\torictwo$,
 which appear in the metric near the poles~\eqref{bound metric near poles toric}. Identifying $A^C=\signpp A_{(3)}$ and  additionally requiring that
\begin{equation}\label{beta choice}
\alpha_3^{(0)}=\alpha_3^{(\pi)}=\frac{\branch}{\labell}\implies \alpha_{2}^{(0)}=\frac{\twist\labell-\branch r_+}{\labell \singm}\,,\quad \alpha_{2}^{(\pi)}=\frac{\branch r_- - \labell}{\labell \singp}\,,
\end{equation}
we conclude  that indeed $A^C_{(0)}-A^C_{(\pi)}$ coincides with~\eqref{anticipating-flux}. 
We emphasize that from the boundary point of view there is no reason to fix  $\alpha_3^{(0)}$ and $\alpha_3^{(\pi)}$. Indeed, 
in principle, they can take any value without affecting the regularity of the gauge field.
However, we have a preferred gauge~\eqref{regular gauge} inherited from the bulk, ensuring that the spinor and the gauge field are regular at the bolt location $q=q_+$. For comparison, the flat connections $\alpha_3^{(0)}, \alpha_3^{(\pi)}$ chosen in~\cite{Inglese:2023tyc} are given by
\begin{equation}
\alpha_3^{(0)}=\alpha_3^{(\pi)}=\frac{\singm+\twist \singp}{\lens}\quad \implies\quad  \alpha_2^{(0)}=\alpha_2^{(\pi)}=\frac{\twist r_--r_+}{\lens}\,,
\end{equation}
which correspond to trivial gauge transformations across the two patches.

We conclude this subsection  discussing the special case when $\lens=1$, which corresponds to the boundary being  topologically a \emph{smooth three-sphere}. In this case
there exists  another set of coordinates in which the torus action is effective. Indeed, we can consider, in each patch, the $2\pi$-periodic coordinates $(\hat\varphi_{1},\hat\varphi_{2})$ given by
\begin{equation}\label{toric coords new}
	\begin{aligned}
		U_-&:\quad \periodicone^-=\frac{1}{\singm}\hat\varphi_1\,,\quad \periodictwo= \singm \hat\varphi_2-\singp \hat\varphi_1\,,
		\\
		U_+&:\quad \periodicone^+=\frac{1}{\singp}\hat\varphi_2\,,\quad \periodictwo= \singm \hat\varphi_2-\singp \hat\varphi_1\,,
	\end{aligned}
\end{equation}
so  that the metric~\eqref{boundary near poles} near the poles $p_\pm$ is
\begin{equation}\label{toric coords}
	\begin{aligned}
		U_-:\quad \dd s^2_{(3)}\underset{p\rightarrow p_-}{\simeq}& \dd R_-^2+R_-^2 \dd\hat\varphi_2^2+\left(\frac{q_+^2-p_-^2}{\singm}\frac{\Delta\angleone}{2\pi}\right)^2\dd\hat\varphi_1^2\,,
		\\
		U_+:\quad \dd s^2_{(3)}\underset{p\rightarrow p_+}{\simeq}& \dd R_+^2+R_+^2 \dd\hat\varphi_1^2+\left(\frac{q_+^2-p_+^2}{\singp}\frac{\Delta\angleone}{2\pi}\right)^2\dd\hat\varphi_2^2\,.
	\end{aligned}
\end{equation} 
This shows that near to the north pole ($p=p_-$) the space looks like $\mathbb{R}^2\times S^1_{\hat\varphi_1}$ and near to the south pole ($p=p_+$) it is locally  $\mathbb{R}^2\times S^1_{\hat\varphi_2}$. The new coordinates $(\hat{\varphi}_1,\hat{\varphi}_2)$ in~\eqref{toric coords new} are related to the effective toric basis~\eqref{effective-coordinates} and~\eqref{true new toric coords} by
\begin{equation}
E_1=\partial_{\toricone}=\singm \partial_{\hat{\varphi}_1}+\singp\partial_{\hat{\varphi}_2}\,,\quad E_2=\partial_{\torictwo}=-r_+ \partial_{\hat{\varphi}_1}+r_- \partial_{\hat{\varphi}_2}\,.
\end{equation}
These are indeed the coordinates used in section 5 of\footnote{The relation between our variables and the ones defined in~\cite{Inglese:2023tyc} is given by $\varphi_i |_{\text{there}}=\hat{\varphi}_i|_{\text{here}}$, $n_1|_{\text{there}}=\singm|_{\text{here}}$, $n_2|_{\text{there}}=\singp|_{\text{here}}$, $t_2|_{\text{there}}=r_-|_\text{here}$ and $t_1|_{\text{there}}=-r_+|_\text{here}$.}~\cite{Inglese:2023tyc}. In order to compare again to our results, we should write the gauge field $A^{C}$ considered there (\cf\ (5.15) in~\cite{Inglese:2023tyc}) in patches
\begin{equation}
	\begin{aligned}
		U_-:\quad A_{(0)}^{C}&=\frac{1}{2}\left(\frac{b_s}{f(\hat{\theta})}+\alpha_{2}^{(0)}\right)\dd\hat\varphi_1+\frac{1}{2}\left(\frac{\mathfrak{s}_\omega}{f(\hat{\theta})}+\alpha_3^{(0)}\right)\dd \hat\varphi_2\,,
		\\
		U_+:\quad A_{(\pi)}^{C}&=\frac{1}{2}\left(\frac{b_s}{f(\hat{\theta})}+\alpha_{2}^{(\pi)}\right)\dd\hat\varphi_1+\frac{1}{2}\left(\frac{\mathfrak{s}_\omega}{f(\hat{\theta})}+\alpha_3^{(\pi)}\right)\dd \hat\varphi_2\,,
	\end{aligned}
\end{equation}
where $b_{s,c}$ are  constants representing the squashing of the $S^3$ and $\mathfrak{s}_\omega=\pm 1$ for twist and anti-twist respectively, so that we identify $\mathfrak{s}_\omega=\twist$. Moreover, $f(\hat{\theta}=0)=-b_c$ and $f(\hat{\theta}=\pi)=-b_s$. Finally $\alpha_2^{(0,\pi)}$ and $\alpha_3^{(0,\pi)}$ implement gauge transformations in the two patches, and these must be fixed to the values 
\begin{equation}
	\alpha_3^{(0)}=\twist\,,\quad \alpha_2 ^{(\pi)}=1\,,
\end{equation}
for $A^C_{(0,\pi)}$ to be well defined in the respective patch, where one $\hat\varphi_i$ becomes ill-defined at a pole (see~\eqref{toric coords}). As before, $\alpha_3^{(\pi)}$ and $\alpha_2 ^{(0)}$ remain undetermined. The transition function is then
\begin{equation}
	A^C_{(0)}-A^C_{(\pi)}=\frac{1}{2}\left(\alpha_{2}^{(0)}-1\right)\dd\hat\varphi_{1}+\frac{1}{2}\left(\twist-\alpha_3^{(\pi)}\right)\dd\hat\varphi_{2}\,,
\end{equation}
and it can be compared to~\eqref{anticipating-flux}, with $\periodictwo$ expressed in terms of $\hat\varphi_i$ from~\eqref{toric coords new} (with $\lens=1$)
\begin{equation}
	(A_{(3)}\bigr)_- -(A_{(3)}\bigr)_+=\frac{\signpp}{2}\biggl[\Big(\frac{\branch- \twist\labell\singp}{\labell \singm}\Big)-1\biggr]\dd\hat\varphi_1+\frac{\signpp}{2}\biggl[\twist-\Big(\frac{\branch-\labell\singm}{\labell\singp}\Big)\biggr]\dd\hat\varphi_{2}\,.
\end{equation}
Upon identifying $A^C=\signpp A_{(3)}$, the transition functions are the same for the specific choice 
\begin{equation}
	\alpha_{2}^{(0)}=\frac{\branch- \twist\labell\singp}{\labell \singm}\,,\quad \alpha_3^{(\pi)}=\frac{\branch-\labell\singm}{\labell\singp}\,.
\end{equation}
Notice that whilst in~\cite{Inglese:2023tyc} the gauge was fixed to $\alpha_2^{(0)}\equiv\alpha_2^{(\pi)}=1$ and $\alpha_3^{(\pi)}\equiv\alpha_3^{(0)}=\twist$ (such that $A_{(0)}^C-A_{(\pi)}^C=0$), we do not have this freedom in our set-up because, again, we must use the  preferred gauge~\eqref{regular gauge}.  Correspondingly, the spinor~\eqref{boundary-spinor} becomes
\begin{equation}
	\begin{aligned}
		U_-:\,\,\,\,\,	\chi_{(3)}=\begin{pmatrix}
			\sqrt{\Pm}\\
			\sqrt{\Pp}\\
		\end{pmatrix}\ee^{\frac{\ii\signpp}{2}[\alpha_2^{(0)}\hat\varphi_1+\twist \hat\varphi_2]}\,,\quad
		U_+:\,\,\,\,	\chi_{(3)}=\begin{pmatrix}
			\sqrt{\Pm}\\
			\sqrt{\Pp}\\
		\end{pmatrix}\ee^{\frac{\ii\signpp}{2}[\hat\varphi_{1}+\alpha_3^{(\pi)}\hat\varphi_2]}\,.
	\end{aligned}
\end{equation}
As usual, we can finally check that the spinor is smooth in $p=p_\pm$ by rotating appropriately the original frame~\eqref{boundary frame} in the two patches as
\begin{equation}
	\begin{aligned}
		U_-:\quad	\begin{pmatrix}
			\tilde{e}^1_{(3)}\\
			\tilde{e}^2_{(3)}\\
			\tilde{e}^3_{(3)}
		\end{pmatrix}=&
		\begin{pmatrix}
			\cos\hat\varphi_2& -\sin\hat\varphi_2&0 \\
			\sin\hat\varphi_2& \cos\hat\varphi_2&0\\
			0&0&1
		\end{pmatrix}	\begin{pmatrix}
			{e}^1_{(3)}\\
			{e}^2_{(3)}\\
			{e}^3_{(3)}
		\end{pmatrix}\,,		
		\\  
		U_+:\quad	\begin{pmatrix}
			\tilde{e}^1_{(3)}\\
			\tilde{e}^2_{(3)}\\
			\tilde{e}^3_{(3)}
		\end{pmatrix}=&
		\begin{pmatrix}
			\cos\hat\varphi_1& -\sin\hat\varphi_1&0 \\
			\sin\hat\varphi_1& \cos\hat\varphi_1&0\\
			0&0&1
		\end{pmatrix}	\begin{pmatrix}
			{e}^1_{(3)}\\
			{e}^2_{(3)}\\
			{e}^3_{(3)}
		\end{pmatrix}\,.
	\end{aligned}
\end{equation}
For example, in  the twist case, the resulting action on the spinor near the poles gives
\begin{equation}
	\twist=+1:\quad \begin{aligned}
		U_-:\quad	\chi_{(3)}'(p_-)&=\begin{pmatrix}
			\sqrt{\mathcal{P}_- (p_-)}(1+\signpp)\ee^{-\frac{\ii(1-\signpp)}{2} \hat\varphi_2}\\
			\sqrt{\mathcal{P}_+(p_-)}(1-\signpp)\ee^{\frac{\ii(1+\signpp)}{2} \hat\varphi_2}\\
		\end{pmatrix}\ee^{\frac{\ii\signpp}{2}\alpha_2^{(0)}\hat\varphi_1}\,,
		\\
		U_+:\quad	\chi_{(3)}'(p_+)&=\begin{pmatrix}
			\sqrt{\mathcal{P}_- (p_+)}(1+\signpp)\ee^{-\frac{\ii(1-\signpp)}{2} \hat\varphi_1}\\
			\sqrt{\mathcal{P}_- (p_+)}(1-\signpp)\ee^{\frac{\ii(1+\signpp)}{2} \hat\varphi_1}\\
		\end{pmatrix}\ee^{\frac{\ii\signpp}{2}\alpha_3^{(\pi)}\hat\varphi_2}\,,
	\end{aligned}
\end{equation}
showing  that for a choice of $\signpp=\pm1$ the spinor is well-defined near $p_\pm$. The anti-twist case works similarly. 

\vspace{3mm}

In  conclusion, we have shown  that  the main significant difference with the analysis in~\cite{Inglese:2023tyc}, both for the squashed three-sphere and for the (possibly branched) lens space, is a distinct choice of gauge field, which is imposed on us from the global analysis in the bulk.

\subsection{Quantization conditions}\label{subsect:Quantization}
In this section we complete the proof of the existence of the solution, \ie\ we solve the \enquote{quantization conditions}~\eqref{quantization_cond_spindle},~\eqref{bound-lens} and~\eqref{spindle-cond1} showing that~\eqref{solution:metric} and~\eqref{solution:graviphoton} describe a supersymmetric solution with topology $\CC/\ZZ_{\labell}\hookrightarrow\mathcal{O}_{}(-\lens)\rightarrow\spindlesing$. It will follow that the field strength associated to~\eqref{solution:graviphoton} is correctly quantized. To this end, it proves useful to express the parameters of the solution ($E,N,M,P,Q,\alpha$) and ($\Delta\angles^{\pm}$) in terms of the integers $(\lens,\singpm,\labell)$. Recall also that supersymmetry~\eqref{susy} reduces the former down to four independent, namely $(N,P,Q,\alpha)$. In turn, $(N,Q,\alpha)$ are exchanged for the real roots of the metric functions $(p_\pm,q_+)$ and three signs $(\signpp,\twist,\branch)$ in~\eqref{N-alpha}-\eqref{Q}. These relevant signs are related to the ones introduced around~\eqref{N-alpha} via
\begin{equation}
	\eta=\sign \text{Re}(P)\,,\quad \twist=\signpp \signpm\,,\quad \branch =\signpp \signqp\,,
\end{equation}
where the value of $\twist$ will reflect the twist or anti-twist on the spindle bolt. We consider first the generic case with $N\neq 0$, and later we devote a section to $N=0$. Since, as already observed, the twist case is always simpler, we start the discussion with $\twist=1$, turning then to $\twist=-1$. 
\subsubsection{Twist} \label{subsubsect: TwistNneq0}
Using the expressions~\eqref{N-alpha} and~\eqref{Q} with $\twist=1$, it is easy to take the ratio of~\eqref{quantization_cond_spindle}. Defining the rescaled parameter 
\begin{equation}
	\freetwist=\frac{P}{p_+ + p_-}\,,
\end{equation}
the ratio of~\eqref{quantization_cond_spindle} takes the simple form
\begin{align}\label{necessary-condition}
	\frac{\singp}{\singm}=\frac{q_+ ^2 -p _+ ^2}{q_+ ^2-p_- ^2}\,\frac{p_- +\signpp \freetwist}{p_+ + \signpp\freetwist}\,.
\end{align}
We can parametrize the roots as
\begin{align}\label{parameterisation}
	p_\pm = w (1\pm x)\,,\quad q_+= \frac{p_+ + p_-}{2}\tilde{q}_+=w \free\,,
\end{align}
with
\begin{align}\label{effective-parameterisation}
	0<w<q_+\,,\quad x>0\,,\quad \tilde{q}_+ >1+x\,,
\end{align}
where these constraints on $w$ and $x$ come from $p_+\pm p_->0$ and $q_+>p_+$. 
From~\eqref{necessary-condition} we get 
\begin{align}\label{thew}
	w=-\signpp\freetwist  \frac{\big[\free^2 - (1 - x)^2\big]\singp - \big[\free^2 - (1 + x)^2\big] \singm}{(1 + x) \big[\free^2 - (1 - x)^2\big]\singp -  \big[\free^2 - (1 + x)^2\big] (1 - x) \singm}\,.
\end{align}
Then, from both~\eqref{quantization_cond_spindle} and~\eqref{spindle-cond1} we obtain
\begin{equation}\label{twist-angles}
	\begin{aligned}
		\frac{\Delta}{2\pi}&=\frac{1}{8\signpp\freetwist}\,\frac{(1 + x) \big[\free^2 - (1 - x)^2\big]\singp -  \big[\free^2 - (1 + x)^2\big] (1 - x) \singm}{x^2 \singp\singm}\,,
		\\
		\frac{\Delta\angleone}{2\pi}&=
		\frac{1}{2\labell(\free - \branch)}\,\frac{1}{w^2\big[2\signpp\freetwist+w(\free ^2 + 1 - x^2)\big]}\,.
	\end{aligned}
\end{equation}
Finally~\eqref{bound-lens} is now simply
\begin{align}{\label{lens-twist}}
	\lens=\frac{\labell(\free-\branch)}{x}(\singm -\singp)\,,
\end{align}
which can be easily inverted to
\begin{align}
	x=(\free - \branch)\frac{(\singm - \singp)\labell}{\lens}\,.
\end{align}
Notice that not all the values of $\lens$ are admissible from~\eqref{lens-twist}, due to the constraint $\free>1+x>1$ and $N\neq 0$. In particular, we show in appendix \ref{appendix: Values} that $\lens\ge2$. 
	
We are now in the position to show that the field strength is correctly quantized.
We start performing a gauge transformation 
\begin{align}
	U_\pm:\quad	A_\pm '=A+ q_+ Q(\parqp \dd \angleone^\pm-\parppm\dd\angletwo^{\pm})\,,
\end{align}
such that on $\loci_2=\{q=q_+\}$ the gauge field reads simply
\begin{align}
	U_\pm:\quad	A'\big|_{\loci_{2}}=-\parppm\frac{(q_+ ^2-p_\pm^2)(Q q_+ - P p)}{q_+ ^2 - p^2}\dd\angletwo^{\pm}\,.
\end{align}
It is then possible, using~\eqref{N-alpha},~\eqref{Q} and the previous quantized quantities to see that 
\begin{align}\label{twist}
	\frac{1}{2\pi}\int_{\loci_{2}}  F=\frac{\parppm\signpp}{2}\Bigl(\chi_\spindle-\branch\frac{\lens/\labell}{\singm \singp}\Bigr)\,,
\end{align}
meaning that $2A$ is a connection one-form on the line bundle $\bundlep$ over $\spindlesingbolt$, and, in turn, that we have a twist.

The additional term $\lens/(\labell\singm\singp)$ in~\eqref{twist} comes from the non trivial fibration on the spindle bolt in~\eqref{metric-for-spindle}, more precisely from the normal bundle to $\loci_{2}$. To see this, we focus on $q=q_+$ in~\eqref{metric-for-spindle} making the change $q=q_+ + \frac{\dQQp}{4}r^2$. The result is
\begin{align}\label{NHL}
	U_{\pm}:\quad\dd s_4 ^2 \underset{q_+}{\simeq} \dd s_{\spindle_{q_+}}^2 +(q_+^2-p^2)\biggl[\dd r^2+\frac{r^2}{\labell^2}\big(\dd \periodicone^{\pm}+ \mu_f^{\pm} \big)^2\biggr]\,,
\end{align}
where $\mu_f^{\pm}$ is the fibration one-form, given by
\begin{align}
	U_\pm:\quad	\mu_f^{\pm} =\parqp\parppm\labell\frac{2 q_+ (q_+^2-p_\pm^2)\PP+(p^2-p_\pm^2)(q_+^2-p^2)\dQQp}{2(q_+^2-p^2)^2}\,\dd\angletwo^{\pm}\,.
\end{align}
Integrating it we get immediately
\begin{align}
\frac{1}{2\pi}\int_{\loci_{2}} \,\dd \mu_f^{\pm}=\parqp\parppm\frac{\lens}{\singm\singp}\,,
\end{align}
which gives correctly the same value as in~\eqref{lens-condition}. 

In summary, we started from six quantities $(N,P,Q,\alpha,\Delta,\Delta\angleone)$ subject to four topological constraint for $(\lens,\singpm,\labell)$. We are left with a solution with \emph{two free continuous parameters}, that we choose to be $(\freetwist,\free)\in\RR$.

\subsubsection{Anti-twist  } \label{subsubsect: Anti-twistNneq0}
We now focus on the other case, when $\twist=-1$. The ratio of~\eqref{quantization_cond_spindle} is much more complicated
\begin{align}\label{wrong}
	\frac{\singp}{\singm}=\frac{q_+ ^2 -p _+ ^2}{q_+ ^2-p_- ^2}\,\frac{\bigl[2 P^2 -(p_+ ^2-p_- ^2)\signpp P - p_- (p_+ + p_-)^3\bigr]}{\bigl[\signpp P-p_+ (p_+ + p_-)\bigr]\bigl[2 \signpp P +(p_+ + p_- )^2\bigr]}\,,
\end{align}
but it can be simplified redefining
\begin{align}\label{further}
	\signpp P=(p_+ + p_-)\signpp\freetwist=(p_++p_-)\frac{p_+ - p_- + \sqrt{(p_+ - p_-)^2-8(p_+ + p_-)\signpp\freeanti}}{4}\,.
\end{align}
With the parametrization~\eqref{parameterisation}, we obtain formally the same expression for $w$ as in~\eqref{thew} but with $\freeanti$ instead of $\freetwist$, specifically
\begin{align}\label{thew-anti}
	w=-\signpp\freeanti  \frac{\big[\free^2 - (1 - x)^2\big]\singp - \big[\free^2 - (1 + x)^2\big] \singm}{(1 + x) \big[\free^2 - (1 - x)^2\big]\singp -  \big[\free^2 - (1 + x)^2\big] (1 - x) \singm}\,.
\end{align}
The other parameters $(\Delta\angleone,\Delta,\lens)$ can be obtained similarly, but the expressions do not have a compact form. In particular, $\lens=\lens(x)$ is not invertible. For future reference, we report it here 
\begin{equation}\label{lens-anti}
	\begin{aligned}
		\lens=&\labell\frac{ \bigl[\free ^2-(x+1)^2\bigr]\singm - \bigl[\free ^2-(x-1)^2\bigr] \singp}{ \bigl[\free ^2-(x+1)^2\bigr]\bigl[\free ^2-(x-1)^2\bigr]}\times
		\\
		&\frac{\free w \bigl[w(\free^2 + x^2-1)-2 x \signpp\freetwist\bigr]+\branch\bigl[-2\freetwist^2 x -w^2 x (\free ^2+x^2-1)+\signpp\freetwist w(\free ^2+3x^2-1)\bigr]}{w^2 x}\,,
	\end{aligned}
\end{equation}
where $w$ and $\tilde{P}$ are given by~\eqref{thew-anti} and~\eqref{further}, respectively. Notice that~\eqref{lens-anti} is not written in a completely explicit form. However, when expanded, it does not depend on $\freeanti$, a fact that will play a role later on. Moreover, here as before, the parameter $\lens$ can not take any value, and in particular it can not happen $\lens=1$.
From now on the analysis continues as before, but it is more involved. Nevertheless, it can be proved that
\begin{align}\label{anti-twist}
	\frac{1}{2\pi}\int_{\loci_{2}}  F=\frac{\parppm\signpp}{2}\Bigl(\frac{\singm-\singp}{\singm\singp}-\branch\frac{\lens/\labell}{\singm \singp}\Bigr)\,.
\end{align}
Even though we can not obtain $x$ from~\eqref{lens-anti}, we regard to this implicit expression as a constraint, which leaves us again a solution with \emph{two free continuous parameters} $(\freeanti,\free)\in\RR$.
\vspace{5mm}

The main result of this section is the following formula which encapsulates simultaneously~\eqref{twist} and~\eqref{anti-twist}
\begin{align}\label{twist-anti-twist}
	\frac{1}{2\pi}\int_{\loci_2}  F=\frac{\parppm\signpp}{2}\Bigl(\frac{\singm+\twist\singp}{\singm\singp}-\branch\frac{\lens/\labell}{\singm \singp}\Bigr)\,,
\end{align}
and state that $2A$ is a connection on the bundle $\bundletwist$ over the spindle bolt $\spindlesingbolt$.  It is interesting to notice that both type of twist can be realized in minimal gauged supergravity, as opposite to the black hole solution~\cite{Ferrero:2020twa} with a spindly horizon where only the anti-twist is realized. In conclusion, we have constructed 
two different families of globally-defined solutions, with the same underlying geometry given by the orbifold line bundle
$\CC/\ZZ_{\labell}\hookrightarrow\mathcal{O}_{}(-\lens)\rightarrow \spindlesingbolt$, but with a distinct type of twist for the graviphoton gauge field on the spindle bolt, according to the value of $\twist=\pm1$. The integral of the first Chern class of the line bundle can be computed from the relations $\rho_\pm =\dd P_\pm$ given in~\eqref{Ricci potentials}, and we get
\begin{align}
	\frac{1}{2\pi}\int_{\loci_{2}}  {\rho}_\pm  =\frac{\singm+\singp}{\singm\singp}\pm\frac{\lens/\labell}{\singm \singp}\,.
\end{align}
Moreover, both families of solutions depend on \emph{two free continuous parameters}, that are $(\freetwist,\free)\in\RR$ for the twist case, and $(\freeanti,\free)\in\RR$ for the anti-twist case. Notice that~\eqref{twist-anti-twist} is identical to the expression of the gauge field transition functions~\eqref{anticipating-flux}. From this formula we can deduce some constraints on the possible manifolds on which we can uplift our solutions to eleven dimensions. Following~\cite{Martelli:2012sz}, a brief study of this issue is reported in appendix~\ref{appendix:Uplift}.

\subsubsection{$N=0$}\label{subsubsect: N=0}
As explained at the end of section~\ref{subsubsect: Regularity in the boundary}, the case $N=0$ is of particular interest, because we can have a boundary topology $\spindle\times S^1$ only if $N=0$. It is interesting to notice that, when $\acc=0$, $N=0$ is is equivalent to turning off the NUT parameter. In particular we require~\cite{Griffiths:2005qp}
\begin{equation}\label{CP no nut}
	E=P^2-\alpha-\omega^2\,,\quad N=0\,,
\end{equation}
where as usual we fix $\omega=1$ for convenience.
This situation is excluded by the general analysis, which assumes that all the parameters in the solution are different from zero. When~\eqref{CP no nut} holds, supersymmetry~\eqref{acceleration-susy} can be realized in two distinct ways 
\begin{equation}\label{CP-no-nut-susy}
	\begin{aligned}
		\text{(i)}&:\quad &&P=0\,,\quad &&Q=\pm\frac{M}{1+\twist\sqrt{\alpha}}\,,
		\\
		\text{(ii)}&:\quad &&M=0\,,\quad &&P=\pm(1+\twist \sqrt{\alpha})\,,
	\end{aligned}
\end{equation}
where $\twist=\pm1$ as usual. These supersymmetry conditions are consistent (in Lorentzian signature) with\footnote{\label{afootnote}In comparing with the Kerr-Newman-Ads section in~\cite{Klemm:2013eca} attention must be paid to the Wick rotation, which can not be performed without introducing the twisting parameter $\omega_L$ before. The correct definitions in~\cite{Klemm:2013eca} are
	\begin{equation*}
		\varepsilon_L=\omega_{L}^2+ a^2\,,\quad a^2\equiv\hat{\alpha}_L-P_{L} ^2\,,\quad \Delta(r)=(r^2+a^2)(\omega_{L}^2+r^2)-2 m r+Q_{L}^2+P_L ^2\,,
	\end{equation*}
	and
	\begin{equation*}
		\mathcal{B}_{\pm}^{L}=m^2-\biggl[\omega_L^2+a^2\pm 2 \sqrt{P_{L}^2(1-\omega_L^2)+\hat{\alpha}_L \omega_L^2}\biggr](P_L ^2+Q_L ^2)\,,
	\end{equation*}
	where $a$ is the Kerr rotation parameter and in particular $\hat{\alpha}_L >0$. The Wick rotation to Euclidean signature is accomplished by requiring
	\begin{equation*}
		(\varepsilon,P,m)_L=(\varepsilon,P,M)_E\,,\quad (\omega,Q)_L=\ii (\omega,Q)_E\,,\quad \hat{\alpha}_L=2P_E ^2-\alpha_E\,.
	\end{equation*}
}~\cite{Klemm:2013eca}. Gathering these, the metric functions~\eqref{metricfunc} result into
\begin{equation}
	\begin{aligned}
		\text{(i)}:\quad \PP&=(p^2-1)(p^2-\alpha)\,,\quad \QQ=(q^2-1)(q^2-\alpha)-2M q-\frac{M^2}{(1+\twist\sqrt{\alpha})^2}\,,
		\\
		\text{(ii)}:\quad \PP&=(p^2-1)(p^2+1+2\twist\sqrt{\alpha})\,,\quad \QQ=(q^2+\twist\sqrt{\alpha})^2-Q^2\,.
	\end{aligned}
\end{equation}
Notice that we do not have to take $\alpha>0$, as long as all the relevant roots are positive. We now analyze separately these cases, which have different features.
\subsubsection*{(i)}
Let us start by considering the case\footnote{Both $\alpha=1$ and $\alpha=0$ are degenerate cases. For $\alpha=1$ the function $\PP$ is never negative, and for $\alpha=0$ there is a double root in $p=0$}  $\alpha>1$, for which
\begin{equation} \label{i-first-choice}
	p_-=1\,,\quad p_+=\sqrt{\alpha}\,.
\end{equation}
From the quantization conditions~\eqref{spindle-conditions-boundary} one obtains
\begin{equation}
	q_+=\alpha^{1/4}\frac{\sqrt{\sqrt{\alpha}\singm-\singp}}{\sqrt{\singm-\sqrt{\alpha}\singp}}\,,\quad \frac{\Delta}{2\pi}=\frac{1}{\singm-\sqrt{\alpha}\singp}\,,
\end{equation}
as well as
\begin{equation}
	M=-(q_+ +\delta)(\delta q_++\twist\sqrt{\alpha})(1+\twist\sqrt{\alpha})\,,
\end{equation}
with $\delta=\pm 1$ and from~\eqref{spindle-cond1}
\begin{equation}
	\frac{\Delta\angleone}{2\pi}=\frac{1}{\labell\bigl[2 q_+ ^3-q_+ (\alpha+1)-M\bigr]}\,.
\end{equation}
Finally, the condition~\eqref{bound-lens} reads
\begin{equation}\label{lens-i-first-choice}
	\lens=\labell\frac{(\twist+\sqrt\alpha)(\singm-\twist\singp)}{(\alpha-1)(\sqrt{\singm-\sqrt{\alpha}\singp})}\Bigl[2\alpha^{1/4}\sqrt{\sqrt{\alpha}\singm-\singp}+\delta(1+\twist\sqrt{\alpha})\sqrt{\singm-\sqrt{\alpha}\singp}\Bigr]\,.
\end{equation}
Gathering all we then obtain
\begin{equation}
\frac{1}{2\pi}\int_{\loci_2} \dd A=\frac{\parppm\pm\twist}{2}\biggl[\frac{\singm +\twist\singp}{\singm\singp}+\delta\twist \frac{\lens/\labell}{\singm\singp}\biggr]\,,
\end{equation}
that means both twist and anti-twist, again. The case $0<\alpha<1$ is completely analogous, with only $p_- \leftrightarrow p_+$. Notice that all the parameters appearing here are real when $\alpha>1$, so that the metric is real as well. This solution does not have a Lorentzian interpretation, since when $P=0$ we have from the footnote~\eqref{afootnote} $0<\hat{\alpha}_L=-\alpha_E$, but here $\alpha_E$ is necessarily positive (and different from one).

When $\alpha<0$, the roots are
\begin{equation}\label{i-second-choice}
	p_-=-1\,,\quad p_+=1\,,
\end{equation}
so that $\lens=0$. Moreover it can be seen now that~\eqref{spindle-conditions-boundary} gives
\begin{equation}
	\singm=\singp\equiv 1\,,\quad \frac{\Delta}{2\pi}=\frac{q_+^2-1}{(1-\alpha)}\,,
\end{equation}
and $M$ and $\Delta\angleone$ take the same values as before. Notice that $M\in \CC$, but this parameter does not appear in the spindle metric~\eqref{spindle_bolt}; we are then in the general set-up described at the end of section~\ref{subsect:Killing spinors}. It follows that
\begin{equation}
\frac{1}{2\pi}\int_{\loci_2} \dd A=0\,,
\end{equation}
which is the no-twist condition. Notice that here $\twist$ does not represent the value of the twist, contrarily to the $\alpha>1$ case. This solution with no-twist is the Euclidean counterpart of the Kerr-Newman-AdS black hole~\cite{Klemm:2013eca}, for which the action is given in~\cite{Cassani:2019mms}. We will confirm this statement at the end of section~\ref{subsubsect:Holographic renormalization}, where the on-shell action for the CP solution will be computed.

\subsubsection*{(ii)}
We start by requiring $\PP\in\RR$, for which we need to take $\alpha>0$. Notice that when $M=0$ the Lorentzian solutions always possess a naked singularity~\cite{Caldarelli:1998hg,Klemm:2013eca}, as it can be seen from the footnote~\eqref{afootnote}. Then, the following cases have significance only in the Euclidean setting and have not been examined before.

If $\twist=1$, then the roots are $p_+=-p_-=1$ and hence $\lens=0$. From~\eqref{spindle-conditions-boundary} and~\eqref{spindle-cond1} we get
\begin{equation}
	\singm=\singp\equiv 1\,,\quad \frac{\Delta}{2\pi}=\frac{q_+^2-1}{2(1+\sqrt{\alpha})}\,,\quad Q=\delta(q_+^2+\sqrt{\alpha})\,,\quad \frac{\Delta\angleone}{2\pi}=\frac{1}{2\labell q_+(q_+^2+\sqrt{\alpha})}\,,
\end{equation} 
so that
\begin{equation}
\frac{1}{2\pi}\int_{\loci_2} \dd A=\pm 1\,.
\end{equation}
This means that we have the topological twist, \ie\ the twist when the spindle is replaced by $S^2$. Notice that for $\alpha>0$ all the parameters are real and this twist solution is consequently real. This solution can be interpreted as the Euclidean counterpart of a Kerr-Newman black hole with topological twist and rotation. In Lorentzian signature it has no meaning, and indeed in minimal gauged supergravity
a black hole with twist can have only horizons with the topology of a Riemann surface  of genus $\mathrm{g}>1$~\cite{BenettiGenolini:2023ucp}. We shall refer to this solution as to Euclidean 
\enquote{Topological Kerr-Newman-AdS}.

When $\twist=-1$, we can take without loss of generality $\alpha>1$ so that 
\begin{equation}
	p_-=1\,,\quad p_+=\sqrt{-1+2\sqrt{\alpha}}\,,
\end{equation}
and the quantization conditions~\eqref{spindle-conditions-boundary} give
\begin{equation}
	q_+=\frac{\sqrt{\singm (-1+2\sqrt{\alpha})-\sqrt{-1+2\sqrt{\alpha}}\singp}}{\sqrt{\singm-\sqrt{-1+2\sqrt{\alpha}}\singp}}\,,\quad \frac{\Delta}{2\pi}=\frac{1}{\singm-\sqrt{-1+2\sqrt{\alpha}}\singp}\,.
\end{equation}
The other conditions~\eqref{spindle-cond1} and~\eqref{bound-lens} are solved by
\begin{equation}
	Q= \delta (q_+^2-\sqrt{\alpha})\,,\quad \frac{\Delta\angleone}{2\pi}=\frac{1}{2q_+ \labell (q_+^2-\sqrt{\alpha})}\,,\quad 	\lens=\labell q_+ \frac{\singm+\sqrt{-1+2\sqrt{\alpha}}\singp}{\sqrt{-1+2\sqrt{\alpha}}}\,,
\end{equation}
so that
\begin{equation}
\frac{1}{2\pi}\int_{\loci_2} \dd A =\mp\frac{\parppm 1}{2}\biggl[\frac{\singm-\singp}{\singm\singp}\pm\delta \frac{\lens/\labell}{\singm\singp}\biggr]\,.
\end{equation}
For $\alpha>0$ the solution is again real. Notice that in principle it seems possible to define an \enquote{extremal limit} of this solution by imposing $Q=0$, such that $\QQ=(q_+^2-\sqrt{\alpha})^2$ and $q_+= \alpha^{1/4}$. However this constraint can be solved only if $\alpha=1$, which is excluded by the requirement $p_+ \neq p_-$.

\vspace{5mm}

Summarizing this section, we have that both $\lens=0$ and $\lens\neq 0$ are possible for the values of the parameters in both (i) and (ii), Moreover the solutions are always real, with the only exception of the no-twist case in (ii), where $\lens=0$ and the spindle is replaced by $S^2$. In appendix~\ref{appendix: Values} we show that, when $\lens\neq 0$, only $\lens\ge 4$ is possible. Notice also that the case (ii) contains the self-dual solution~\cite{Martelli:2011fu}. Indeed we have $N=M=0$, and requiring also $P=Q$ implies $\sqrt{\alpha}=q_+=1$, which is degenerate because the bolt disappears and the topology is the one of AdS$_4$~\cite{Martelli:2011fu}.

\subsection{Limits to old solutions}\label{subsect:Limits to the old solutions}
As anticipated, our solution~\eqref{solution:metric}-\eqref{metricfunc} generalizes in various ways the ones presented in~\cite{Martelli:2012sz,Martelli:2013aqa}. In this section we study the limits which  reproduce those results.

\subsubsection*{$U(1)\times U(1)$-invariant solution of~\cite{Martelli:2013aqa}}
Let us start from the case~\cite{Martelli:2013aqa}, which is simpler to be recovered since the local form of the solution is exactly the same. The main difference from this reference is that therein the regularity led to $P=Q$, $N=M$ and $q_+=p_+$ for which the boundary is topologically a three-sphere and the bulk contain a nut-type singularity in $q_+=p_+$. Moreover, exactly when $N=M$ and $P=Q$ the Weyl tensor is self-dual and $F$ is an instanton. The condition $q_+=p_+$, in terms of the parametrization~\eqref{parameterisation}, reads 
\begin{equation}
	q_+\rightarrow p_+ \iff \free \rightarrow 1+x\,.
\end{equation}
Moreover, in this limit $q_+$ and $p_+$ must be zeros of the same type of function~\eqref{susy-func-dec}, which means
\begin{equation}
	\signpp=\signqp \implies\branch=1\,,
\end{equation}
Finally, since the spindle bolt disappears, we require
\begin{equation}
	\singm-\singp\rightarrow0\,.
\end{equation}
Notice that in this degenerate situation, contrarily to the general case, $\text{Re}(\signpp P)<0$ due to~\eqref{singm-singp} supplemented with~\eqref{constraint}. With these requirements, it can be seen that $N-M\propto P-Q\rightarrow 0$, as expected. For future reference, we write also the value of the fibration parameter 
\begin{equation}\label{old-lens}
	\begin{aligned}
		\twist=1&:\quad &&\lens\rightarrow 0\,,
		\\
		\twist=-1&:\quad &&\lens\rightarrow \frac{-2}{x}\labell\,.
	\end{aligned}
\end{equation}
Clearly in this limit the interpretation of $\lens$ breaks down (and the negative sign is an artefact due to $\text{Re}(\signpp P)$ not being positive anymore). In particular, the Type I and Type II solutions of~\cite{Martelli:2013aqa} have boundary topology of $S^3$, for which more properly $\lens=1$ in both cases.

\subsubsection*{$SU(2)\times U(1)$-invariant solution of~\cite{Martelli:2012sz}}
Now we move to the $SU(2)\times U(1)$ invariant solution presented in~\cite{Martelli:2012sz} and summarized in our conventions in appendix~\ref{appendix:Old-bolt-solutions}. In the spirit of~\cite{Farquet:2014kma} (see appendix C therein), we construct a scaling limit under which~\eqref{solution:metric} and~\eqref{solution:graviphoton} become~\eqref{old-sol:metric} and~\eqref{old-sol:graviphoton}, at least locally. Notice that in our case the solution is not (anti) self-dual, so that~\cite{Farquet:2014kma} must be extended. The limit is obtained for $\epsilon\rightarrow0^{+}$ on the coordinates
\begin{equation}\label{scaling-limit:coords}
	\begin{aligned}
		&p=
		s(1-\epsilon x\cos\hat\theta)\,,\quad \tau=s\biggl[2\hat\psi+\frac{\hat\varphi}{\epsilon x}\biggr]\,,\quad &&\sigma=\frac{\hat\varphi}{ s }\frac{1}{\epsilon x}\,,
	\end{aligned}
\end{equation}
as well as on the parameters
\begin{equation}\label{scaling-limit:pars}
	\begin{aligned}
		&N=-s(4s^2-1)\,,\quad E=1-6s^2\,,\quad  &&\alpha=\hat{P}^2-s^2\bigl[3s^2-1+\epsilon^2x^2\bigr]\,,
	\end{aligned}
\end{equation}
where we employed the parametrization~\eqref{parameterisation} for the roots and $\hat\theta\in[0,\pi]$.  Moreover, we identify
\begin{equation}\label{identifications}
	\begin{aligned}
		q=r&\,,\quad(M,P,Q)=(\hat M,\hat P,\hat Q)\,,\quad w= s\,,
	\end{aligned}
\end{equation}
where $s$ is the squashing parameter as in~\eqref{squashing}. In the limit, the BPS conditions in~\eqref{susy} boil down to~\eqref{old-BPS}, as expected.  Notice that under this scaling limit we get $p_+\rightarrow p_-$, which is in a certain sense the complementary situation to $q_+\rightarrow p_+$ of the previous subsection. As $p_+\rightarrow p_-$, the spindle degenerates and the parametrization variable $w$ is unconstrained, since $\singp/\singm=1+O(\epsilon)$. However, the limit procedure instructs us that $w=s+O(\epsilon)$, and applying this on~\eqref{thew} one obtains
\begin{equation}\label{twist-limit-integers}
	\twist=1:\quad w=s+O(\epsilon) \iff \singpm=\frac{1}{2 (2s^2+\signpp\freetwist)}\mp\frac{s^2(r_0^2+s^2+\signpp\freetwist)}{(r_0^2-s^2)(2s^2+\signpp\freetwist)^2}\,\epsilon x+O(\epsilon^2)\,.
\end{equation}
Even if it looks odd, using the $1/4$-BPS conditions~\eqref{old-BPS} the first term is exactly $1$, as expected for a spindle which becomes a sphere. This expression makes~\eqref{twist-angles} infinite
\begin{equation}\label{requested-angle2}
	\Delta= \frac{2\pi (r_0^2-s^2)}{s }\frac{1}{\epsilon x}+O(\epsilon)\,,
\end{equation}
in the precise way such that $\text{vol}(\spindle_{q_{+}})<\infty$ and~\eqref{NHL} coincides with (4.7) in~\cite{Martelli:2012sz}. Plugging the values~\eqref{twist-limit-integers} in~\eqref{lens-twist} and inverting the parametrization~\eqref{parameterisation} to
\begin{equation}
	\twist=1:\quad	\freetwist=\frac{P}{2 s}\,,\quad\free=\frac{r_0}{s}\,,
\end{equation}
we obtain correctly that~\eqref{lens-twist} boils down to $\lens=\labell\hat{\lens}+O(\epsilon)$, respectively, for the appropriate class of regularity conditions~\eqref{old-regularity-quarter} with $\branch=\pm1$. 
The procedure can be now repeated also for the anti-twist case, for which
\begin{equation}\label{anti-limit-integers}
	\twist=-1:\quad \singpm= 1\mp\frac{2s\bigl[s+(r_0^2-s^2)(2s\mp\sqrt{4s^2-1})\bigr]}{r_0^2-s^2}\,\epsilon x+O(\epsilon^2)\,,
\end{equation}
where we already plugged~\eqref{old-BPS}, and with
\begin{equation}
	\twist=-1:\quad \freeanti=-\frac{P(-2 s^2 \epsilon x+\signpp P)}{4s^3}\,,\quad \free=\frac{r_0}{s}\,,
\end{equation}
one gets again $w=s+O(\epsilon)$ and $\lens=\labell\hat{\lens}+O(\epsilon)$.

Finally, let us comment on the relation with the supersymmetric nuts of~\cite{Martelli:2012sz}. These must correspond to the even more degenerate case in which $p_-$, $p_+$ and $q_+$ coincide, thus finishing all the possibilities. We shall refrain to construct explicitly this limit, since it was already observed in~\cite{Martelli:2013aqa} that the supersymmetric $1/2$ and $1/4$-BPS nuts of~\cite{Martelli:2012sz} are subcases of the nuts in~\cite{Martelli:2013aqa}. In addition, the nuts of~\cite{Martelli:2012sz} must come from a limit in which $r_0\rightarrow s$ on the Bolts solutions, so that the function $\Omega(r)$ in~\eqref{old-sol:metricfunc} has a double root in $r_0=s$ and the nut singularity becomes \enquote{naked}.  Therefore, one can think to perform the aforementioned limit in two steps, moving $q_+$ to $p_+$ before, and collapsing $p_-$ on $p_+$ subsequently. In the next section we will compute the on-shell action for our solution, showing later that it reproduces the Type I and $1/4$-BPS on-shell actions for $\twist=1$, and Type II and $1/2$-BPS for $\twist=-1$.

	\section{On-shell action}\label{subsect:On-shell action}

	In this section we compute the renormalized on-shell action for the solutions~\eqref{solution:metric}-\eqref{metricfunc}. At this point it should be clear that the twist case $\twist=1$ discussed in section~\ref{subsect:Quantization} is simpler, and indeed we will be able to express the on-shell action in a compact form, depending only on $(\lens,\labell,\singpm)$. Even if the solution has two free continuous parameters $(\freetwist,\free)\in\RR$, these do not appear in the action. It is the same phenomenon already observed in the Type I solutions of~\cite{Martelli:2013aqa} and in the $1/4$-BPS Bolt solutions of~\cite{Martelli:2012sz}, which we summarized in section~\ref{subsect:A list of notable solutions}.
	Different is the case for the anti-twist $\twist=-1$, where the action will be a function of one (of the two) free real parameters, $\free$. Again, this is familiar from the old Type II~\cite{Martelli:2013aqa}, $1/2$-BPS~\cite{Martelli:2012sz} and black hole \cite{Cassani:2021dwa} solutions. Using the results of section~\ref{subsect:Limits to the old solutions}, we will show that the on-shell action computed here reduces to~\eqref{Type I and Type II actions} and~\eqref{bolt actions} in the appropriate limit. We will then demonstrate how to reproduce our result starting from the localization of the action~\cite{BenettiGenolini:2019jdz,BenettiGenolini:2023kxp}, thus furnishing a non-trivial check of the validity of the method. We will  conclude this section with an intriguing observation about the extremization of the action.

	 \subsection{Holographic renormalization}\label{subsubsect:Holographic renormalization}
	 	
		The computation is a standard procedure in literature
	 \cite{Emparan:1999pm,Skenderis:2002wp} so we will be as brief as possible.  It consists in adding the standard boundary term and the counter-terms to remove the singularities that appears as $q\rightarrow\infty$. In particular we consider the renormalized action
	 \begin{equation}
	 	S_{\text{ren}}=S_{{E}\Lambda}+S_{F}+S_{GH}+S_{\text{ct}}\,,
	 \end{equation}
	 evaluated on the solution~\eqref{solution:metric} and~\eqref{solution:graviphoton}, where the bulk contributions are
	 \begin{equation}\label{actions}
	 	\begin{aligned}
	 		S_{E\Lambda}=\frac{-1}{16\pi G_{4}}\int \dd^{4}x \sqrt{g} \,(R^{(g)}+6)\,,
	 		\quad
	 		S_{F}=\frac{-1}{16\pi G_{4}}\int \dd^{4}x \sqrt{g} \,(F^2)\,,
	 	\end{aligned}
	 \end{equation}
	 whilst the boundary ones are
	 \begin{align}
	 	S_{GH}=\frac{-1}{8\pi G_{4}} \int_{q=q_0} \dd^{3}x \sqrt{\gamma}\,(\gamma^{\mu\nu}K_{\mu\nu}^{(\gamma)})\,,\quad	S_{ct}=\frac{1}{8\pi G_{4}}  \int_{q=q_0} \dd^{3}x \sqrt{\gamma}\,(2+\frac{1}{2}R^{(\gamma)})\,.
	 \end{align}
	 The hypersurface $S_{{0}}=\{q=q_0\}$ with induced metric $\gamma_{ij}$ acts as a cut-off, and in the limit $q_0 \rightarrow+\infty$ the total action should be finite. Here we have introduced the second fundamental form $2K_{\mu\nu}\equiv (\mathcal{L}^{(g)} _{n} \gamma)_{\mu\nu}$, with $n^{\mu}$ a unit vector normal to $S_0$ and such that $g_{\mu\nu}= n_\mu n_\nu + \gamma_{\mu\nu}$. In the case at hand, we have 
	 \begin{equation}
	 	n= \sqrt{\frac{\QQ}{q^2- p^2}}\,\partial_{q}\,,\quad K^{(\gamma)}=\nabla_{\mu}^{(g)}n^{\mu}=\frac{2 q \QQ  + (q^2-p^2)\mathcal{Q}'(q)}{2 (q^2 -p^2)^{\frac{3}{2}}\sqrt{\QQ}}\,.
	 \end{equation}
	 Whilst the contribution of the gauge field in~\eqref{actions} is already finite
	 \begin{align}\label{non-accelerating gauge action}
	 	S_{F}&=(16\pi G_{4} ) \Delta\tau \Delta\sigma(p_+ - p_-) q_+ \biggl[\frac{(P-Q)^2}{(q_+ - p_+)(q_+ - p_-)}+\frac{(P+Q)^2}{(q_+ + p_+)(q_+ + p_-)}\biggr]\,,
	 \end{align}
	 the Einstein$-\Lambda$ term is indeed divergent in the limit $q_0 \rightarrow \infty$
	 \begin{equation}
	 	\begin{aligned}
	 		S_{E\Lambda}&=(32\pi G_{4}) \Delta\tau \Delta\sigma\int_{p_-}^{p_+}\dd p\big[q_0 ^3 -q_+ ^3+ 3(q_+- q_0)p^2\big]\,.
	 	\end{aligned}
	 \end{equation}
	 Moreover
	 \begin{equation}
	 	\begin{aligned}
	 		S_{GH}+S_{\text{ct}}=(32\pi G_{4} ) \Delta\tau \Delta\sigma\int_{p_-}^{p_+}\dd p\big[-q_0 ^3 + 
	 		3q_0 p^2+M\big]\,,
	 	\end{aligned}
	 \end{equation}
	 so that the terms containing $q_0$ cancel.
	 Then the renormalized on-shell action does not display infinities and reads
	 \begin{equation}{\label{renormalized-action}}
	 	\begin{aligned}
	 		S_{\text{ren}}=\frac{\Delta\tau \Delta\sigma (p_+ - p_-)q_+}{8 \pi G_4}\biggl[&p_+ ^2+p_+ p_-+p_-^2+\frac{M}{q_+}-q_+ ^2
	 		\\
	 		&+\frac{(P-Q)^2}{2(q_+ - p_+)(q_+ - p_-)}+\frac{(P+Q)^2}{2(q_+ + p_+)(q_+ + p_-)}\biggr]\,,
	 	\end{aligned}
	 \end{equation}
	 where recall that the product $\Delta\tau \Delta\sigma$ is equal in the patches $U_\pm$ and is given by~\eqref{period-product}. Let us anticipate that in the limit $P\rightarrow Q$ and $q_+\rightarrow p_+$ the action~\eqref{renormalized-action} returns to that of~\cite{Martelli:2013aqa} (see (5.7) therein), thus confirming our expectations.
	 
	For the twist case ($\twist=1$), employing all the results of section~\ref{subsect:Quantization}, we can express the renormalized action simply as
	 \begin{equation}
	 	\begin{aligned}{\label{easy-OS}}
	 		\text{twist}:\quad S_{\text{ren}}
	 		=\frac{ \pi }{8 G_4 \labell} \biggl[{2\chi_\spindle}{}-\branch\frac{\lens/\labell}{\singm\singp}-\branch \frac{\labell(\singm-\singp)^2}{\lens\,\singm\singp}\biggr]\,,\quad \lens\ge 2\,.
	 	\end{aligned}
	 \end{equation}
	 In particular, notice that   the on-shell action is fixed in terms of the integers $(\lens,\labell,\singpm)$ and $(\freetwist,\free)$ do not appear. From this expression we can easily recover the Type I~\eqref{Type I and Type II actions} and $1/4$-BPS~\eqref{bolt actions} renormalized actions using with the limits explained in section~\ref{subsect:Limits to the old solutions}.  For the Type I case, $\singpm\rightarrow 1$ and $\lens\rightarrow 0$ from~\eqref{old-lens}, for which only the first contribution in~\eqref{easy-OS} survives. For the other case $\singpm\rightarrow 1+O(\epsilon)$ from~\eqref{twist-limit-integers} but $\lens=\labell \hat{\lens}+O(\epsilon)$ remains non-zero. Thus only the last term in~\eqref{easy-OS} vanishes. From this point of view it is not a case that the two actions in the first lines of~\eqref{Type I and Type II actions},~\eqref{bolt actions} coincide for $\hat{\lens}=0$. Notice that it is still left a factor $\labell^{-1}$ in front of the action, which generalizes the old results by the presence of a transverse singularity to the bolt.

	 The $\twist=-1$ is, as usual, more involved. Looking explicitly at~\eqref{renormalized-action}, it proves useful to define
	 \begin{equation}\label{q-trick}
	 	\begin{aligned}
	 		\twist=-1:\quad\mathfrak{q}&=\frac{q_+^2-p_+^2}{p_+^2-p_-^2}\,\frac{(p_+-p_-)\big[\signpp P+p_-(p_++p_-)\big]}{\bigl[(p_+-p_-)\signpp P-(p_++p_-)(q_+^2-p_+ p_-)\bigr]}
	 		\\
	 		&=\frac{\bigl[\free ^2-(1+x)^2\bigr]\bigl[w(x-1)-\signpp\freetwist\bigr]}{2w(\free^2+x^2-1)-4 x \signpp\freetwist}\,.
	 	\end{aligned}
	 \end{equation}
	 Notice that~\eqref{q-trick} seems to depend also on $\freeanti$, but if one solves for $\freetwist$ and $w$ through~\eqref{further} and~\eqref{thew-anti}, $\freeanti$ disappears. Recall also that $x=x(\free)$ is fixed (non-explicitly) via~\eqref{lens-anti} which in this context reads
	 \begin{equation}\label{lens-charge}
	 	\begin{aligned}
	 		\twist=-1:\quad t=&\labell \bigl[(\free^2-(x+1)^2)\singm -(\free ^2-(x-1)^2)\singp\bigr]\times
	 		\\
	 		& \frac{-\free^3+\free \bigl[1+2(1+2\mathfrak{q})x+x^2\bigr]+\branch\bigl[\free ^2-(1+x)^2+2\mathfrak{q}(-1+\free ^2+x^2)\bigr]}{x \bigl[1-\free ^2+2(1+2\mathfrak{q})x+x^2\bigr]^2}\,.
	 	\end{aligned}
	 \end{equation}
	 Therefore, as $x(\free)$, also $\mathfrak{q}=\mathfrak{q}(\free)$. 
	 The renormalized action for $\twist=-1$ can then be written as a function of $\free$ only
	 \begin{equation}
	 	\begin{aligned}\label{anti-twist-ren}
	 		\twist=-1:\quad S_{\text{ren}}(\free)=&\frac{\pi \bigl[(\free^2-(x+1)^2)\singm -(\free ^2-(x-1)^2)\singp\bigr]^{2}}{8 G_4 \lens\, \singm\singp}\times
	 		\\
	 		& \frac{2\free (1+x+2\mathfrak{q})x-\branch\bigl[\tilde{q}_+ ^2+(1+x)^2+2\mathfrak{q}(1+\free ^2+x^2)\bigr]}{x^2 \bigl[1-\free ^2+2(1+2\mathfrak{q})x+x^2\bigr]^2}\,.
	 	\end{aligned}
	 \end{equation}
	 We see that in the anti-twist case $S_{\text{ren}}$ not only depends on the choice of the integers $(\lens,\labell,\singpm)$, but contains a real degree of freedom $\free$ as well. Again, we can obtain the Type II~\eqref{Type I and Type II actions} and $1/2$-BPS~\eqref{bolt actions} with the limits of section~\ref{subsect:Limits to the old solutions}. With $\lens=-2\labell/x$ from~\eqref{old-lens} and $\singpm\rightarrow1$, we get 
	 	\begin{equation}\label{typeii}
	 	\begin{aligned}
	 		\text{Type II}&:\quad S_{\text{ren}}=\frac{\pi}{ 2 G_4}\frac{1}{1-4[Q/(p_++p_-)^2]^2}\,.
	 	\end{aligned}
	 \end{equation}	
	 This is equivalent to the one presented in~\eqref{Type I and Type II actions} when $p_++p_-=1$, a condition fixed by the authors of~\cite{Martelli:2013aqa} taking advantage of the scaling symmetry \eqref{scaling}. Here we have restored that factor, since $p_+ +p_-\neq1$ for us.  The limit to the $1/2$-BPS action in more complicated, since in this limit $\mathfrak{q}$ is infinite. However, $\mathfrak{q}/\singpm$ is finite and using~\eqref{anti-limit-integers} and $\lens=\labell\hat{\lens}+O(\epsilon)$ we obtain indeed the second line of~\eqref{bolt actions}.
	 
	 We point out that the $1/2$-BPS and $1/4$-BPS nuts of~\cite{Martelli:2012sz} are subcases of the ones in~\cite{Martelli:2013aqa}. The action of the $1/4$-BPS nut in~\eqref{nut actions} is already equal to the one of Type I in~\eqref{Type I and Type II actions}. For $1/2$-BPS, we can rewrite the Type II  action~\eqref{typeii} using the parametrization~\eqref{parameterisation} as
	 \begin{equation}
	 	\text{Type II}:\quad S_{\text{ren}}=\frac{2\pi}{G_4}\frac{w^4}{4 w^4-P^2}\,,
	 \end{equation}
	 where recall that $Q=P$ for Type I and Type II. Then, since $w=s$ and $P=P(s)$ from~\eqref{identifications} and~\eqref{old-BPS}, we reproduce exactly the second line of~\eqref{nut actions}.
	 
	 We conclude by writing (some)  of the on-shell actions for the solutions with $N=0$ in section~\ref{subsubsect: N=0}. We have 
\begin{equation}\label{N=0 actions}
\begin{aligned}
&\text{(i) with twist}:\quad &&S_{\text{ren}}=\frac{ \pi }{8 G_4 \labell} \biggl[{2\chi_\spindle}{}+\delta\frac{\lens/\labell}{\singm\singp}+\delta \frac{\labell(\singm-\singp)^2}{\lens\,\singm\singp}\biggr]\,,\quad &&& \lens\ge 4\,,
\\
&\text{(i) with anti-twist}:\quad 	&&S_{\text{ren}}=\frac{\pi(1+\delta q_+)^2}{G_4 \labell \bigl[1+2 \delta q_+(1-\twist \sqrt{\alpha})
	-\alpha\bigr]}\,,\quad&&& \lens=0\,,
\\
&\text{(ii) with twist}:\quad && S_{\text{ren}}=\frac{\pi}{2 G_{4} \labell }\,,\quad&&& \lens=0\,,
\end{aligned}
\end{equation}	 
The anti-twist actions in (i) and (ii) with $\lens\neq 0$ are as usual very complicated, and we do not report them here. Some comments are now in order. As anticipated, the action in the second line refers to the Euclidean counterpart of the Kerr-Newman-AdS black hole~\cite{Klemm:2013eca,Cassani:2019mms}. Indeed it can be rewritten in the form
\begin{equation}
	S_{\text{ren}}=\pm\frac{1}{2G_4 \labell} \frac{\varphi^2}{z}\,,\quad \varphi-\frac{1}{2}z=\pm\pi\,,
\end{equation}
where we have introduced
\begin{equation}
	\varphi=\pm2\pi \frac{(1+\delta q_+)}{1-\twist\sqrt{\alpha}}\,,\quad z=\pm2\pi\frac{\bigl[1+2\delta q_+ (1-\twist \sqrt{\alpha})-\alpha\bigr]}{(1-\twist\sqrt{\alpha})^2}\,.
\end{equation}
This is a particular case of the general action for the accelerating black hole\footnote{Notice the absence of the $\ii$ factors with respect to~\eqref{action of the BH} or~\cite{Cassani:2019mms}. This is due to the fact that here we are working in Euclidean signature.}~\eqref{action of the BH}. When $\singm=\singp$ the magnetic charge of the spherical black hole is zero, \ie\ $Q_m=0$, and $\chi(S^2)=2$, so that the second term in~\eqref{action of the BH} is absent. Finally, the action of the Euclidean \enquote{Topological Kerr-Newman-AdS} solution in the last line of~\eqref{N=0 actions} agrees with both the Type I and $1/4$-BPS actions  (when $\hat{\lens}=0$ and $\labell=1$) in~\eqref{Type I and Type II actions},~\eqref{bolt actions}. Moreover, it is also compatible with the $\mathrm{g} = 0$ case of  a topologically twisted black hole with a $\Sigma_{\mathrm{g}}$ horizon~\cite{BenettiGenolini:2023ucp}. In Lorentzian signature the solution with $\mathrm{g}=0$ does not exist, as explained at the end of section \ref{subsubsect: N=0}, but it is perfectly well-defined in Euclidean.

	  \subsection{Equivariant localization}\label{subsubsect:Equivariant localization}
	
	We now want to recover these values of the on-shell actions using the results of~\cite{BenettiGenolini:2019jdz,BenettiGenolini:2023kxp}, with reference to figure~\ref{fig:non-compact polytope non-zerot}, normal vectors~\eqref{toric-fan} and toric weights~\eqref{weights}. To this end, we elaborate briefly on the general setting, rewriting the main result of~\cite{BenettiGenolini:2019jdz} in a way which is more useful for our purposes. Moreover, we will make a connection with the gravitational blocks for the supersymmetric black holes proposed in~\cite{Faedo:2021nub}, which correspond to figure~\ref{fig:non-compact polytope zerot} and $\lens=0$.
	
	 In~\cite{BenettiGenolini:2019jdz} the action is computed on-shell by using the fact that for supersymmetric solutions with a supersymmetric Killing vector $\vec{	\epsilon}=(\epsilon_1,\epsilon_2)$ the bulk metric can be written as a $U(1)$ circle fibration generated by $\vec{	\epsilon}$ over a non trivial three-dimensional base. The action reduced on this base generalizes~\cite{Gibbons:1979xm} by the presence of a gauge field and is then evaluated explicitly employing supersymmetry and the bilinears constructed on top of it. The final formula turns out to be dependent only on quantities at the fixed points of $\vec{	\epsilon}$, which resembles the results of equivariant localization, \ie\ the Duistermaat-Heckman, Berline-Vergne or Atiyah-Bott theorems \cite{Duistermaat:1982vw,berline1982classes,Atiyah:1984px}. In fact, the procedure presented in~\cite{BenettiGenolini:2023kxp,BenettiGenolini:2023ndb} is based on the equivariant integration of a certain equivariant closed poly-form $\Phi$ with respect to the supersymmetric Killing vector\footnote{We use the the notation $\vec{\epsilon}$ when the supersymmetric Killing vector is written in an effective toric basis $\vec{E}_i$, so that in this basis $\epsilon=\vec{	\epsilon}\cdot\vec{E}_i$.}
	 \begin{equation}
	 	{\epsilon}=\varepsilon^{\dagger}\Gamma^{\mu}\Gamma_{\star}\varepsilon\,\partial_{\mu}\,,
	 \end{equation}
	 that is $(\dd-i_{\vec{	\epsilon}})\Phi=0$, and depends on the chirality of $\vec{	\epsilon}$ at the fixed points as well. 
	 In particular, the \enquote{off-shell} action (\ie\ the action for a generic \enquote{Reeb} vector $\vec{\epsilon}$ ) is computed as
	 \begin{equation}{\label{Genolini}}
	 	I_{\text{off-shell}}(c_A,\vec{\epsilon}\,)=\frac{\pi}{2 G_{4}}\biggl\{\sum_{A=1}^{2} -c_A \frac{\big[b_1^{(A)}-c_A b_2^{(A)}\big]^2}{4 \, d_A \,b_{1}^{(A)}b_{2}^{(A)}}\biggr\}\,, 
	 \end{equation}
	 where $c_A$ are signs linked to the chirality of the spinor at the fixed point $p_A$~\eqref{fixed-points}, on which the sum runs, and $b_{i}^{(A)}$ are the weights of $\vec{	\epsilon}$ on each copy of $\CC_{i}\subset T_{p_{A}}\mathcal{M}_4 $. Since, more precisely, the tangent space at the fixed points is not just $\CC\oplus\CC$ but there are orbifolds singularities, a factor of $d_A$ (the order of the singularity at $p_A$) must be included~\cite{BenettiGenolini:2024xeo}, so that now~\eqref{Genolini} holds also for orbifold solutions.
	 For the toric case, the weights of $\vec{\epsilon}$ can be simply expressed in terms of the toric weights~\eqref{weights} as~\cite{BenettiGenolini:2019jdz}
	 \begin{equation}\label{toric-relation}
	 	b^{(A)}_{1,2}\equiv\vec{\epsilon} \cdot \vec{\mu}^{\,(A)}_{1,2}\,,
	 \end{equation}
	 where for our case they are given by~\eqref{weights}.
	 
	 A word of caution is needed here: our solutions are \enquote{bolt} in the sense that $K_{q_{+}}$ introduced in section~\ref{subsubsect:Regularity in the bulk} has a two-dimensional fixed point set $S_+=\{q=q_+\}$, but are \enquote{two-nuts} solutions with respect to the supersymmetric Killing vector which degenerates only in $(q_+,p_\pm)$  as we will see. Therefore we use the \enquote{nuts} result in~\cite{BenettiGenolini:2023kxp}, which is indeed~\eqref{Genolini}. Moreover~\eqref{Genolini} is exactly the result found previously in~\cite{Farquet:2014kma} in a different way for (anti-)self-dual solutions with the topology of the four-ball, \ie\ with a single nut. There it is shown that the bulk is regular only if $\epsilon_2/\epsilon_1>0$ or $\epsilon_2=-\epsilon_1$, and these two distinct behaviours will play a role later on. We note  that at the moment it is not known if, for a generic choice of $\vec{	\epsilon}$, there exists a supergravity solution with such a specific vector. However, if it exists, its renormalized on-shell action should be given by~\eqref{Genolini}. We will come back to this observation in~\ref{subsubsect:Comments on extremization}.
	 
	 In~\cite{BenettiGenolini:2023kxp} it is showed that~\eqref{Genolini}, when computed on the supersymmetric Killing vector $\vec{\epsilon}_{\ast}$, reproduces the supergravity computation for the renormalized action, that is
	 \begin{equation}
	 	I_{\text{on-shell}}\equiv I_{\text{off-shell}}(c_A, \vec{\epsilon}^{}_{\ast})=S_{\text{ren}}\,,
	 \end{equation}
	 in our language. To show this explicitly in our case, let us present a more convenient expression for $I_{\text{off-shell}}$ in the toric case, when~\eqref{toric-relation} holds. Since for a generic set-up of two normal vectors $\vec{v}_{1,2}$, the weights can be written as in~\eqref{generic-weights}, it turns out that~\eqref{Genolini} can be rewritten in a symmetric form as 
	 \begin{equation}\label{general}
	 	I_{\text{off-shell}}(c_A,\mathrm{Q})= \frac{\pi}{8 G_4}\biggl[ 2 \Big[\frac{1}{d_{1,2}}+\frac{1}{d_{2,3}}\Big]-\frac{ c_{1}}{d_{1,2}}\Bigl[\mathrm{Q}^2+\frac{1}{\mathrm{Q}^2}\Bigr] -\frac{ c_{2}}{d_{2,3}}\Bigl[\tilde{\mathrm{Q}}^2+\frac{1}{\tilde{\mathrm{Q}}^2}\Bigr]  \biggr]\,,
	 \end{equation}
	 where $d_{a,b}=\det(\vec{v}_a,\vec{v}_b)$ and we defined the \enquote{off-shell} quantities
	 \begin{equation}{\label{off-shell-charges}}
	 	\mathrm{Q}^2=-\frac{\det(\vec{\epsilon},\vec{v}_1)}{\det(\vec{\epsilon},\vec{v}_2)}\,,\quad\tilde{\mathrm{Q}}^2=-\frac{\det(\vec{\epsilon},\vec{v}_3)}{\det(\vec{\epsilon},\vec{v}_2)}\,.
	 \end{equation}
	 Notice that~\eqref{general} can be also written as
	 \begin{equation}
	 	I_{\text{off-shell}}(c_A,\mathrm{Q})= -\frac{\pi}{8 G_4}\biggl[  \frac{c_1}{d_{1,2}} \left( {\mathrm{Q}}-  \frac{c_1}{{\mathrm{Q}}}\right)^2  +\frac{ c_{2}}{d_{2,3}}\left({\tilde{\mathrm{Q}}}-\frac{c_2}{{\tilde{\mathrm{Q}}}}\right)^2  \biggr]\,.
	 \end{equation}
and that in general the action can be complex.
The formula for the localized action~\eqref{Genolini} is invariant under a generic $SL(2,\ZZ)$ transformation of the basis of the torus. This is evident from~\eqref{general}, which depends only on the $SL(2,\ZZ)$-invariant determinants contained in~\eqref{off-shell-charges}. It is also independent on the overall normalization of the vector $\vec{\epsilon}$. Moreover these quantities are not independent, since
	 \begin{align}\label{vect-id}
	 	d_{2,3} \mathrm{Q}^2 + d_{1,2}\tilde{\mathrm{Q}}^2=-d_{1,3}\,,
	 \end{align}
	 where we used that for a generic vector $\vec{v}$ the following four-vectors identity holds
	 \begin{align}
	 	d_{1,2} d_{v,3}+ d_{2,3} d_{v,1}=d_{1,3}d_{v,2}\,.
	 \end{align}
	 Specifically, it is clear that in general the on-shell action depends on the  supersymmetric Killing vector $\vec{\epsilon}=(\epsilon_1,\epsilon_2)$ only through the combination $z\equiv \epsilon_2/\epsilon_1$. To make this dependence explicit  it is useful to insert the toric data~\eqref{toric-fan}.
After  defining the (a priori complex) variables
  \begin{align}\label{newvars}
	 	\beta^2_+ \equiv \frac{1}{\labell}\left(\frac{\singp}{z}+ r_- \right) \,,\qquad 
		\beta^2_- \equiv \frac{1}{\labell}\left( \frac{\singm}{z}-r_+ \right) \, ,
	 \end{align}
 the action takes the suggestive form
 \begin{equation}
 \label{nicestesosa}
I _{\text{off-shell}}(\twist,z) =-\frac{c_2\pi}{8 G_4\labell }\biggl[ \frac{1}{\singm}\left( \beta_- - \frac{c_2}{\beta_-}\right)^2- \frac{\twist}{\singp} \left( \beta_+ + \frac{\twist c_2}{\beta_+}\right)^2   \biggr] \,,
\end{equation}
where $c_1c_2=\twist$ and
 \begin{equation}
\singm \beta^2_+ - \singp \beta^2_- = \frac{\lens}{\labell}\,.
\end{equation}

In the next section we will indeed show that the on-shell actions of our solutions, for twist or anti-twist cases, take precisely this form.  In addition, in~\eqref{nicestesosa} the  dependence on $z$ is particularly simple and we expect that this form  should be suitable for comparisons with large $N$ calculations in the dual field theories.  Next, let us consider the case $\lens=0$,   that is relevant for supersymmetric black holes. This corresponds to the rectangular polytope in figure~\ref{fig:non-compact polytope zerot},  and may be obtained from~\eqref{nicestesosa} simply  setting $r_\pm=0=\lens$. 
		The resulting expressions for the twist and anti-twist  read
 \begin{align}\label{equivariant t=0:AT}
	  	\text{anti-twist}: & \quad  I^\text{anti-twist}_{\text{off-shell}} (z)=-\frac{c_2\pi}{8 G_4}\biggl[    \left( \frac{\sqrt{z}}{\singm}  - \frac{c_2}{\labell\sqrt{z}}\right)^2  +   \left( \frac{\sqrt{z}}{\singp}  - \frac{c_2}{\labell\sqrt{z}}\right)^2\, \biggr]\,,\, &&\lens=0\,, \\
	  	\label{equivariant t=0:T}
	  	\text{twist}:&	 \quad  I^\text{twist}_{\text{off-shell}} (z)=\frac{\pi \chi_\spindle }{8 G_4} \biggl[    \frac{2}{\labell}  +c_2 \left(\frac{1}{ \singp} - \frac{1}{\singm}\right) z \biggr]\,, \,&& \lens=0\,.
	  	\end{align}
		It is then straightforward to see that the two expressions above reproduce the gravitational block form of the black hole 
		entropy functions for minimal $d = 4$  gauged supergravity,  proposed\footnote{These expressions have been proved in supergravity using  different methods,  considering near-horizon geometries with a spindle factor~\cite{Boido:2022iye,Boido:2022mbe,Martelli:2023oqk,BenettiGenolini:2024kyy}.  See also  \cite{Hristov:2024cgj}  for a discussion in the context of higher derivative supergravity. 
		In the dual field theories these are reproduced by the large $N$ limit of the spindle index~\cite{Colombo:2024mts}.}  in~\cite{Faedo:2021nub}. Indeed, defining 
		 \begin{align}		
		\mathfrak{n}=\frac{\singp+\twist\singm}{4\singm\singp}\,, \qquad 2\varphi-\frac{\singp-\twist\singm}{2\singm\singp}z=\frac{1}{\labell}\,,
\end{align}
we have\footnote{\label{footnp}We have adapted the overall normalization and inserted a $\labell$ factor, to compare to our results. Moreover in both cases $c_2=-1$.} 
 \begin{equation}
 \label{t=0osas}
 I^\text{anti-twist}_{\text{off-shell}} (z) = F^+ (\varphi,z;\mathfrak{n}) =\frac{\pi}{G_4}\biggl[  \frac{\varphi^2}{z}+ \mathfrak{n}^2 z \biggr]\,,\quad   I^\text{twist}_{\text{off-shell}} (z)= F^- (\varphi,z;\mathfrak{n})=\frac{\pi \mathfrak{n}\varphi}{G_4}2\,,
 \end{equation}
 with
	 \begin{equation}
	 F^{-\twist}(\varphi,z;\mathfrak{n})=\frac{1}{z}\Bigl[\mathcal{F}_3 (\varphi+\mathfrak{n} z)-\twist \mathcal{F}_3 (\varphi -\mathfrak{n} z)\Bigr]\,, 	
  \end{equation}
 where  $\mathcal{F}_3$ is proportional to the $S^3$ off-shell free energy of the ABJM theory (\cf\ Table 2 in~\cite{Faedo:2021nub}) and $\twist=\pm 1$ is as usual for the twist or the anti-twist, respectively.   Notice that for the twist on the sphere, with $c_2=c_1$ and $\singm=\singp=1$, the on-shell action does not depend any more on the supersymmetric Killing vector, and the result is simply $I=\pi/(2 \labell G_4)$.

	 \subsection{Comparison}

	 We will now  compare the general results obtained from localization  with the  gravitational on-shell actions~\eqref{easy-OS} and~\eqref{anti-twist-ren} computed performing holographic renormalization. Again we will refer to the polytopes of figure~\ref{fig:non-compact polytope}, normal vectors~\eqref{toric-fan} and toric weights~\eqref{weights}.
	From the explicit Killing spinors~\eqref{four-spinor} we obtain the supersymmetric Killing vector\footnote{We have normalized differently the spinor $\varepsilon$ than before, but the overall normalization will not be important since the action~\eqref{nicestesosa} depends only on $z$.}
	 \begin{equation}\label{susy-killing-vector}
	 	{\epsilon}_{\ast}=\sqrt{\alpha}\partial_\tau+\partial_\sigma\, .
	 \end{equation}
   Notice that this degenerates only at the points  $(q_+,p_\pm)$ (as visible from~\eqref{Killing-norm}), so that the fixed points $p_A$ are \enquote{nuts} for $	{\epsilon}_{\ast}$.
	 From~\eqref{four-spinor-fixed-twist} and~\eqref{four-spinor-fixed-antitwist} we recall that
	 \begin{equation}\label{easy-chirality}
	 	\twist=1:\quad 
	 	\begin{aligned}
	 		\Gamma_{\star} \varepsilon &= \branch \varepsilon  \text{ at }p_1\,,
	 		\\
	 		\Gamma_{\star} \varepsilon &=\branch\varepsilon \text{ at } p_2\,,
	 	\end{aligned}
	 \end{equation}
	 and
	 \begin{equation}\label{non-easy-chirality}
	 	\twist=-1:\quad 
	 	\begin{aligned}
	 		\Gamma_{\star} \varepsilon &=   -\branch\varepsilon &&\text{ at }p_1\,,
	 		\\ 
	 		\Gamma_{\star} \varepsilon &= \branch \varepsilon &&\text{ at } p_2\,,
	 	\end{aligned}
	 \end{equation}
	 where the chirality matrix has been defined in~\eqref{gammas}.  The chirality itself is fixed by the value of $\branch=\pm 1$, so that the correct assignment for the signs $c_A$ is
	 \begin{equation}\label{prescription}
	 	\begin{aligned}
	 		\twist=1&:\quad &&c_2=c_1= \branch\,,
	 		\\
	 		\twist=-1&:\quad &&c_2=-c_1=\branch\,.
	 	\end{aligned}
	 \end{equation}
That the spinor must be chiral at the fixed points comes from general arguments, in particular as a consequence of the bilinear $\varepsilon^{\dagger}\varepsilon$ being nowhere vanishing ~\cite{Ferrero:2021etw,BenettiGenolini:2023kxp}.
	 In the coordinates~\eqref{effective-coordinates}, which recall are written in the patch $U_-$, the supersymmetric Killing vector~\eqref{susy-killing-vector} reads 
	 
	 \begin{equation}\label{explicit-Reeb}
	 	\vec{	\epsilon}_{\ast}=\frac{1}{q_+^2-p_-^2}\biggl\{\biggl[{(p_- ^2+\sqrt{\alpha})}\frac{2\pi}{\Delta\angleone}+\frac{r_+}{\singm} {(q_+^2+\sqrt{\alpha})}\frac{2\pi}{\Delta\angletwo^{-}}\biggr]E_1+{(q_+^2+\sqrt{\alpha})}\frac{2\pi}{\Delta\angletwo^{-}}E_2\biggr\}\,.
	 \end{equation}
	 It is now a matter of algebra to plug explicitly the vector $\vec{\epsilon}_{\ast}$~\eqref{explicit-Reeb} in~\eqref{general} by using~\eqref{off-shell-charges}. After some work, using the B\'ezout’s lemma~\eqref{Bezout} and eliminating everywhere $\Delta\theta_1$ from~\eqref{explicit-Reeb} in favour of the fibration parameter $\lens$ using~\eqref{bound-lens}, the result takes the form 
	 \begin{equation}\label{Genolini-action}
	 	\begin{aligned}
	 		I_{\text{on-shell}}(c_A, \mathtt{q}_\ast)=
	 		&\frac{\pi}{8G_4 \labell}\biggl[  2\chi_\spindle -\frac{c_1}{\singp}\Bigl[\frac{\mathtt{q}_{\ast}^2\, \lens}{\labell \singm}+\frac{\labell\singm}{\mathtt{q}_{\ast}^2\, \lens}\Bigr]-\frac{c_2}{\singm}\Bigl[\frac{\tilde{\mathtt{q}}_{\ast}^2 \,\lens}{\labell \singp}+\frac{\labell\singp}{\tilde{\mathtt{q}}_{\ast}^2\, \lens}\Bigr]\biggr]\,,
	 	\end{aligned}
	 \end{equation}
	 where we have defined the on-shell reduced quantities
	 \begin{equation}\label{reduced-charges}
	 	\begin{aligned}
	 		\tilde{{\mathtt{q}}}_{\ast}^2=\frac{q_+^2 -p_+^2}{p_+^2-p_-^2}\,\frac{p_-^2+\sqrt{\alpha}}{q_+^2+\sqrt{\alpha}}\,,\quad
	 		{\mathtt{q}}_{\ast}^2=-\frac{q_+^2 -p_-^2}{p_+^2-p_-^2}\,\frac{p_+^2+\sqrt{\alpha}}{q_+^2+\sqrt{\alpha}}\,,\quad \tilde{\mathtt{q}}_{\ast}^2+\mathtt{q}_{\ast}^2=-1\,.
	 	\end{aligned}
	 \end{equation}
	 such that $(\mathrm{Q}^2_\ast,\tilde{\mathrm{Q}}^2_\ast)=(\lens/\labell\singm\singp)(\singp\mathtt{q}_\ast^2,\singm\tilde{\mathtt{q}}_\ast^2)$. As a consequence of the reality discussion at the end of section~\ref{subsect:Killing spinors}, the parameter $\sqrt{\alpha}$ here can be complex and in turn the action itself can be complex. It should be no surprise at this point that
	 the equal-signs case ($c_2=c_1=\branch$) is particularly easy to handle. From the point of view of the equivariant localization, this simplification comes from $\tilde{\mathtt{q}}_{\ast}^2+\mathtt{q}_{\ast}^2=-1$ which enters directly in~\eqref{Genolini-action}. Indeed we have
	 \begin{align}\label{on-shell-charge-twist}
	 	\text{twist}:\quad ({\mathtt{q}}_{\ast}^2,\tilde{\mathtt{q}}_{\ast}^2)=\frac{1}{\singm-\singp} (\singm,-\singp)\,,
	 \end{align}
	 and as a consequence
	 \begin{equation}\label{Ios-twist}
	 	\text{twist}:\quad I_{\text{on-shell}}(c_2,\mathtt{q}_\ast)=\frac{ \,\pi }{8 G_4 \labell} \biggl[{2\chi_\spindle}{}-c_2\frac{\lens/\labell}{\singm\singp}-c_2 \frac{\labell(\singm-\singp)^2}{\lens\,\singm\singp}\biggr]\,,\quad\lens\ge 2\,,
	 \end{equation}
	 which reproduces precisely~\eqref{easy-OS} for $c_2=\branch$ as evident from~\eqref{easy-chirality} and~\eqref{prescription}.  Correspondingly, the Reeb vector~\eqref{explicit-Reeb} is fixed
	 \begin{equation}\label{fixed-Reeb}
	 	\text{twist}:\quad\vec{	\epsilon}_\ast= f(\freetwist,\free)\bigl[(r_++r_-)E_1+(\singm-\singp)E_2\bigr]\,,
	 \end{equation}
	 where $f$ is a complicated but unimportant function of the free real parameters of the solution $(\freetwist,\free)$.
	 
	 When $c_2=-c_1=\branch$, using the parametrization~\eqref{N-alpha} with $\twist=-1$, it is very easy to recognize that $\mathfrak{q}=\mathtt{q}_{\ast}$ in~\eqref{q-trick}. Then, gathering this fact, the expression~\eqref{lens-charge} and the non-trivial identity  
	 \begin{equation}
	 	\twist=-1:\quad\frac{\singp}{\singm}=-\frac{\tilde{\mathtt{q}}_{\star}^2\bigl[2 \tilde{\mathtt{q}}_{\star}^2(-1+\free^2-2x+x^2)-(2-x)(\free^2-(x+1)^2)\bigr]}{(1-\tilde{\mathtt{q}}_{\star}^2)\bigl[2\tilde{ \mathtt{q}}_{\star}^2(-1+\free^2+2x+x^2)+x(\free^2-(x+1)^2)\bigr]}\,,
	 \end{equation}
	 it is possible to show that indeed~\eqref{Genolini-action} matches~\eqref{anti-twist-ren}. In this case the ratio of the entries of the Reeb vector~\eqref{explicit-Reeb} is not fixed, and depends on $\free$ but not on $\freeanti$, as the action~\eqref{anti-twist-ren}. This is then a highly non-trivial check of the validity of the equivariant procedure of~\cite{BenettiGenolini:2023kxp}. We have also checked  that the on-shell actions for $N=0$ case 
	 in  (\ref{N=0 actions})  are also correctly reproduced by the  expressions  obtained with equivariant localization.

	   \subsection{Extremization}\label{subsubsect:Comments on extremization}
	 
Since we have  an \emph{off-shell} action as a function of the equivariant parameter $z\equiv {\epsilon}_2/{\epsilon}_1$, it is natural to wonder whether  the on-shell actions corresponding to the explicit solutions extremize this. In fact, we will  show that the extremization of~\eqref{nicestesosa}  with respect $z$ reproduces the gravitational results for the renormalized on-shell action computed previously \emph{only in the twist case} $\twist=1$. Choosing again $c_2=c_1=\branch$ for the twist, there are two values of $z$ that extremize $I$, namely
	 \begin{equation}
	 \label{critcalreebtwsit}
	 	z_{\pm}^{\ast}=\frac{\pm \singm - \singp}{\pm r_+ + r_-}\,.
	 \end{equation}
	 It is then straightforward to compute the action at the extremizing values, using B\'ezout’s lemma~\eqref{Bezout}. At the saddle $z^{\ast}_{+}$ we have
	 \begin{align}
		  I _{\text{off-shell}}(c_A,z^*_+) = S_{\text{ren}}^{\,(\twist=1)}\,,
	 \end{align}
	 where the latter is given for example in~\eqref{easy-OS}. Indeed, the value of $z_+ ^\ast$ is constant and compatible with~\eqref{fixed-Reeb}. Notice that in the limit $\singm\rightarrow\singp$, $z_+^\ast\rightarrow\labell$ from~\eqref{lens-twist} with $\free=1+x$ and $\branch=1$ as discussed in~\ref{subsect:Limits to the old solutions}. For $\labell=1$, this is the expected result for the $1/4$-BPS spherical bolt (see~\eqref{Killing vectors spherical bolts}).
	 For the other extremum, with the same choice of signs, the action reads
	 \begin{align}\label{mystery}
	 	I_{-}^{\ast}(\singpm,\labell)\equiv I(\singpm,\labell, z_{-}^{\ast})= \frac{ \pi }{8 G_4 \labell} \biggl[{2\chi_\spindle}{}-\branch\frac{\lens/\labell}{\singm\singp}-\branch \frac{\labell(\singm+\singp)^2}{\lens\,\singm\singp}\biggr]\,.
	 \end{align}
	 This equation does not reproduce any known result for $\singm=\singp=1$. Moreover, whilst the vector $\vec{\epsilon}_+=-(1,z_+^\ast)$ always lies inside the polytope, that is $\vec{	\epsilon}_+\cdot\vec{\mu}^{\,(A)}_{1,2}>0$,  $\vec{\epsilon}_-=-(1,z_-^\ast)$ is always outside it. 
	 
	On the other hand,  in the anti-twist case, the on-shell actions match the off-shell one  for generic values of $z$, depending on continuous parameters, and therefore there is \emph{no extremization}  taking place in this case. At present we do not have an understanding of  the reason underlying this difference\footnote{One first observation is that this  is consistent with the behaviour observed in~\cite{Farquet:2014kma}. Even if there the focus was on solutions with a single nut in the bulk geometry, the authors observed that the solutions were everywhere regular only if the components of the Reeb vector satisfy $\epsilon_2/\epsilon_1>0$ or $\epsilon_2=-\epsilon_1$ (with some conventional choice on the sign of one between $\epsilon_i$). }.  In the case $t=0$, with anti-twist, we noted that the on-shell action reduces to the black hole entropy functions (\ref{t=0osas}) which indeed matches with the complexified black hole off-shell action for generic values of the parameter $z$. 
	However, in that  case it is known that a Legendre transform yields the extremal BPS black hole.  Assuming that in the twist case solutions must extremize the off-shell action, we then conclude from (\ref{t=0osas}) that in the minimal gauged supergravity there cannot be supersymmetric black holes with twist. 
	This is consistent with the fact that supersymmetric accelerating black holes with twist have been found only the STU supergravity~\cite{Ferrero:2021ovq,Ferrero:2021etw,Couzens:2021cpk}. 
	However, for $m_+=m_-=1$ the off-shell action becomes a constant
and indeed we noted  that the Topological Kerr-Newman-AdS solution realizes this possibility.
	This  may then be interpreted as the gravitational saddle dual to the  large $N$ limit of the topologically twisted index 
\cite{Benini:2015noa} for the sphere.

It is interest to note  that the equivariant volume introduced in~\cite{Martelli:2023oqk} reproduces the action~\eqref{Genolini} \emph{only} in the twist case. To see this, we write the fixed-point formula for the equivariant volume presented   in~\cite{Martelli:2023oqk}  in our conventions (already specialized to the planar case) as\footnote{No sum over $A$ in the exponent is intended.}
	 \begin{align}
	 	\mathbb{V}(\lambda_a,\vec{\epsilon})=\frac{\pi}{4 G_4}\sum_{A=1}^{2}\frac{1}{d_{A}}\prod_{i=1}^{2}\frac{\mathrm{e}^{\lambda^{A}_{a_{i}}b^{(A)}_{i}}}{b^{(A)}_{i}}\,,
	 \end{align}
	 where the $b^{(A)}_i$ are defined in~\eqref{toric-relation}, the K\"{a}hler parameters for each divisor are identified as $\lambda^1 _1 \equiv\lambda_1$, $\lambda^1 _2=\lambda^2_1\equiv \lambda_2$, $\lambda^2 _2\equiv\lambda_3$ and the coefficient $(\pi/4G_4)$ has here been tuned \emph{a posteriori} to have agreement with the supergravity computations.
	 We now Taylor expand this formula around $\lambda_a\sim 0$ and pick the second order term
	 \begin{equation}
	 	\begin{aligned}
	 		\frac{4 G_4}{\pi}	\mathbb{V}(\lambda_A,\vec{\epsilon})\Big|_{\lambda_a^2}&=\sum_{A=1}^{2}\frac{1}{d_A}\biggl[\lambda^{A}_{1}\lambda^{A}_{2}+\frac{1}{2}\frac{b^{A}_1 \lambda^{A}_{1}}{b^{A}_2}+\frac{1}{2}\frac{b^{A}_2 \lambda^{A}_{2}}{b^{A}_1}\biggr]
	 		\\
	 		&=\frac{1}{d_1}\biggl[\lambda_1\lambda_2+\frac{1}{2}\frac{b^{(1)}_1 \lambda_1}{b^{(1)}_1}+\frac{1}{2}\frac{b^{(1)}_2 \lambda_2}{b^{(1)}_1}\biggr]+\frac{1}{d_2}\biggl[\lambda_2\lambda_3+\frac{1}{2}\frac{b^{(2)}_1 \lambda_2}{b^{(2)}_2}+\frac{1}{2}\frac{b^{(2)}_2 \lambda_3}{b^{(2)}_1}\biggr]\,.
	 	\end{aligned}
	 \end{equation}
	 To write each term in square brackets in the form of~\eqref{Genolini}, we need $\lambda_1=\lambda_2=\pm1$ but also $\lambda_2=\lambda_3=\pm1$. But then $\lambda_1=\lambda_2=\lambda_3\equiv-\branch=\pm 1$ and the (second order of the) equivariant volume becomes
	 \begin{equation}
	 	\mathbb{V}(\branch,\vec{\epsilon}_{})\Big|_{\lambda^2}=\frac{\pi}{2 G_4}\sum_{A=2}^{3} \frac{1}{d_A}\biggl[-\branch  \frac{\big[b_1^{(A)}-\branch \,b_2^{(A)}\big]^2}{4 \,b_{1}^{(A)}b_{2}^{(A)}}\biggr]\,.
	 \end{equation}
	 This is indeed~\eqref{Genolini} for $c_A=\branch$, which is the twist case~\eqref{prescription}.

	\section{Accelerating solutions}\label{sect:Accelerating solutions}
	
	We now come back to the study of the general class of Pleba\'nski-Demianski solutions~\eqref{accelerating:metric}-\eqref{metricfunc}.  Since the analysis will be considerably more difficult, but also similar, to sections~\ref{subsect:Regularity} and~\ref{subsect:Toric data}, we will recap only the key ingredients and results.  Before studying the regularity of the boundary and the bulk metrics, we will compute the holographically renormalized on-shell action of the general non supersymmetric PD solutions. Indeed, during the procedure we will obtain the correct boundary metric, which will be analyzed subsequently. Although we will  not attempt here to solve explicitly the supersymmetry equation~\eqref{KSE}, we will be able to infer the supersymmetric Killing vector indirectly by comparing  the solution at the boundary with the general analysis of~\cite{Inglese:2023wky,Inglese:2023tyc}.

		\subsection{General on-shell action}\label{subsect:General on-shell action}
	
	We start by computing the general on-shell action for the non-supersymmetric solutions. Such general result is not present in the literature, although it has been computed in numerous sub-cases (see \eg~\cite{Papadimitriou:2005ii,Martelli:2013aqa,Anabalon:2018qfv,Cassani:2021dwa}).  We consider again the renormalized action, which is obtained by the procedure explained in section~\ref{subsect:On-shell action}. For convenience, we recap here the various terms
	\begin{equation}
		\begin{aligned}
			S_{E\Lambda}+
			S_{F}+S_{GH}+S_{ct}=&-\frac{1}{16\pi G_{4}}\int \dd^{4}x \sqrt{g} \,\bigl[R^{(g)}+6-F^2\bigr]
			\\
			&+\frac{1}{8\pi G_4}\int \dd^{3}x \sqrt{\gamma}\, \bigl[2+\frac{1}{2}R^{(\gamma)}-K^{(\gamma)}\bigr]\,,
		\end{aligned}
	\end{equation}
	where the second integral must be computed on the boundary, which is now located at $q=(\acc p)^{-1}$ instead of at $q=+\infty$. As usual, the contribution of the gauge field from $S_F$ is already finite, and we can anticipate the result
	\begin{equation}\label{accelerating-gauge-contribution}
		\begin{aligned}
			\frac{(16\pi G_{4})S_{F}}{ (\newpplus-\newpminus) \Delta\tau \Delta\sigma}=&\newqplus \biggl[\frac{(P-Q)^2}{(\newqplus -\omega  \newpplus)(\newqplus- \omega\newpminus)}+\frac{(P+Q)^2}{(\newqplus + \omega\newpplus)(\newqplus +\omega \newpminus)}\biggr]\,,
			\\
			&+\acc (\newpplus-\newpminus)\frac{2\acc\omega P Q(\newpplus^2+\newpminus^2)-(P^2+Q^2)(1+\acc^2\omega^2\newpplus^2\newpminus^2 )}{(1-\acc^2\omega^2\newpplus^4)(1-\acc^2\omega^2\newpminus^4)}\,.
		\end{aligned}
	\end{equation}
	Here we integrated simply $q$ between $\mathbf{q}_+$ and  $(\acc p)^{-1}$ and $p$ between $\mathbf{p}_-$ and $\mathbf{p}_+$. Notice that the first line, which is non-vanishing for $\acc\rightarrow 0$, coincides with~\eqref{non-accelerating gauge action}. To compute correctly the on-shell action, we need to bring the metric in the Fefferman-Graham form~\cite{AST_1985__S131__95_0}, following standard literature (see for example~\cite{Papadimitriou:2005ii,Anabalon:2018qfv}), which is
	\begin{equation}
		\dd s^2 \underset{r\rightarrow0^+}\simeq\frac{\dd r^2}{r^2}+\frac{\dd s^2_{3}}{r^2}\,,
	\end{equation}
	where
	\begin{equation}
		\dd s^2_{3}= \dd s^2_{(0)}+ \dd s^2_{(1)} \,r+ \dd s^2_{(2)}\, r^2+ \dd s^2_{(3)}\,r^3+O(r^4)\,,
	\end{equation}
	and the leading contribution is (a conformal representative of) the boundary metric, $\dd s_{(0)}^2\equiv \dd s^{2}_{\gamma}$. To this end we make the following change of coordinates\footnote{By inspection, we noticed that $f_{5}(x)$ and $g_5(x)$ do not enter in $\dd s^{2}_{(3)}$.} 
	\begin{equation}\label{FG-change}
		q=\frac{1}{\acc x}-\sum_{i=1}^{4} f_{i}(x) r^i \,, \quad p=x+ \sum_{i=1}^{4} g_{i}(x) r^i\,,
	\end{equation}
	where all the functions except for $f_1(x)$ are determined by requiring the absence of mixed $(\dd r\,\dd x)$ terms in the metric\footnote{In particular, for the procedure to be correct, we have to expand the metric until $\dd s^2 _{(5)}$ and require that the Ricci scalar of the metric in $(r,x)$ coordinates is $R^{(g)}=-12+\mathcal{O}(r^4)$.} and that the coefficient of $\dd r^2$ is $1$. As an example, we get 
	\begin{equation}
		g_1(x)=\frac{\mathbf{P}(x)}{\acc x^2\mathbf{Q}\bigl(\frac{1}{\acc x}\bigr)}f_1(x)\,,
	\end{equation}
	where the following identity holds
	\begin{equation}\label{identity-for-the-boundary}
		\mathbf{Q}(\acc^{-1} p^{-1})=\frac{\Pfunc}{\acc^2 p^4}+\frac{p^{-4}-\acc^2 \omega^2}{\acc^4}\,.
	\end{equation}
	In passing, let us notice that $f_1(x)$ defines a conformal class for the boundary metric, and we expect that the finite terms are independent of it. For future reference, let us define a specific boundary metric $\dd s^2 _{b}$ given by
	\begin{equation}
		\dd s^2 _{(0)}=\frac{\mathbf{Q}\big(\frac{1}{\acc x}\big)}{f_{1}(x)^2 (1-\omega^2\acc^2 x^4)^2 }\dd s_{b}^2\,,
	\end{equation}
	where 
	\begin{equation}\label{accelerating-boundary-metric}
		\begin{aligned}
			\dd s_{b}^2 =\frac{(1-\omega^2\acc^2 x^4)^2\dd x^2}{-\mathbf{P}(x)}&-\frac{\acc^8 x^8 \mathbf{Q}\big(\frac{1}{\acc x}\big)^2 \mathbf{P}(x)}{\big[1+\acc^2 \mathbf{P}(x)\big]}\dd \sigma^2
			\\
			&+\acc^4 x^4 \big[1+\acc^2 \mathbf{P}(x)	\big]\mathbf{Q}\big(\frac{1}{\acc x}\big) \biggl[\dd\tau+\frac{\omega x^2}{1+\acc^2 \mathbf{P}(x)}\dd\sigma\biggr]^2\,.
		\end{aligned}
	\end{equation}
	The strategy now is to perform a mixed analysis, where the bulk integrals will be computed in the original coordinate system $(p,q)$, whilst the boundary ones will be in $(x,r)$ coordinates, taking due care to the Jacobian for~\eqref{FG-change}. The bulk contribution is
	\begin{equation}
		S_{E\Lambda}=-\frac{\Delta\tau\Delta\sigma}{16\pi G_4}\int_{\newpminus}^{\newpplus} \dd p\int_{\newqplus}^{q_{\epsilon}(p)}\dd q \sqrt{g}\,(-12+6)\,,
	\end{equation}
	where $q_{\epsilon}(p)$ is determined as follows. The expansion for $p$ in~\eqref{FG-change} can be inverted order by order to give
	\begin{equation}\label{xforpandr}
		x(p,r)=p+\sum_{i=1}^{4}\tilde{g}_{i}(p) r^i+\mathcal{O}(r^5)=p-g_1(p)r+\bigl[g_1'(p)g_1(p)-g_2(p)\bigr] r^2+\ldots+O(r^5)\,,
	\end{equation}
	where the derivatives are taken with respect to $p$. From this mixed expression for $x(p,r)$ one easily gets $q=q(p,r)$ by inserting~\eqref{xforpandr} in the $q$ expansion~\eqref{FG-change}. Finally, we define the limit of integration to be $q_{\epsilon}(p)\equiv q(p,r)|_{r=\epsilon}$.  From this computation we get divergent terms, proportional to $\epsilon^{-3},\epsilon^{-2},\epsilon^{-1}$, and a finite contribution. The infinite parts should be cancelled from the boundary, which gives
	\begin{equation}
		\begin{aligned}
			S_{ct}+S_{GH}
			&=\frac{ \Delta\tau\Delta\sigma}{8\pi G_4} \int_{\mathbf{p}_-}^{\mathbf{p}_+}\dd p |\partial_{p} x(p,\epsilon)|\sqrt{\gamma} \biggl[2+\frac{1}{2}R^{(\gamma)}-K^{(\gamma)}\biggr]\bigg|_{x=x(p,\epsilon)}\,,
		\end{aligned}
	\end{equation}
	where $\partial_{p} x(p,\epsilon)$ can be computed from~\eqref{xforpandr}. The terms that diverge as $\epsilon^{-3}$ for $\epsilon\rightarrow0^{+}$ are cancelled in the sum before computing the integral in $\dd p$. Differently, both the $\epsilon^{-2}$ terms are (separately) integrated to zero, because they  both can be written as (proportional to)
	\begin{equation}
		\partial_{p}\biggl[\frac{\acc ^8 p^4 \Pfunc \mathbf{Q}^2(\acc^{-1} p^{-1})}{(1-\omega^2 \acc^2 p^4)^2 f_{1}(p)^2}\biggr]\frac{1}{\epsilon^2}\,,
	\end{equation}
	Since $\mathbf{P}(\newpplusminus)=0$ by definition, they do not contribute to the final answer. Finally, the terms proportional to $\epsilon^{-1}$, after being summed up, are again of the form $\partial_{p}\bigl[\Pfunc F(p)\bigr]$ for some function $F(p)$. After some work we can integrate the finite part of the computation, finding
	\begin{equation}\label{acceleratingOSaction}
		\begin{aligned}
			S_{\text{ren}}=\frac{\Delta\tau\Delta\sigma}{\acc 16\pi G_4}\biggl[&\frac{-\newqplus^2}{(1-\acc p \newqplus)^2}-\frac{\omega^2 (1-2\acc p \newqplus)}{\acc^2 \newqplus^2 (1-\acc p \newqplus)^2}+\frac{\omega^2 p^2 (1+16 \omega^2\acc^2 p^4)}{1-\omega^2\acc^2 p^4}
			\\
			&-\frac{\acc^2 \bigl[64\omega^4 p^6+4\omega^2 p^3 \mathbf{P}'(p)-[\mathbf{P}'(p)]^2\bigr]}{4(1-\omega^2 \acc^2 p^4)}-\frac{\mathbf{P}''(p)}{6}
			\biggr]_{\newpminus}^{\newpplus}+S_{F}\,,
		\end{aligned}
	\end{equation}
	with $S_F$ as in \eqref{accelerating-gauge-contribution}. Despite appearances, after the result is made explicit by substituting $\Pfunc$, the expression is finite for $\acc\rightarrow 0$ as expected. Indeed, in this limit we get
	\begin{equation}{\label{non-accelerating-renormalized-action}}
		\begin{aligned}
			S_{\text{ren}}=\frac{\Delta\tau \Delta\sigma (\newpplus - \newpminus)\newqplus}{8 \pi G_4}\biggl[&\omega^2\bigl[\newpplus ^2+\newpplus \newpminus+\newpminus^2\bigr]+\frac{M}{\newqplus}-\newqplus ^2
			\\
			&+\frac{(P-Q)^2}{2(\newqplus -\omega \newpplus)(\newqplus -\omega \newpminus)}+\frac{(P+Q)^2}{2(\newqplus + \omega \newqplus)(\newqplus +\omega \newpminus)}\biggr]\,,
		\end{aligned}
	\end{equation}
	which coincides with~\eqref{renormalized-action} for $\omega=1$.
	
	 It is also interesting to compare our result~\eqref{acceleratingOSaction} to the on-shell action of~\cite{Cassani:2021dwa}. Such an accelerating black hole is obtained from the general (Lorentzian) PD solution by turning off the NUT parameter. In our parametrization of the solution we should impose (see \eg~\cite{Podolsky:2022xxd})
	\begin{equation}\label{no-nut conditions}
		\begin{aligned}
			N&=-\omega \acc M \,.
			\\
			E&=\frac{N \omega}{\acc M }+\frac{\acc}{\omega M N }\bigl[(M^2+N^2)(-P^2+\alpha)\omega^2+N^2 (P^2-Q^2)\bigr]
			\,,
		\end{aligned}
	\end{equation}
	With these conditions imposed we have
	\begin{equation}
		\Pfunc=(p^2-1)\Bigl\{P^2-\alpha-2\acc M p + \Bigl[\acc^2\bigl[P^2-Q^2-(P^2-\alpha)\omega^2\bigr]+\omega^2\Bigr]p^2\Bigr\}\,,
	\end{equation}
	so that the roots are taken to be $\newpplusminus=\pm 1$ \cite{Griffiths:2005qp,Ferrero:2020twa}.
	Moreover, to compare with~\cite{Cassani:2021dwa} we identify
	\begin{equation}\label{wicktoBH}
		\newqplus=r_{+}\,,\quad \acc=\alpha\,,\quad Q=-\ii e\,,\quad P=g\,,\quad M=m\,,\quad \omega=-\ii a\,,\quad \alpha=P^2-1 \,, 
	\end{equation}
	where the $\ii$ factors are due to the analytic continuation from our Euclidean solution. Upon these identifications,~\eqref{acceleratingOSaction} matches perfectly the one presented in~\cite{Cassani:2021dwa}. 
	\vspace{5mm}
	
	Summarizing,~\eqref{acceleratingOSaction} is the most general on-shell action for the PD solution, it is very general and is valid independently on the way in which the roots of $\Pfunc$ are chosen.
	
	\subsection{Supersymmetric Killing vector}\label{subsect:Supersymmetric Killing vector}
	
	Even if we do not want to solve the supersymmetry equation~\eqref{KSE}, we will be interested in the supersymmetric Killing vector: from this, as in section~\ref{subsubsect:Equivariant localization}, we will recover our on-shell action~\eqref{acceleratingOSaction}, once evaluated on some specific cases in the next sections. In particular we can obtain the supersymmetric Killing vector using the boundary metric~\eqref{accelerating-boundary-metric} and the boundary gauge field
	\begin{equation}\label{acceleratin_boundary_gauge}
		A_{b}=\frac{\acc p (P\omega \acc  p^2-Q)\dd\tau+(p P-Q\omega\acc p^3)\dd\sigma}{1-\omega^2\acc^2p^4}\,,
	\end{equation}
	obtained from~\eqref{accelerating_graviphoton} evaluated at the boundary $q=(\acc p)^{-1}$. Following~\cite{Inglese:2023wky,Inglese:2023tyc}, we write the boundary fields as\footnote{Notice an overall minus sign with respect to~\cite{Inglese:2023wky}. Moreover, we have performed a specific gauge transformation to compare with~\eqref{acceleratin_boundary_gauge}.}
	\begin{equation}
		\begin{aligned}
			\dd s^2 _b &= f(p)^2 \dd p^2+h_{11}(p)\dd\sigma^2+2h_{12}\dd\sigma\dd\tau+ h_{22}\dd\tau^2\,,
			\\
			A_{b}&=\frac{-v(p)^3}{4 f(p)\sqrt{h(p)}}\biggl[\frac{1}{\mathrm{w}}\partial_{p}\Bigl[\frac{h_{11}(p)}{v(p)^2}\Bigr]\dd\sigma-\partial_{p}\Bigl[\frac{h_{22}(p)}{v(p)^2}\Bigr]\dd\tau\biggr]+\frac{M P-N Q}{2 \acc (P^2-Q^2)}\dd\sigma+\frac{N P-M Q}{2(P^2-Q^2)}\dd\tau\,,
		\end{aligned}
	\end{equation}
	where $h(p)\equiv \det(h_{ij})$ and
	\begin{equation}
		v(p)^2=h_{11}(p)+2 \mathrm{w} h_{12}(p)+\mathrm{w}^2 h_{22}(p)^2\,.
	\end{equation}
	In this formalism, the supersymmetric 
	killing vector reads\footnote{The overall normalization does not play any role in the following discussions.}
	\begin{equation}\label{accelerating:susy vector}
		{\epsilon}_{\ast}\propto \partial_{\sigma}+\mathrm{w}\partial_{\tau}\,,
	\end{equation}
	and by comparison we obtain the relative weight $\mathrm{w}$ as
	\begin{equation}\label{accelerating:weight}
		\mathrm{w}=\frac{\acc \rho}{\omega(MQ-NP)}\frac{\alpha\omega (2\acc M N \rho^2+\omega \mu^2)-E \rho(N^2 \rho+\omega^2P^2 \mu)}{2\acc N P \rho^2+\omega (M P-N Q)\mu +2\omega^2\acc P^2 (MQ-N P)\rho}\,,
	\end{equation}
	where for convenience we defined $\rho=P^2-Q^2$ and $\mu=M^2-N^2$. This formula still needs to be evaluated on the supersymmetric relations~\eqref{accelerating-susy-easy} or~\eqref{accelerating-susy-complicated}. In the simpler case of~\eqref{accelerating-susy-easy}, we get
	\begin{equation}\label{easy-accelerating-weight}
		\mathrm{w}=\frac{1}{\acc}\frac{MP-NQ}{NP-MQ}\,,
	\end{equation}
	which is ill-defined in the limit $\acc\rightarrow0$. Using instead~\eqref{accelerating-susy-complicated}, we obtain a complicated formula which reduces to
	\begin{equation}
		\mathrm{w}\underset{\acc\rightarrow 0}{\rightarrow} \sqrt{\alpha}\,,
	\end{equation}
	as expected from the direct computation in the non-accelerating case~\eqref{susy-killing-vector}. 
	
	\subsection{Global analysis and toric data}\label{subsect:Regularity and toric data}
	
	For the regularity analysis, we follow the same procedure of section~\ref{subsect:Regularity}. In particular, we introduce again two patches $U_{\pm}$ on which we change coordinates
	\begin{equation}{\label{acceleration:change:tau-theta}}
		U_\pm:\quad	\tau^{\pm} = \parqp^{} \newqplus ^2\angleone^{\pm} -\parppm  \omega  \newpplusminus^2\angletwo^{\pm}\,,\quad \sigma^{\pm}=-\parqp^{} \omega \angleone^{\pm}+\parppm^{}\angletwo^{\pm}\,,
	\end{equation}
	for which $(\Delta\tau\Delta\sigma)^{\pm}=(\mathbf{q}_+^2-\omega^2\mathbf{p}_{\pm}^2)\Delta\angleone\Delta\angletwo^{\pm}$. Upon this change, the boundary metric~\eqref{accelerating-boundary-metric} becomes\footnote{Since at the leading order $p=x+\mathcal{O}(r)$ from~\eqref{xforpandr}, we can take the boundary metric in $(p,q)$ as written in~\eqref{accelerating-boundary-metric} with $x=p$.}
	\begin{equation}
		\begin{aligned}
			U_\pm:\quad \dd s^2 _b=&\frac{(1-\omega^2\acc^2 p^4)^2}{-\Pfunc} \dd p^2-\frac{ (\newqplus^2-\omega^2\newpplusminus^2)^2 \Pfunc \bigl[1-\omega^2\acc^2 p^4+\acc^2\Pfunc\bigr]^2}{\bigl[(\newqplus^2-\omega^2 p^2)^2+(\acc^2 \newqplus^4-\omega^2)\Pfunc\bigr]}(\dd\angletwo^{\pm})^2
			\\
			&+\bigl[1-\omega^2\acc^2 p^4+\acc^2\Pfunc\bigr]\bigl[(\newqplus^2-\omega^2 p^2)^2+(\acc^2 \newqplus^2-\omega^2)\Pfunc\bigr]\bigl[\dd \angleone^{\pm}+\mu_{f}^{\pm}\bigr]^2\,,
		\end{aligned}
	\end{equation}
	where $\mu_{f}^{\pm}$ is a fibration one-form and reads
	\begin{equation}
		U_\pm:\quad \mu^{(\pm)}=\omega\frac{(\newqplus^2-\omega^2 p^2)(p^2-\newpplusminus^2)+(1-\acc^2\newpplusminus^2\newqplus^2)\Pfunc}{(\newqplus^2-\omega^2 p^2)^2 +(\acc^2\newqplus^4-\omega^2)\Pfunc}\frac{\parppm}{\parqp}\dd\angletwo^{\pm}\,.
	\end{equation}
	Not surprisingly, this is the accelerating version of the boundary metric~\eqref{boundary-metric}. As a consequence, we define again
	\begin{equation}
		\Delta=(\mathbf{q}_+^2-\omega^2\mathbf{p}_{+}^2)\Delta\angletwo^{+}=(\mathbf{q}_+^2-\omega^2\mathbf{p}_{-}^2)\Delta\angletwo^{-}\,,
	\end{equation}
	and impose 
	\begin{equation}\label{acclelerating: spindle and lens}
		U_\pm:\quad \pm\frac{\mathbf{P}'(\newpplusminus)}{2(\newqplus^2-\omega^2\newpplusminus^2)}\Delta=\frac{2\pi}{\newsingpm}\,,\quad \newlens=\omega\,\newsingm\newsingp\frac{\newpplus^2-\newpminus^2}{(\newqplus^2-\omega^2\newpplus^2)(\newqplus^2-\omega^2\newpminus^2)}\frac{\Delta}{\Delta\angleone}\,.
	\end{equation}
	The regularity condition from the bulk is identical to~\eqref{spindle-cond1}
	\begin{equation}\label{accelerating: labell condition}
		\frac{\mathbf{Q}'(\newqplus)}{2}  \Delta\angleone=\frac{2\pi}{\newlabell}\,,
	\end{equation}
	and collecting all these expressions we can write
	\begin{equation}\label{accelerating: period-product}
		\Delta\tau^{\pm}\Delta\sigma^{\pm}=\Delta\angleone\Delta=\omega\frac{\newsingm\newsingp}{\newlens}\Delta^2\frac{\newpplus^2-\newpminus^2}{(\newqplus^2-\omega ^2 \newpminus^2)(\newpplus^2-\omega^2 \newpminus^2)}\,,
	\end{equation}
	similarly to~\eqref{period-product}.
	Once everything is imposed, the topology is again $\CC/\ZZ_{\labell}\hookrightarrow\mathcal{O}_{}(-\newlens)\rightarrow\spindlesing$. Consequently, the polytope describing this accelerating Bolt solution is the same as in figure~\ref{fig:non-compact polytope}, with normal singularity to $\loci_2=\{q=\newqplus\}$ given by $\newlabell$. In particular we can use the same basis~\eqref{effective-coordinates}, and all the other expressions~\eqref{normalized Killing vectors}-\eqref{Killing in toric basis} remain
	 unchanged\footnote{See also footnote-\ref{symnotoric}.}. The only (small) difference lies in the definition of the Killing vectors, that should be taken as
	\begin{equation}
U_{\pm}:\quad	K_{\mathbf{q}_{+}}=\parqp(\mathbf{q}_+^2 \partial_{\tau^{\pm}}-\omega\partial_{\sigma^{\pm}})=\partial_{\angleone^{\pm}}\,,\quad 
	K_{\mathbf{p}_{\pm}}=-\parppm(\omega \mathbf{p}_\pm^2 \partial_{\tau^{\pm}}-\partial_{\sigma^{\pm}})=\partial_{\angletwo^{\pm}}\,.
	\end{equation}

	\subsection{A special class}\label{subsect: A special class}

As already observed, the accelerating solution is richer but also tremendously more complicated. For this reason, we will impose a specific (but  arbitrary) constraint between the parameters with the sole purpose of simplifying the computations. Without turning to zero any parameter we remain in the generic situation, and indeed we will find that both the twist and the anti-twist are admitted. It is worth noting 
 that requiring~\eqref{no-nut conditions} and $\alpha=P^2-1$ as in~\eqref{wicktoBH} we recover  the complex Euclidean version of the accelerating black hole. However, since this is discussed extensively in \cite{Cassani:2021dwa}, we will not consider this further. On the other hand, to our knowledge no example of this special class has been analyzed in the literature before, therefore in this section we will not recover  any previous solutions.

The special class of accelerating solutions that  we will consider is obtained by setting\footnote{In the following equations we are taking a specific branch for the roots, such that $\sqrt{P^2}=P$ and $\sqrt{Q^2}=Q$. Recall also that the parameter $\omega$, as introduced for example in~\cite{Griffiths:2005qp}, is a scaling parameter. Thus we are only imposing a single simplifying condition here.}
\begin{equation}\label{accelerating:twist-simple-example}
	\alpha=P^2\,,\quad \acc= \frac{\omega}{Q}\,.
\end{equation}
In this case the supersymmetry constraints~\eqref{accelerating-susy-complicated} collapse to
\begin{equation}\label{simple-susy-twist}
	E=\frac{N^2}{Q^2}+2 P\,,\quad M=\frac{NP}{Q}\,,
\end{equation}
which are formally very similar to the non-accelerating ones~\eqref{susy}. The strong simplification is due  to the fact that the square root in the first line of~\eqref{accelerating-susy-complicated} becomes a perfect square.  Further, the metric functions simplify to
\begin{equation}
	\begin{aligned}
		\Pfunc&=\frac{1}{\omega^2 Q^2}\mathbf{P}_{+}(p)\mathbf{P}_{-}(p)\,,\quad &&\mathbf{P}_{\pm}(p)=\omega^2 P p^2- \omega N p+ Q^2(1\pm 1)\,,
		\\
		\Qfunc&=\frac{1}{P^2 Q^2}\mathbf{Q}_{+}(q)\mathbf{Q}_{-}(q)\,,\quad&&\mathbf{Q}_{\pm}(q)=PQ q^2-N Pq+ P Q(P\pm Q)\,,
	\end{aligned}
\end{equation}
so that the four roots of $\Pfunc$ are
\begin{equation}
	\begin{aligned}
		\mathbf{P}_+ (p)&:\quad r_1=\frac{N-\sqrt{N^2-8 P Q^2}}{2 \omega P}\,,\quad && r_2=\frac{N+\sqrt{N^2-8 P Q^2}}{2 \omega P}\,,
		\\
		\mathbf{P}_- (p)&:\quad r_3=0\,,\quad &&r_4=\frac{N}{\omega P}\,.
	\end{aligned}
\end{equation}
From these roots we can generate instances of solutions  with the twist or the anti-twist realization of supersymmetry, according to the paring we choose. In particular, as in section~\ref{subsect:Quantization}, we will end up with a twist if $\newpplusminus$ are roots of the same parabola, \ie\ $\mathbf{P}_+(\newpplusminus)=0$ or $\mathbf{P}_-(\newpplusminus)=0$; otherwise we will get an anti-twist. However, we can not have a case with $\newlens=0$ because it is not possible to choose the parameters such that $\newpminus=-\newpplus$.
Even if the analysis can be carried out (in principle) in full generality, meaning by parametrizing the various parameters in terms of the roots $\newpplusminus$ and $\newqplus$ as in section \ref{subsect:Killing spinors}, we take a more direct approach. More specifically, we choose the parameters of the solution such that the largest roots of $\Pfunc$ are given by
\begin{equation}{\label{twist-simple-roots}}
	\newpminus=r_3=0\,,\quad \newpplus=r_4= \frac{N}{\omega P}\,,
\end{equation}
and as a consequence we will have a twist. The case $\newpminus=r_4$ and $\newpplus=r_3$ is as well possible, and can be obtained by changing the sign of the parameters.
Then from the first two conditions in~\eqref{acclelerating: spindle and lens} we have
\begin{equation}\label{A1N}
	N= \frac{P \newqplus\sqrt{\newsingm-\newsingp}}{\sqrt{\newsingm}}\,,\quad \frac{\Delta}{2\pi} = \frac{\omega \newqplus}{P \sqrt{\newsingm} \sqrt{\newsingm-\newsingp}}\,,
\end{equation}
and from $\mathbf{Q}_{\lambda}(\newqplus)=0$ (with $\lambda=\pm 1$ determining if $\newqplus$ is solution of $\mathbf{Q}_+(q)$ or $\mathbf{Q}_-(q)$) we obtain
\begin{equation}
	\newqplus= \frac{\sqrt{Q}\sqrt{P+\lambda Q} \,\newsingm^{1/4}}{\sqrt{P\sqrt{\newsingm-\newsingp}-\sqrt{\newsingm}Q}}\,.
\end{equation}
For $\newpplus>\newpminus=0$ and $\Delta>0$ we should require $(N,P,\omega)$ to be positive. Moreover, for $\newpplusminus$ to be the largest roots we must have then that $N^2 - 8 P Q^2<0$, so that $r_{1,2}\in \CC$.
Finally, from the the second condition in~\eqref{acclelerating: spindle and lens} and~\eqref{accelerating: labell condition} we get
\begin{equation}{\label{twist-simple-lens}}
	\begin{aligned}
		\frac{\Delta\angleone}{2\pi}
		=\frac{\lambda}{\newlabell(N-2 Q \newqplus)}\,, \quad
		\newlens=\frac{\lambda \sqrt{\newsingm-\newsingp}\bigl[P\sqrt{\newsingm-\newsingp}-2 Q \sqrt{\newsingm}\bigr]\newlabell}{P}\,.
	\end{aligned}
\end{equation}
Here $P\sqrt{\newsingm-\newsingp}>2 Q \sqrt{\newsingm}$ strictly, due to the condition $\mathbf{Q}'(\newqplus)>0$. Consequently, as anticipated, $\newlens\neq 0$ in every case\footnote{The case $\newsingm=\newsingp$ is excluded because it would imply $N=0$ from~\eqref{A1N}, but in turn this means $\newpplus=\newpminus$ from~\eqref{twist-simple-roots}, which is degenerate because of the presence of a double root. Moreover, when $N=0$ the relevant roots should not be $r_3$ and $r_4$, but for example $\newpminus=r_3=r_4=0$ and $\newpplus=r_2$, for which $\newlens\neq 0$ again. It follows that this class of solutions is not directly related with the twist case with  $\newlens=0$ in section~\ref{subsubsect: N=0}.  }. Moreover in appendix~\ref{appendix: Values} we show that for this solution $\lens\ge 1$, so that the boundary can have the topology of the (squashed) $S^3$. The last can be easily inverted to give, for example, $Q=Q(P,\newlens)$. 
With these results, it is immediate to show that
\begin{equation}
\frac{1}{2\pi}\int_{\loci_2}\dd A=\frac{\parppm 1}{2}\biggl[\frac{\newsingm+\newsingp}{\newsingm\newsingp}+\lambda \frac{\newlens/\newlabell}{\newsingm\newsingp}\biggr]\,,
\end{equation}
confirming that $2A$ is a connection on the twisted line bundle $\bundlep$ over $\spindlesing$. Notice that in principle one could have taken the parameters $(N,P,\omega)\in\CC$ such that $\newpplus\in\RR_{+}$. However, requiring $(\newqplus,\Delta,\Delta\angleone,\newlens)\in\RR_{+}$ implies $(N,P,\omega,Q)\in\RR$. This is consistent with the statement that the solutions with twist are real. Moreover we can now also insert all the \enquote{quantized} expressions in the on-shell action~\eqref{acceleratingOSaction}. Using explicitly these results and~\eqref{accelerating: period-product}, it is immediate to show that
\begin{equation}{\label{accelerating: easy-OS}}
	\begin{aligned}
		\text{twist}:\quad S_{\text{ren}}
		=\frac{ \pi }{8 G_4 \newlabell} \biggl[{2\chi_\spindle}{}+\lambda\frac{\newlens/\newlabell}{\newsingm\newsingp}+\lambda \frac{\newlabell(\newsingm-\newsingp)^2}{\newlens\,\newsingm\newsingp}\biggr]\,,
	\end{aligned}
\end{equation}
which is equal, for example, to~\eqref{easy-OS}. This result is expected, since the polytope of the solution is the same as for the non-accelerating case (see figure~\ref{fig:non-compact polytope non-zerot}), and the solution is of the twist type. Consequently, as observed in section~\ref{subsubsect:Comments on extremization}, the extremization of the action~\eqref{Genolini} should give the same result, independently of the specific form of the solution. We can provide further confirmation of this by plugging explicitly the supersymmetric vector~\eqref{accelerating:susy vector},~\eqref{accelerating:weight} in the action~\eqref{Genolini}. However, since~\eqref{simple-susy-twist} makes $\mathrm{w}$ infinite in~\eqref{easy-accelerating-weight}, we can not use~\eqref{Genolini-action},~\eqref{reduced-charges} directly. Indeed this means that for the specific sub-case under examination the supersymmetric Killing vector is aligned only with the $\tau$ direction, $\epsilon_{\ast}\propto \partial_\tau$ (see also~\cite{Klemm:2013eca}). As a consequence we need to write the single Killing vectors $\partial_\tau$ and $\partial_\sigma$
in the toric basis~\eqref{effective-coordinates}
\begin{equation}
	U_-:\quad
	\begin{aligned}
		\partial_{\tau^{-}}=&\biggl[\frac{1}{\newqplus^2-\omega^2 \newpminus^2}\frac{2\pi}{\Delta\angleone}+\omega\frac{r_+}{\newsingm}\frac{2\pi}{\Delta}\biggr]E_1 + \omega \frac{2\pi}{\Delta}E_2\,,
		\\
		\partial_{\sigma^{-}}=&\biggl[\frac{\omega \newpminus^2}{\newqplus^2-\omega^2 \newpminus^2}\frac{2\pi}{\Delta\angleone}+\newqplus^2\frac{r_+}{\newsingm}\frac{2\pi}{\Delta}\biggr]E_1 + \newqplus^2 \frac{2\pi}{\Delta}E_2\,.
	\end{aligned}
\end{equation}
Inserting the first line in~\eqref{Genolini} with $c_2=c_1=\lambda$ and using~\eqref{twist-simple-roots}-\eqref{twist-simple-lens}, we find again~\eqref{accelerating: easy-OS}. The situation here is completely analogous to that of section~\ref{subsubsect:Holographic renormalization}. Indeed in the CP solution we started with six parameters plus the periodicities $(\Delta\angleone,\Delta)$, minus two constraints from supersymmetry and four regularity conditions, remaining with two free parameters $(\free,\freetwist)$. In passing from CP to PD we add two parameters $(\omega,\acc)$, but we also fixed two simplifying relations~\eqref{accelerating:twist-simple-example}. Then we are similarly left with \emph{two free continuous parameters} ($\omega,P$), that however do not appear in the on-shell action.

Certainly other choices of roots are possible, but the remaining twist case is excluded because there are no parameter values for which $r_1$ and $r_2$ can be the largest roots. Instead all the alternative combinations of roots which give the anti-twist are admissible. In these cases we have checked that the gravitational on-shell action~\eqref{acceleratingOSaction} is reproduced by~\eqref{nicestesosa} with $c_2=-c_1$, as expected. However we will not present an explicit expression for the anti-twist action, since as in section~\ref{subsubsect:Holographic renormalization} it is very complicated and not particularly illuminating.

We conclude our general analysis noticing that since we have imposed the simplifying constraint \eqref{accelerating:twist-simple-example}, from this special class of solutions it is impossible to reach smoothly the non-accelerating CP families. However, since in all cases we discussed the  form of the on-shell action is always consistent with the predictions of equivariant localization, combined with observations we made on  its extremization, we believe that in principle the general PD family should give rise to \emph{two} distinct families of solutions, corresponding to twist and anti-twist, each depending on three  continuous parameters, in addition to the topological data $(\singpm,\labell,\lens)$, that contain all the classes discussed in this paper. Moreover, in principle there could be an even larger class of solutions, depending on a larger number of continuous parameters, containing the entire PD family.

\section{Discussion}\label{sect:Discussion}

In this paper we constructed several new supersymmetric  solutions of four-dimensional minimal gauged supergravity.
These  have been obtained starting  with the  family of local Pleba\'nski-Demianski solutions, which comprises the most  general Petrov type D  metric of the Maxwell-Einstein-$\Lambda$ theory
and possesses an  $U(1)\times U(1)$ isometry.
Despite these local solutions have been known for almost fifty years, many different global completions, either in Lorentzian, or in Euclidean, gave rise to a plethora of solutions over the years, with very different physical 
interpretations. For example,  the  local K\"ahler-Einstein foliations  used in the construction of the $L^{a,b,c}$ Sasaki-Einstein manifolds~\cite{Cvetic:2005ft} were obtained from a scaling limit of the Carter-Pleba\'nski sub-family~\cite{Martelli:2005wy}. 
Moreover,  the various supersymmetric 
NUTs and Bolts solutions constructed in the series of papers~\cite{Martelli:2011fu,Martelli:2011fw,Martelli:2012sz,Martelli:2013aqa}  arise as special cases or limits 
of this sub-family. 
Significant supersymmetric solutions lying in the more general Pleba\'nski-Demianski family are the  accelerating black hole~\cite{Ferrero:2020twa} and its complex non-extremal deformation~\cite{Cassani:2021dwa}.  
The \emph{global solutions} that we have constructed in this paper generalize all these solutions and include them as special cases or limits. A compendium of the old and new solutions is given in Table \ref{tablesummary}.

\begin{table}[]
	\begin{center}
\begin{tabular}{|ccc|c|c|c|}
	\hline
	\multicolumn{3}{|c|}{Parameters}                                                                                       & Twist         & Bolt       & Notable cases                                                       \\ \hline
	\multicolumn{1}{|c|}{\multirow{6}{*}{$\acc=0$}} & \multicolumn{2}{c|}{\multirow{2}{*}{$N\neq 0$}}                      & $\twist=+1$   & $\spindle$ & Type I \cite{Martelli:2013aqa}, $1/4$-BPS bolts \cite{Martelli:2012sz}    \\ \cline{4-6} 
	\multicolumn{1}{|c|}{}                          & \multicolumn{2}{c|}{}                                                & $\twist=-1$   & $\spindle$ & Type II \cite{Martelli:2013aqa}, $1/2$-BPS bolts \cite{Martelli:2012sz} \\ \cline{2-6} 
	\multicolumn{1}{|c|}{}                          & \multicolumn{1}{c|}{\multirow{4}{*}{$N=0$}} & \multirow{2}{*}{$P=0$} & $\twist=\pm1$ & $\spindle$ & -                                                                   \\ \cline{4-6} 
	\multicolumn{1}{|c|}{}                          & \multicolumn{1}{c|}{}                       &                        & $\twist=-1$   & $S^2$      & Kerr-Newman-AdS \cite{Cacciatori:2009iz, Cassani:2019mms}         ($\lens=0$)                       \\ \cline{3-6} 
	\multicolumn{1}{|c|}{}                          & \multicolumn{1}{c|}{}                       & \multirow{2}{*}{$M=0$} & $\twist=+1$   & $S^2$      & Topological Kerr-Newman-AdS                ($\lens=0$)                          \\ \cline{4-6} 
	\multicolumn{1}{|c|}{}                          & \multicolumn{1}{c|}{}                       &                        & $\twist=-1$   & $\spindle$ & -                                                                   \\ \hline
	\multicolumn{1}{|c|}{\multirow{3}{*}{$\acc\neq 0$}} & \multicolumn{2}{c|}{\multirow{2}{*}{$\alpha=P^2, \omega=\acc Q$}}    & $\twist=+1$   & $\spindle$ & -                                                                   \\ \cline{4-6} 
	\multicolumn{1}{|c|}{}                          & \multicolumn{2}{c|}{}                                                & $\twist=-1$   & $\spindle$ & -                                                                   \\ \cline{2-6} 
	\multicolumn{1}{|c|}{}                          & \multicolumn{2}{c|}{\eqref{no-nut conditions}}                          & $\twist=-1$   & $\spindle$ & Accelerating black hole \cite{Ferrero:2020twa,Cassani:2021dwa}   ($\lens=0$)                       \\ \hline
\end{tabular}
\end{center}
\caption{A summarizing table of the CP and PD solutions of sections~\ref{sect:Non-accelerating solutions} and~\ref{sect:Accelerating solutions}, respectively. Since $N=0$ is the only case for which $\lens=0$, we have split the table according to the value of $N$. Further, for $N=0$ supersymmetry can be realized in two different ways, with $P=0$ or $M=0$, respectively (see section~\ref{subsubsect: N=0}). Only the spherical bolts and the accelerating black hole have $\lens=0$. The last row is dedicated to the PD solution, with $\acc\neq 0$.
	The second column shows the admissible realization of supersymmetry on the spindle (or the sphere). Finally, the last column indicates notable cases, which include some old solutions and the  Euclidean Topological Kerr-Newman-AdS, which was missed in the literature.}
	\label{tablesummary}
\end{table}

All the solutions that we have discussed comprise a metric on some orbifold line bundle over the spindle, with the latter appearing (in the bulk) as the zero section, which is usually called a \enquote{bolt}~\cite{Gibbons:1979xm}. 
The topology of the conformal boundary can be that of  a smooth lens space, but generically it displays orbifold singularities, which  we refer to  as branched lens spaces. As particular cases we can also have the topology of the
three-sphere ($\lens=1$), as well as that of the direct product $\spindle\times S^1$ ($\lens=0$). The various classes of solutions display both types of twist for the graviphoton gauge field, characterised 
by the flux/curvature integrated over the spindle bolt, see \eg~\eqref{twist-anti-twist}.  As the solutions have $U(1)\times U(1)$ isometry, the underlying orbifolds 
may be described in terms of toric geometry\footnote{Our solutions are in fact explicit examples of the general framework introduced in~\cite{BenettiGenolini:2024hyd}.} employing the approach described in~\cite{Faedo:2024upq}.  In particular, we have derived the toric data for all the solutions, consisting in two-dimensional labelled polyhedral cones, as depicted in figure~\ref{fig:non-compact polytope}. Incidentally, these may be obtained from the compact polytopes (quadrilaterals) described in~\cite{Faedo:2024upq} by removing one of the four edges and indeed, the local form of the co-homogeneity two metrics discussed in~\cite{Faedo:2024upq} is remarkably similar (but not diffeomorphic) do the metrics of  Pleba\'nski-Demianski, that we discussed presently.

Obtaining explicit solutions corresponding to non-compact toric orbifolds with more than two fixed points would be  impressive, but we believe that with present days technology, this  is  improbable. Thus our solutions are likely to remain the state-of-the-art of explicit examples of toric gravitational instantons of minimal gauged supergravity for some time.  The regularity analysis of the Pleba\'nski-Demianski solutions with generic values of the parameters appears technically  cumbersome. However, it may be possible that with some canny reparametrization and/or changes of variables (see \eg~\cite{Podolsky:2022xxd,Astorino:2024bfl,Ovcharenko:2024yyu})  one can make analytic progress. 
We leave the problem of improving our global analysis of the Pleba\'nski-Demianski solutions to someone else   starting afresh. It might also be interesting to investigate spindle bolt solutions of the type we discussed here in five or higher dimensions, extending the few known examples of topological solitons in dimensions higher than four~\cite{Chong:2005hr,Cassani:2015upa,Cvetic:2005zi}.

Although the main motivation for this paper was holography, and therefore we focussed on asymptotically hyperbolic non-compact solutions, from a mathematical viewpoint it may be 
interesting to perform a systematic analysis analogous to our one in the compact setting. As we noted around equation~\eqref{minimalSUGRAeomMATH}, 
solutions to four dimensional Einstein-Maxwell-$\Lambda$ theory have been actively investigated in the mathematics literature in the past decade, starting with the work of~\cite{LeBrun:2008kh}, where it was pointed out the relation of these to constant scalar curvature metrics. Interestingly, in this reference it was also noted  that 
any constant scalar curvature K\"ahler metric on a compact 4-manifold is actually   a solution to the Einstein-Maxwell-$\Lambda$ equations, for a specific choice of curvature $F$. 
Examples of this correspondence, for smooth compact 
manifolds as well as for orbifolds,  are discussed in~\cite{2015arXiv151206391A,LeBrun_2015,LeBrun_2016,2015arXiv151106805K,2017arXiv170801958F,2017arXiv170607953F}. It would be interesting to study how this correspondence works in the sub-set of supersymmetric solutions of minimal four-dimensional supergravity. 
Moreover, it is also natural to wonder whether  existence and uniqueness of the non-compact asymptotically hyperbolic solutions
may be addressed employing the  ideas of volume  extremization~\cite{2017arXiv170607953F}.

Perhaps the most significant general lesson that we drew from our results is that the geometry of  the conformal boundary encodes in a subtle way the  information about how the solution
	\enquote{fills} this in the bulk. Generically, the three-dimensional space at the boundary is a Seifert orbifold, with the topology of a lens space  $L(\lens,1)$, possibly with orbifold singularities. In addition, however, the three-dimensional 
	Killing spinors and gauge field are characterised by further data, namely the type of twist $(\sigma=\pm 1)$, as well as the the positive integers $m_\pm$. These data then determine the  topology of the four-dimensional filling, which correspondingly comprises an orbifold line bundle over a spindle $\spindle_{[m_+,m_-]}$, with twist or anti-twist for the graviphoton.  We noted that this dichotomy explains   the existence of the old classes of  $1/4$-BPS and $1/2$-BPS Bolt solutions of~\cite{Martelli:2012sz}, which arise as limits of the two different twists on the spindle. Specifically, the $1/4$-BPS Bolt solutions have a topological twist for the graviphoton, while the $1/2$-BPS Bolt solutions have no twist. Compare  \eg\ equation~\eqref{twist-anti-twist} with $\singp=\singm=1$ with the integrated fluxes in \eqref{bolt actions}. We have shown that after fixing  all the topological data as well as the type of twist, in general there  exist multiple solutions which are not
	diffeomorphic.  For example, picking $\sigma=1$, and setting $(\singp,\singm,\labell,\lens)$ to particular values  there exist both non-accelerating as well as accelerating solutions (see appendix~\ref{appendix: Values}). As discussed at the end of section~\ref{sect:Accelerating solutions} this is an indication that for fixed topological data $(\singp,\singm,\labell,\lens)$ and type of twist $\twist=\pm 1$, there should exist a family  with several continuous parameters, encapsulating the different sub-families that  we analyzed.

For all our solutions, we have computed the holographically renormalized on-shell action, uncovering an intriguing pattern. In all cases the on-shell action takes the general form~\eqref{Genolini}, predicted by equivariant localization~\cite{BenettiGenolini:2019jdz,BenettiGenolini:2023kxp,BenettiGenolini:2024hyd}. In particular, it can be expressed entirely in terms of the data at the two fixed points of the orbifolds (the poles of the spindle) \emph{and} the choice of supersymmetric Killing vector $\vec{\epsilon}$. However, we have found that the form of this Killing vector for the explicit solutions depends in an interesting way on the type of twist for the graviphoton field of the specific solution, leading to two markedly different types of expressions for the on-shell action in the twist and anti-twist cases.  Specifically, we uncovered that in the presence of twist, the supersymmetric Killing vector field of the explicit solutions coincides with the value determined by \emph{extremizing} the \enquote{off-shell action} with respect to the vector field itself.  Thus, the \enquote{critical} vector field does depend on the topological data of  the underlying space -- see~\eqref{critcalreebtwsit}.  This is completely analogous to the Sasaki-Einstein~\cite{Martelli:2006yb} and GK~\cite{Couzens:2018wnk} extremization problems and leads to an universal expression for the on-shell action of all the solutions with twist,  see \eg\ equations~\eqref{easy-OS} and \eqref{accelerating: easy-OS}.  On the other hand, in all the anti-twist cases the supersymmetric Killing vector field is arbitrary, and may depend on the continuous parameters of the solutions, so it is not determined by extremizing anything. This is consistent with 
the behaviour of  the supersymmetric non-extremal deformations of the AdS$_4$ black holes~\cite{Cassani:2019mms,Cassani:2021dwa}, 
that indeed correspond to anti-twist cases and have generic values of the supersymmetric Killing vectors\footnote{In these cases the extremal solutions correspond to 
	extremizing the \enquote{off-shell entropy}, namely performing the Legendre transform of the action.}. In the light of our findings, 
the 1/4-BPS and 1/2-BPS spherical Bolt solutions of~\cite{Martelli:2012sz} are earlier examples of this  
pattern\footnote{In the self-dual subcases, a similarly  different behaviour of the 1/4-BPS and 1/2-BPS NUT solutions of~\cite{Martelli:2012sz} was observed in~\cite{Farquet:2014kma}, 
	and it was related to fact that generically there exist two complementary fillings the unit sphere in $\mathbb{C}^2$.}.
Understanding the origin of the different behaviour of twist and anti-twist  is perhaps the most intriguing open problem stemming from our work. This is likely to be related to  two different types
of boundary conditions in the supergravity  variational problem: in the anti-twist case the supersymmetric Killing vector should be a fixed boundary datum of the problem, while in the  twist case this is not fixed and as a result the off-shell action must be extremized with respect to different choices of this vector.

Via holography, this problem can be reformulated in terms of properties of the dual three-dimensional supersymmetric field theories placed on the geometry of the boundary \cite{Inglese:2023tyc}. 
	In particular, we have pointed out that for a given topology of the three-dimensional boundary, \eg\ a lens space $L(\lens,1)$, there exist \emph{two choices}, labelled by a 
	sign $\sigma=\pm   1$, 
	that  lead to two different behaviours of the localized partition functions.  We refer to these two choices as to twist and anti-twist, despite the fact that there are no two-cycles in the three dimensional geometry.
	Moreover, we expect that the additional choices of flat connections   depending on  $m_\pm \in \mathbb{N}$, will determine the large $N$ limit of the partition function, through a mechanism analogous to that governing the large $N$ limit of the spindle index \cite{Colombo:2024mts}.

It will be very interesting to reproduce our predictions from the large $N$ limit of the three dimensional localized partition functions.

\section*{Acknowledgments}
\noindent 
We acknowledge partial support by the INFN. The work of DM is supported in part by a grant Trapezio (2023) of the Fondazione Compagnia di San Paolo. 
We thank  P. Benetti Genolini, D. Cassani, E. Colombo, V. Dimitrov, F. Faedo, J. Gauntlett, A. Pittelli, J. Sparks and A. Zaffaroni for comments and discussions.

		\appendix
		
			\section{Old bolt solutions}\label{appendix:Old-bolt-solutions}
			
		In this section we review briefly some aspects of the spherical Bolt solutions  of~\cite{Martelli:2012sz}, since these are sub-cases of~\eqref{solution:metric}-\eqref{solution:graviphoton}. Throughout this section we shall employ hatted quantities to distinguish them from the parameters in the main body of the paper. Locally, the solution is given by
		\begin{equation}\label{old-sol:metric}
			\begin{aligned}
				\dd s^2 _4&=\frac{r^2-s^2}{\Omega(r)}\dd r^2+(r^2-s^2)(\sigma_1^2+\sigma_2^2)+\frac{4 s^2 \Omega(r)}{r^2-s^2}\sigma_3\,,
			\end{aligned}
		\end{equation}
		with graviphoton 
		\begin{equation}\label{old-sol:graviphoton}
			A=\biggl[\hat{P} \frac{r^2+s^2}{r^2-s^2}-\hat{Q}\frac{2 r s}{r^2-s^2}\biggr]\sigma_3\,.
		\end{equation}
		The metric function is
		\begin{align}\label{old-sol:metricfunc}
			\Omega(r)=(r^2-s^2)^2+(1-4 s^2)(r^2+s^2)-2 \hat{M}r +\hat{P}^2-\hat{Q}^2\,,
		\end{align}
		and the $\sigma_i$ are the $SU(2)$ left-invariant one-forms, that for definiteness we take to be
		\begin{align}
			\sigma_1+\ii \sigma_2 =e^{-\ii\hat{\psi}}(\dd\hat{\theta}+\ii\sin\hat{\theta}\dd\hat{\varphi})\,,\quad \sigma_3=\dd\hat{\psi}+\cos\hat{\theta}\dd\hat{\varphi}\,.
		\end{align}
		As demonstrated in~\cite{Martelli:2012sz}, this is the most general local form of the supersymmetric solutions with $SU(2)\times U(1)$ symmetry, which is enhanced with respect to the $U(1)\times U(1)$ symmetry of the spindle bolt. For large $r$, the metric becomes
		\begin{equation}{\label{squashing}}
			\dd s_4 ^2 =\frac{\dd r^2}{r^2}+r^2 (\sigma_1^2+\sigma_2^2+4 s^2 \sigma_3^2)\,,
		\end{equation}
		so that the space is asymptotically locally $\mathbb{H}^4$ and $s$ is a squashing parameter.
		The BPS conditions then split this solution into two classes\footnote{Notice the insertion of the sign $\signpp=\pm1$ here, which accounts in the two possible choices explained in~\cite{Martelli:2012sz}.}
		\begin{equation}{\label{old-BPS}}
			\begin{aligned}
				\frac{1}{2}\text{-BPS}:\quad &\hat{M}=\signpp\hat{Q}\sqrt{4s^2-1}\,,\quad &&\hat{P}=-\signpp s\sqrt{4s^2-1}\,,
				\\
				\frac{1}{4}\text{-BPS}:\quad &\hat{M}=2 \signpp s \hat{Q}\,,\quad &&\hat{P}=-\signpp\frac{4 s^2-1}{2}\,,
			\end{aligned}
		\end{equation}
		for which, in both cases, the topology can be the one of a non-trivial fibration on a round bolt $\mathcal{M}_4=\mathcal{O}(-\hat{\lens})\rightarrow S^2$, or a nut $\RR^4/\ZZ_{\hat{t}}$, with the origin identified by $r=s$. From symmetry, one can always take the squashing parameter to be positive $s>0$, and the round $S^3$ case is obtained for $s=1/2$. The range of the radial coordinate is taken to be $s\le r_0\le r<+\infty$, where $r_0$ is the largest root of $\Omega(r)$, with $r_0>s$ for the bolt case. The regularity analysis of the metric in the two classes leads to the following conditions
		\begin{equation}\label{old-regularity-half}
			\begin{aligned}
				\frac{1}{2}\text{-BPS}:\quad \hat{Q}=\signpp \frac{(128s^4-16s^2-\hat{\lens}^2)\branch}{64 s^2}\,,\quad r_0 =\frac{1}{8}\biggl[\frac{\hat{\lens}}{s}-4 \branch\sqrt{4s^2-1}\biggr]\,,
			\end{aligned}
		\end{equation}
		and
		\begin{equation}\label{old-regularity-quarter}
			\frac{1}{4}\text{-BPS}:\quad \begin{aligned}
				\hat{Q}&=\signpp \frac{\hat{\lens}^2\pm(\hat{\lens}-16s^2)\sqrt{\mathcal{G}_+(\hat{\lens},s)}}{128s^2}\,,&&r_0 =\frac{\hat{\lens}\pm\sqrt{\mathcal{G}_+(\hat{\lens},s)}}{16s}\,,\quad \branch=1\,,
				\\
				\hat{Q}&=-\signpp \frac{\hat{\lens}^2\mp(\hat{\lens}+16s^2)\sqrt{\mathcal{G}_-(\hat{\lens},s)}}{128s^2}\,,&&r_0 =\frac{\hat{\lens}\mp\sqrt{\mathcal{G}_-(\hat{\lens},s)}}{16s}\,,\quad \branch=-1\,.
			\end{aligned}
		\end{equation}
		where
		\begin{equation}
			\mathcal{G}_\pm(\hat{\lens},s)=(\hat{\lens}\pm16s^2)^2-128s^2\,.
		\end{equation}
		Notice that here, contrarily to~\cite{Martelli:2012sz}, we considered two branches for both the $1/4$-BPS conditions, as pointed out in~\cite{Toldo:2017qsh}.

\section{A spindle Calabi-Yau metric}\label{appendix:A-spindle-Calabi-Yau-metric}

We devote this appendix to the investigation of a Ricci-flat version of the set-up considered in section~\ref{sect:Non-accelerating solutions}. The local form of the metric can be obtained as the Ricci-flat limit of the one presented in~\cite{Apostolov2001TheGO,Martelli:2005wy}, which is generalized to arbitrary dimensions in~\cite{10.4310/jdg/1146169934,Martelli:2007pv}, although the four-dimensional case (corresponding to $n=0$ in~\cite{Martelli:2007pv}) is not discussed in the latter. 
See also~\cite{Ferrero:2024vmz}. In this simpler setting, which is toric in the standard mathematical language, we are able to explicitly write down the symplectic 2-form and the corresponding moment maps, enabling the computation of the toric diagram~\cite{Martelli:2004wu}. This fact, alongside with the total space being Calabi-Yau, which will provide a generalization of the computation in section 3.3.2 of~\cite{Martelli:2023oqk}, serves as main motivation for this appendix.

Explicitly, we introduce the metric
\begin{equation}\label{ricci-flat metric}
	\begin{aligned}
		\dd s^2_{4}=\frac{\Gcofl-\Fcofl}{2}\biggl[\frac{\dd\Gcofl^2}{G(\Gcofl)}+\frac{\dd\Fcofl^2}{-F(\Fcofl)}\biggr]+\frac{2}{\Gcofl-\Fcofl}\biggl[G(\Gcofl)(\dd\angafl+\Fcofl\,\dd\angbfl)^2-F(\Fcofl)(\dd\angafl+\Gcofl\,\dd\angbfl)^2\biggr]\,,
	\end{aligned}
\end{equation}   
where we have defined
\begin{equation}
	\begin{aligned}
		F(\Fcofl)=E\,\Fcofl^2-2 \linparfl\,\Fcofl-P^2\,,\quad G(\Gcofl)=E\,\Gcofl^2-2\linparfl\,\Gcofl-Q^2\,.
	\end{aligned}
\end{equation}		
To ensure the correct Euclidean signature we take $\Gcofl>\Fcofl$, $G(\Gcofl)>0$ and $F(\Fcofl)<0$ in a compact interval. This is possible only if $E>0$, so that\footnote{In a similar fashion as the comment at the beginning of paragraph \ref{subsubsect:Regularity in the bulk} we note that $\Gcofl=\Gcofl_+=\Fcofl_+=\Fcofl$ is a singular point.} $\Fcofl_- \le \Fcofl \le \Fcofl_+<\Gcofl_+ \le\Gcofl$, where we use $\Gcofl_+$ to denote the largest real root of $G(\Gcofl)$ and $\Fcofl_\pm$ the two roots of $F(\Fcofl)$. 
Given the vierbein choice
\begin{equation}\label{vielbein-flat}
	\begin{aligned}
		e^1 &=\sqrt{\frac{\Gcofl-\Fcofl}{-2F(\Fcofl)}}\dd\Fcofl\,,\qquad &&e^2 =\sqrt{\frac{-2F(\Fcofl)}{\Gcofl-\Fcofl}}(\dd\angafl+\Gcofl\,\dd\angbfl)\,, \\
		e^3 &=\sqrt{\frac{2G(\Gcofl)}{\Gcofl-\Fcofl}}(\dd\angafl+\Fcofl\,\dd\angbfl)\,,\qquad  &&e^4 =\sqrt{\frac{\Gcofl-\Fcofl}{2G(\Gcofl)}}\dd\Gcofl\,,
	\end{aligned}
\end{equation}
we can construct the K\"ahler 2-form $\omega_{(2)}$ and the holomorphic $(2,0)$-form $\Omega^{(2,0)}$ as
\begin{equation} \label{G-structures}
	\begin{aligned}
		\omega_{(2)} &=e^1\wedge e^2+ e^4\wedge e^3\,, \\
		\Omega^{(2,0)} &=\ee^{2\ii(E \angafl+\linparfl\angbfl)}(e^1+\ii e^2)\wedge (e^4+\ii e^3)\,.
	\end{aligned}
\end{equation}
They satisfy the algebraic conditions of a $SU(2)$-structure, namely $\omega\wedge \Omega=0$ and $-\frac{1}{2}\omega\wedge\omega=-\frac{1}{4}\Omega\wedge\bar{\Omega}=\mathrm{vol}_4$, and it can be easily checked that they are closed. Moreover, it can be checked that~\eqref{ricci-flat metric} is Ricci-flat,  $R_{\mu\nu}=0$.

\subsection*{Regularity}
The procedure adopted to ensure regularity and find the quantization conditions is analogous to the one of section~\ref{subsect:Regularity}. Likewise, we introduce two patches $U_\pm$ such that $\xi_+\in U_+$, $\xi_-\in U_-$ and $\mathcal{M}_4=U_+\cup U_-$. We then adopt a change of  coordinates
\begin{align}\label{adapted-coords-flat}
	U_\pm:\quad	\angafl^{\pm}= \Gcofl_+ \newangafl^{\pm} -{\parflpm} \Fcofl_\pm \newangbfl^{\pm}\,,\quad \angbfl^{\pm}=-\newangafl^{\pm}+{\parflpm}  \newangbfl^{\pm}\,,
\end{align}
after which the metric reads
\begin{align}\label{bulk-metric-flat}
	\begin{split}
		U_\pm:\quad			\dd s_4 ^2 &=\frac{\Gcofl-\Fcofl}{2 G(\Gcofl)}\dd \Gcofl^2+\frac{\Gcofl-\Fcofl}{-2F(\Fcofl)}\dd \Fcofl^2- \frac{2(\Gcofl-\Fcofl_\pm)^2 F(\Fcofl)}{\Gcofl-\Fcofl} \dd (\newangbfl^{\pm})^2 \\
		&-2(\Gcofl-\Gcofl_+)\frac{ \dd (\newangafl^{\pm}) ^2 (\Gcofl-\Gcofl_+)-2 {\parflpm}  (\Gcofl-\Fcofl_\pm)\dd \newangafl^{\pm} \dd \newangbfl^{\pm}}{\Gcofl-\Fcofl}F(\Fcofl) \\
		&+2\frac{\big[\parflpm(\Fcofl-\Fcofl_\pm)\dd \newangbfl^{\pm}+\dd \newangafl^{\pm} (\Gcofl_+ - \Fcofl)\big]^2}{\Gcofl-\Fcofl}G(\Gcofl)\,.
	\end{split}
\end{align}

The boundary metric, defined as 
\begin{align}
	\lim_{\Gcofl\rightarrow +\infty}\dd s^2_4=\dd \Gcofl^2/(2 E \Gcofl)+(2 E \Gcofl) \dd s^2 _b+ O (1/\Gcofl)\,,
\end{align}
is found to be
\begin{align} \label{boundary-metric-flat}
	\begin{split}
		U_\pm:\quad	\dd s^2 _b =&\frac{-\dd \Fcofl^2}{4 E F(\Fcofl)}-\frac{(\Gcofl_+ -\Fcofl_\pm)^2 F(\Fcofl)}{E(\Gcofl_+ - \Fcofl)^2-F(\Fcofl)}\dd (\newangbfl^{\pm})^2
		\\
		&+\frac{\left[E(\Gcofl_+ - \Fcofl)^2-F(\Fcofl)\right]}{E}\biggl\{\dd \newangafl^{\pm} + \frac{E(\Gcofl_+ - \Fcofl)(\Fcofl-\Fcofl_\pm)+F(\Fcofl)}{\left[E(\Gcofl_+ -\Fcofl)^2-F(\Fcofl)\right]}\parflpm\dd \newangbfl^{\pm}\biggr\}^2\,,
	\end{split}
\end{align}
where the normalization is such that the Ricci scalar reads $R=6 E^2$. As in~\eqref{bound-limit}, we can zoom near the zeros of $F(\Fcofl)$, namely $\Fcofl_\pm$, and impose
\begin{align}\label{bound-spindle-cond-flat-patch}
	U_\pm:\quad		\pm F'(\Fcofl_\pm)\Delta \newangbfl^{\pm} =\frac{2\pi}{\singpmfl}\,,
\end{align}
after which the metric~\eqref{boundary-metric-flat} describes topologically a fibration over a spindle $\spindle^{\infty}_{[\singmfl,\singpfl]}$.

Similarly to~\eqref{bound-base}, the metric on $\spindle^{\infty}_{[\singmfl,\singpfl]}$ is the first line of~\eqref{boundary-metric-flat} and again we can show that $\Delta \newangafl^{+}=\Delta \newangafl^{-}\equiv \Delta \newangafl$. Moreover, introducing
\begin{align}
	\Delta \equiv (\Gcofl_+ - \Fcofl_+)\Delta\newangbfl^{+}=(\Gcofl_+ - \Fcofl_-)\Delta\newangbfl^{-},
\end{align}  
equation~\eqref{bound-spindle-cond-flat-patch} and the well-definiteness of the fibration read
\begin{align}\label{bound-spindle-cond-flat}
	U_\pm:\quad		\pm\frac{F'(\Fcofl_\pm)}{\Gcofl_+-\Fcofl_\pm}\Delta  =\frac{2\pi}{\singpmfl},\quad \lensfl= \singmfl\,\singpfl\frac{(\Fcofl_+  -\Fcofl_-) 
	}{(\Gcofl_+ -\Fcofl_+ )(\Gcofl_+ -\Fcofl_-)}\frac{\Delta}{ \Delta \newangafl}\,.
\end{align}

Proceeding on the same lines to analyze the bulk we find an additional condition (analogous to~\eqref{spindle-cond1})
\begin{align}\label{label-flat}
	G'(\Gcofl_+) \Delta \newangafl=\frac{2\pi}{\labellfl}\,.
\end{align}
After imposing~\eqref{bound-spindle-cond-flat} and~\eqref{label-flat}, the topology is indeed 
$\CC/\ZZ_{\labellfl}\hookrightarrow\mathcal{O}(-\lensfl)\rightarrow \spindle_{[\singmfl , \singpfl]}$.

\subsection*{Quantization}
We now proceed to solve all the quantization conditions~\eqref{bound-spindle-cond-flat} and~\eqref{label-flat}, using the following expressions for the parameters in terms of the roots\footnote{Note that $E$ is required to be positive, implying that $\Fcofl_-$ is negative, which in turn means that $\Fcofl_+ >0$, due to the negative sign in front of $P^2$ and $Q^2$ in $G(\Gcofl)$ and $F(\Fcofl)$.}
\begin{equation}
	\begin{aligned}
		E&=-\frac{1}{\Fcofl_- \Fcofl_+}P^2\,,
		\\
		\linparfl&=-\frac{(\Fcofl_- + \Fcofl_+)}{2\, \Fcofl_- \Fcofl_+}P^2\,,
		\\
		Q^2&=-\frac{\Gcofl_{+}(\Gcofl_{+}-\Fcofl_+ -\Fcofl_-) }{\Fcofl_- \Fcofl_+}P^2\,.
	\end{aligned}
\end{equation}
In particular, employing the parametrization\footnote{For this parametrization to be effective we should take either $(0<w<\Gcofl_{+},\,x>1)$ or $(w<0,\,x<-1)$.} $\Fcofl_\pm=w(1\pm x)$, equation~\eqref{bound-spindle-cond-flat} implies
\begin{align}
	w= \frac{\singmfl - \singpfl}{\singmfl (1+x)-\singpfl (1-x)}\,\Gcofl_+\,,\quad \frac{\Delta} {2\pi}= \frac{(\singmfl - \singpfl)(x^2-1)}{\bigl[\singmfl (1+x)-\singpfl (1-x)\bigr]^2}\frac{\Gcofl_+ ^2}{P^2}\,,
\end{align}
and
\begin{align}
	\frac{\Delta \newangafl}{2\pi}= \frac{(\singmfl - \singpfl)^2 (x^2-1) }{2 x (\singmfl+\singpfl)\bigl[\singmfl (1+x)-\singpfl (1-x)\bigr]}\frac{\Gcofl_+}{\labellfl P^2}\,.
\end{align}
Finally~\eqref{label-flat} results in
\begin{align}\label{fibration-solved-flat}
	\lensfl= \labellfl \,(\singmfl + \singpfl)\,.
\end{align}

\subsection*{Toric data}
Let us now construct the polytope. The symplectic two-form $\omega_{(2)}$ in~\eqref{G-structures}, after plugging in the choice of the vierbein \eqref{vielbein-flat}, can be equivalently rewritten as 
\begin{align}\label{new-moment-map}
	U_\pm:\quad	\omega_{(2)}=\dd(\Gcofl+\Fcofl)\wedge\dd\angafl^{\pm}+\dd(\Gcofl\,\Fcofl)\wedge\dd\angbfl^{\pm}\,.
\end{align}
In the $2\pi$-periodic coordinates 
\begin{equation}
\newangtilafl^{\pm}=\frac{2\pi}{\Delta\newangafl}\newangafl^{\pm}\,,\quad \newangtilbfl^{\pm}=\frac{2\pi}{\Delta\newangbfl}\newangbfl^{\pm},
\end{equation}
where the coordinates $(\newangafl^{\pm},\newangbfl^{\pm})$ were defined in~\eqref{adapted-coords-flat}, the symplectic form reads
\begin{align}
U_\pm:\quad 	\omega_{(2)}=\dd\biggl\{\frac{\Delta \newangafl}{2\pi}[\Gcofl_{+}(\Gcofl+\Fcofl)-\Gcofl\,\Fcofl]\biggr\}\wedge\dd\newangtilafl^{\pm}+\dd\biggl\{-{\parflpm} \frac{\Delta \newangbfl^{\pm}}{2\pi}[\Fcofl_{\pm}(\Gcofl+\Fcofl)-\Gcofl\,\Fcofl]\biggr\}\wedge\dd\newangtilbfl^{\pm} \,.
\end{align}
Without loss of generality we restrict to the patch $U_-$. Then, with respect to the toric basis\footnote{In a similar fashion to~\eqref{trivial-effective-toric-transition-functions}, we can actually argue that the transition functions between $(\phi_1^+,\phi_2^+)$ and $(\phi_1^-,\phi_2^-)$ are trivial, implying that $E_i^-=E_i^+\equiv E_i$.} 
\begin{align}{\label{generic-toric-base}}
U_-:\quad	E_1^-\equiv\partial_{\torcoafl^-}=\partial_{\newangtilafl^{-}}\,,\quad E_2\equiv \partial_{\torcobfl}={\parflm} \partial_{\newangtilbfl^{-}} -\frac{r_+}{\singmfl}\partial_{\newangtilafl^{-}}\,, \\
U_+:\quad	E_1^+\equiv\partial_{\torcoafl^+}=\partial_{\newangtilafl^{+}}\,,\quad E_2\equiv \partial_{\torcobfl}={\parflm} \partial_{\newangtilbfl^{+}} +\frac{r_-}{\singpfl}\partial_{\newangtilafl^{+}}\,,
\end{align}
the symplectic form $\omega$ becomes
\begin{equation}
	\begin{aligned}
		\omega_{(2)}=&\dd\biggl\{\frac{\Delta \newangafl}{2\pi}[\Gcofl_{+}(\Gcofl+\Fcofl)-\Gcofl\,\Fcofl]\biggr\}\wedge\dd\torcoafl+
		\\
		&\dd\biggl\{-\frac{\Delta \newangafl}{2\pi}[\Gcofl_{+}(\Gcofl+\Fcofl)-\Gcofl\,\Fcofl]\,\frac{r_+}{\singmfl}-\frac{\Delta \newangbfl^-}{2\pi}[\Fcofl_-(\Gcofl+\Fcofl)-\Gcofl\,\Fcofl]\biggr\}\wedge\dd\torcobfl \,.
	\end{aligned}
\end{equation}
Therefore, the moment map relative to~\eqref{generic-toric-base} is
\begin{align}\label{moment-map}
	\vec{\mu}(\Fcofl,\Gcofl)=\frac{\Delta\newangafl}{2\pi}\biggl([\Gcofl_{+}(\Gcofl+\Fcofl)-\Gcofl\,\Fcofl]\,,-[\Gcofl_{+}(\Gcofl+\Fcofl)-\Gcofl\,\Fcofl]\frac{r_+}{\singmfl}-\frac{\Delta \newangbfl^-}{\Delta\newangafl}[\Fcofl_-(\Gcofl+\Fcofl)-\Gcofl\,\Fcofl]\biggr)\,.
\end{align}

We define the loci 
\begin{equation}
	\loci_1=\{\Fcofl=\Fcofl_+\}\,,\qquad\loci_2=\{\Gcofl=\Gcofl_+\}\,,\qquad \loci_3=\{\Fcofl=\Fcofl_-\}\,,
\end{equation}
which intersect at the two fixed points
\begin{align}
	p_1=\{\Fcofl=\Fcofl_+,\Gcofl=\Gcofl_+\}\,,\quad p_2=\{\Fcofl=\Fcofl_-,\Gcofl=\Gcofl_+\}\,.
\end{align}
Formally we can also define
\begin{align}
p_3=\{\Fcofl=\Fcofl_-,\Gcofl=\Gcofl_0 > \Gcofl_{+}\}\,,\quad	p_4=\{\Fcofl=\Fcofl_+,\Gcofl =\Gcofl_0 > \Gcofl_{+}\}\,, 
\end{align}
where $\Gcofl_0$ is taken to be much greater than $\Gcofl_+$.

The Killing vectors $K_a$ degenerating on $\loci_a$ and normalized to have unit surface gravity 
\begin{equation}
U_-:\quad
	\begin{aligned}
		K_1 &=\frac{1}{-F'(\Fcofl_+)}(\Fcofl_+\partial_{\phi^-}-\partial_{\psi^-})=\singpfl\left({\parflm} \partial_{\newangtilbfl^-}-\frac{\lensfl}{\singpfl\singmfl}\partial_{\newangtilafl^-}\right)\,,
		\\
		K_2&=\frac{1}{G'(\Gcofl_{+})} (\Gcofl_{+}\partial_{\phi^-}-\partial_{\psi^-})=\labell\, \partial_{\newangtilafl^-}\,,		
		\\
		K_3 &=\frac{1}{F'(\Fcofl_-)}(\Fcofl_-\partial_{\phi^-}-\partial_{\psi^-})=\singmfl{\parflm} \partial_{\newangtilbfl^-}\,,
	\end{aligned}
\end{equation}
in this case become
\begin{equation}\label{toric-killings-flat}
	\begin{aligned}
				K_1&=\singpfl E_2 - r_- E_1\,,
				\\
		K_2&= \labellfl \,E_1\,,
		\\
	K_3&=\singmfl E_2+r_+ E_1\,.
	\end{aligned}
\end{equation}

We can now construct the polytope both from the moment map~\eqref{moment-map} or from the prescription $K_a=\vec{v}_a  \cdot(E_1,E_2)$, which allows one to extract the (non-primitive) vectors $\vec{v}_a$ describing the labelled polytope. 
Using the moment map~\eqref{moment-map} we get
\begin{equation}{\label{polytope-flat}}
	\begin{aligned}
			\vec{\mu}(p_4)-\vec{\mu}(p_1)&=\frac{\Delta \newangafl}{2\pi}(\Gcofl_+-\Gcofl_0)(\Gcofl_+ -\Fcofl_+)(-1,-r_-/\singpfl)\,,
			\\
		\vec{\mu}(p_1)-\vec{\mu}(p_2)&=\biggl(0,\frac{\Delta \newangafl}{2\pi}\frac{\lensfl}{\singmfl \singpfl}(\Gcofl_{+}-\Fcofl_+)(\Gcofl_{+} - \Fcofl_-)\biggr)\,,
		\\
		\vec{\mu}(p_2)-\vec{\mu}(p_3)&=\frac{\Delta \newangafl}{2\pi}(\Gcofl_+-\Gcofl_0)(\Gcofl_+ -\Fcofl_-)(1,-r_+/\singmfl)\,.
	\end{aligned}
\end{equation}

The outward-pointing normal vectors coming from~\eqref{toric-killings-flat} or~\eqref{polytope-flat} are 
\begin{align}
\vec{v}_1 =(r_- , -\singpfl)\,,	\quad  \vec{v}_2=(\labellfl,0)\,, \quad \vec{v}_3=(r_+,\singmfl)\,,
\end{align}
The resulting polytope is exactly the same of figure \ref{fig:non-compact polytope non-zerot}, or also the one in figure 5 of~\cite{Martelli:2023oqk} in the $CY$ context. 
The charges defining the Kähler quotient such that $\sum_{a}^{3}Q^{a} \vec{v}_a =0$ can be taken to be
\begin{align}
	Q^a =(\labellfl\,\singmfl, -\lensfl,\labellfl\,\singpfl )\,.
\end{align}
Since $\lensfl$ in our explicit solution is given by~\eqref{fibration-solved-flat}, we have that $\sum_{a}^{3}Q^a=0$ and, in turn, our space is a $CY_2$\footnote{Since this is two-dimensional 
 toric, Calabi-Yau metric, it should be diffeomorphic to a two-center Gibbons-Hawking metric~\cite{Gibbons:1978tef}. We thank B. Acharya for making this observation.}. This generalizes the computation in~\cite{Martelli:2023oqk}, where the horizontal vector was not associated to a label; the topology considered there was $\CC\hookrightarrow\mathcal{O}(-\lensfl)\rightarrow \spindle_{[\singmfl , \singpfl]}$, while here it is $\CC/\ZZ_{\labellfl}\hookrightarrow\mathcal{O}(-\lensfl)\rightarrow \spindle_{[\singmfl , \singpfl]}$.

\section{Allowed values of the parameters}\label{appendix: Values} 
In this appendix we comment on the allowed lens spaces $L(\lens,1)$ at the boundary and their fillings for the cases presented in the paper, showing some explicit examples. We start from the simplest examples, which are the subcases of CP with $N=0$ discussed in section~\ref{subsubsect: N=0}. Then we consider the general CP solution with twist, concluding finally with the special solution of PD of section~\ref{subsect: A special class}.  Recall that in all the cases, when $\lens=0$ we have $\singm=\singp=1$ and the spindle degenerates to a sphere (see table~\ref{tablesummary}). Summarizing, we say that a solution is acceptable if
\begin{equation} \label{function-requirements}
	\begin{aligned}
\Pfunc&\le 0\,,\quad \Qfunc\ge 0\,,\quad \newqplus^2-\newpplusminus^2\ge 0\,,\quad  \newpminus<\newpplus<\newqplus\,,
\\
& \,\,\,\,\newpminus+\newpplus>0\,,\, \quad \singm>\singp\,,\quad (\singpm,\labell,\lens)\in \NN\,.
\end{aligned}
\end{equation}
The first four requirements come from the fact that the metric must have Euclidean signature, whilst $\newpminus+\newpplus>0$ has been fixed using the scaling symmetry~\eqref{scaling}.

\subsection*{$N=0$, $P=0$, $\alpha>1$ case}

We begin by studying the case (i) of~\ref{subsubsect: N=0}, focusing on $\alpha>1$. In this case $\lens$ can be read from~\eqref{lens-i-first-choice}, namely
\begin{equation}\label{lens-i-first-choice-app}
	\lens=\labell\frac{(\twist+\sqrt\alpha)(\singm-\twist\singp)}{(\alpha-1)(\sqrt{\singm-\sqrt{\alpha}\singp})}\Bigl[2\alpha^{1/4}\sqrt{\sqrt{\alpha}\singm-\singp}+\delta(1+\twist\sqrt{\alpha})\sqrt{\singm-\sqrt{\alpha}\singp}\Bigr]\,.
\end{equation}

Recall that this case admits both twist and anti-twist. Moreover, note that we require $\alpha>1$ and $\singm>\sqrt{\alpha}\,\singp$; these two conditions are sufficient to ensure that all the~\eqref{function-requirements} are satisfied.

\subsubsection*{Anti-twist}
In the anti-twist case, which corresponds to $\twist=-1$,~\eqref{lens-i-first-choice-app} reads
\begin{equation} \label{lens-i-first-choice-app-AT}
	\lens=\labell\frac{(-1+\sqrt\alpha)(\singm+\singp)}{(\alpha-1)(\sqrt{\singm-\sqrt{\alpha}\singp})}\Bigl[2\alpha^{1/4}\sqrt{\sqrt{\alpha}\singm-\singp}+\delta(1-\sqrt{\alpha})\sqrt{\singm-\sqrt{\alpha}\singp}\Bigr]\,.
\end{equation}
In both the cases $\delta=\pm 1$, we are able to prove that
\begin{equation}
	\partial_\alpha \lens(\alpha)>0\,,\quad\forall\alpha\in \left(1,\frac{\singm^2}{\singp^2}\right),
\end{equation}
hence implying that $\lens(\alpha)$ attains its infimum at $\alpha=1$, which is $\inf_\alpha \lens(\alpha)=\labell(\singm+\singp)$. Given that $(\singm,\singp,\labell)\in\mathbb{N}$ with $\singm>\singp$, we have $\lens\geq 4$. Note that the inequality
\begin{equation}
	\lens>\labell(\singm+\singp)>3,
\end{equation}
means that at fixed $\lens$ we have only a finite set of $(\singm,\singp,\labell)$ triples which yields an acceptable solution. For completeness, we note that a possible example for $\lens=4$ is $(\lens=4,\labell=1,\singp=1,\singm=2)$ and $\alpha$ fixed from~\eqref{lens-i-first-choice-app-AT}, namely $\alpha\approx 1.36$ for $\delta=1$ and $\alpha\approx 1.63$ for $\delta=-1$.

\subsubsection*{Twist}
In the twist case, which corresponds to $\twist=1$, $\lens$ reads
\begin{equation} \label{lens-i-first-choice-app-T}
	\lens=\labell\frac{(1+\sqrt\alpha)(\singm-\singp)}{(\alpha-1)(\sqrt{\singm-\sqrt{\alpha}\singp})}\Bigl[2\alpha^{1/4}\sqrt{\sqrt{\alpha}\singm-\singp}+\delta(1+\sqrt{\alpha})\sqrt{\singm-\sqrt{\alpha}\singp}\Bigr]\,.
\end{equation}
When $\delta=-1$, for example, we can prove in a similar fashion that
\begin{equation}
	\partial_\alpha \lens(\alpha)>0\,,\quad\forall\alpha\in \left(1,\frac{\singm^2}{\singp^2}\right),
\end{equation}
resulting again in
\begin{equation}
	\lens>\labell(\singp+\singm),
\end{equation}
which implies $\lens\geq 4$ and an analogous finite bound on the number of $(\singm,\singp,\labell)$ triples at fixed $\lens$. A possible example is $(\lens=4,\labell=1,\singp=1,\singm=2)$ and $\alpha$ fixed from~\eqref{lens-i-first-choice-app-T}, namely $\alpha\approx 1.81$.

\subsection*{$N=0$, $M=0$, $\alpha>1$ case}
Moving now to the case (ii) of~\ref{subsubsect: N=0} with $\alpha>1$, which is an anti-twist case, $\lens$ is given by
\begin{equation} \label{lens-ii-app-AT}
	\lens=\labell\,\frac{\sqrt{-\singp+\singm\sqrt{-1+2\sqrt{\alpha}}}}{\sqrt{\singm-\singp\sqrt{-1+2\sqrt{\alpha}}}}\,\frac{\singm+\singp\sqrt{-1+2\sqrt{\alpha}}}{(-1+2\sqrt{\alpha})^{1/4}}.
\end{equation}
Here we have to ensure that $\alpha>1$ and $\singm>\singp\sqrt{-1+2\sqrt{\alpha}}$; these conditions are sufficient to meet the requirements in~\eqref{function-requirements}.

We can show that 
\begin{equation}
	\partial_\alpha \lens(\alpha)>0\,,\quad\forall\alpha\in \left(1,\frac{(\singm^2+\singp^2)^2}{4\singp^4}\right).
\end{equation}

In this case the main implication is as well
\begin{equation}
	\lens>\labell(\singp+\singm),
\end{equation}
resulting in $\lens\geq 4$ and a finite number of $(\labell,\singm,\singm)$ triples at fixed $\lens$. As usual, we provide an example: an acceptable solution is given by ($\lens=4,\labell=1,\singp=1,\singm=2$) and $\alpha$ determined by~\eqref{lens-ii-app-AT}, that is $\alpha\approx 1.57$.

\subsection*{Generic non-accelerating solutions}

We can try a similar analysis on the solutions presented in section~\ref{subsubsect: TwistNneq0}.
In the twist case it can be shown easily that $\lens > 1$. Indeed, the expression~\eqref{lens-twist} for $\lens$ is simply
\begin{align}
	\lens=\frac{\labell(\free-\branch)}{x}(\singm -\singp)\,.
\end{align}
Since $\singm>\singp$ and $\free>1+x$ from~\eqref{effective-parameterisation}, this clearly implies $t>\labell\geq 1$.
There are, however, acceptable configurations for $\lens=2$, such as $(\lens=2,\labell=1,\singm=2,\singp=1)$, given $(\tilde{q}_+=4,\branch=1,x=3/2)$. In the anti-twist case of section~\ref{subsubsect: Anti-twistNneq0} a general analysis is much more complicated, due to the form of $\lens$ in~\eqref{lens-anti}. We will refrain from performing such an analysis, but we note that there exist infinite values of the parameters for which all the conditions~\eqref{function-requirements} are met. As an example we can take ($\lens=10,\labell=1,\singm=4,\singp=1)$ as well as ($\free=2, x \approx0.39, \branch=-1,\bar{P}\in\mathbb{R}$).

\subsection*{Accelerating solutions}
For the solution discussed in section~\ref{subsect: A special class}, we can produce various examples, such as
\begin{equation}
	\lambda=1,\newlens=1,\newlabell=1,\newsingm=3,\newsingp=1, Q=1,P=2\sqrt{6},\omega =1\,.
\end{equation} 
For CP we can not have $\newlens=1$, which can occurs only in the $U(1)\times U(1)$-invariant solution of~\cite{Martelli:2013aqa}. This means that \emph{only} in the accelerating case we can have the squashed $S^3$ at the boundary.

\emph{En passant}, we also note that the two following solutions are admissible:
\begin{equation}
	\begin{aligned}
		\text{CP}&:\quad \lambda=-1,\newlens=5,\newsingm=5,\newsingp=1,\newlabell=1,Q=\frac{9 P}{4\sqrt{5}},P>0, \omega> \frac{2}{\sqrt{5}}\,,
		\\
		\text{PD}&:\quad \lambda=\eta=1, \newlens=5, \newsingm=5,\newsingp=1,\newlabell=1,\free=1+\frac{5 x}{4},\freetwist>0,x>\frac{8}{9}\,.
	\end{aligned}
\end{equation}
This shows that the same topological data $(\newsingpm,\newlabell,\newlens)$ can indeed correspond to two different non-diffeomorphic solutions.

\section{Uplift to M-theory}\label{appendix:Uplift}
In this appendix we comment on the conditions under which our solution to~\eqref{minimalSUGRAeom} and~\eqref{KSE} uplifts to a M-theory solution.  In particular,
using the results in~\cite{Gauntlett:2007ma, Gauntlett:2009zw} one can uplift locally  any supersymmetric solution on a seven-dimensional Sasaki-Einstein manifold, but there can be global obstructions.

We begin our analysis recalling the bosonic part of the Euclidean 11d supergravity action\footnote{Here and in what follows we define, for any $p$-form $\omega$, $\omega^2 = \frac{1}{p!}\omega_{\mu_{1}..\mu_{p}}\omega^{\mu_{1}..\mu_{p}}$.}\textsuperscript{,}\footnote{We adopt the following (classical) convention for the indices: capital Latin letters $\{M, N,\ldots\}$ are reserved for the eleven-dimensional metric and take values in $\{1,\ldots,11\}$, small Greek letters $\{\mu,\nu,\ldots\}$ are used for the four-dimensional external space, and small Latin letters $\{m,n,\ldots\}$ are left for the internal seven-dimensional metric.} 
\begin{equation}
	\begin{aligned}
		-(16\pi G_{11} )S_{11}&=\int\Bigl[R\star 1-\frac{1}{2}\star G_{(4)}\wedge G_{(4)}\Bigr]
		-\frac{1}{6}\int G_{(4)}\wedge G_{(4)}\wedge C_{(3)}
		\\
		&=\int \dd^{11}x \sqrt{g}\Big[R-\frac{1}{2}|G_{(4)}|^2\Big]-\frac{\tilde{\epsilon}^{\,M_{1}\ldots M_{11}}}{(3! \,2)^4}\int\dd^{11}x G_{M_{1}\ldots M_{4}}G_{M_{5}\ldots M_{8}}C_{M_{9}\ldots M_{11}}\,,
	\end{aligned}
\end{equation}
whose equations of motion read
\begin{equation} \label{11dSUGRAeom}
	\begin{aligned}
		R_{MN}&=\frac{1}{12}\Big[G_{M R_{1}\ldots R_{3}}G_{N}{}^{R_{1}\ldots R_{3}}-\frac{1}{12}g_{MN} G_{R_{1}\ldots R_{4}}G^{R_{1}\ldots R_{4}}\Big]\,,\quad\dd\star G_{(4)}&=-\frac{1}{2}G_{(4)}\wedge G_{(4)}\,,
	\end{aligned}
\end{equation}
where $G_{(4)}=\dd C_{(3)}$ and $\star$ is the Hodge star map with respect to the eleven-dimensional metric. It is possible to show (see \cite{Gauntlett:2007ma,Gauntlett:2009zw}) that the Ansatz
\begin{equation}
	\begin{aligned}
		\dd s^{2}_{11}&=L^2\biggl\{\frac{1}{4}\dd s^{2}_{4\text{d}}+\Big[\contform+\frac{1}{2} A_{4\text{d}}\Bigr]^2+\dd s^{2}_{T}\biggr\}\,,
		\\
		G_{(4)}&=L^3\biggl\{\frac{3}{8}\star_{4\text{d}}1-\frac{1}{4}\star_{4\text{d}} F_{4\text{d}}\wedge \dd \eta\biggr\}\,,
	\end{aligned}
	\label{UpliftAnsatz}
\end{equation}
with $R_{mn}^{(7)}=6 g_{mn}^{(7)}$, indeed leads to a solution of~\eqref{11dSUGRAeom} for any solution of~\eqref{minimalSUGRAeom}. Here we adopt the conventions of~\cite{Martelli:2012sz}, where the effective radius $L$ of $\AdS_4$ is determined by the quantization of $\star G_{(4)}$ through $Y_7$, which is a Sasaki-Einstein seven-manifold with contact one-form $\eta$ and transverse K\"ahler-Einstein metric $\dd s_{T}^2$.

The uplift of our spindle Bolt solution parallels the one in section 6.3 of~\cite{Martelli:2012sz}. We basically just need to remark that the spindle Bolt solution is topologically $\mathcal{M}_{4}\equiv\CC/\ZZ_{\labell}\hookrightarrow \mathcal{O}(-\lens)\rightarrow \spindle_{[m_+,m_-]}^{q_+}$, while the gauge field is quantized according to~\eqref{twist-anti-twist}, which reads
\begin{align} \label{FUpliftquantization}
	\frac{1}{2\pi}\int_{\loci_2}  F_{4\mathrm{d}}=\frac{\parppm\signpp}{2}\Bigl(\frac{\singm+\twist\singp}{\singm\singp}-\branch\frac{\lens/\labell}{\singm \,\singp}\Bigr)\,,
\end{align}
where $\loci_2$ is the surface of the spindle bolt $\spindle^{q_+}$.
Let us focus on the case of a regular Sasaki-Einstein internal manifold $Y_7$. By definition, in this case, $Y_7$ can be regarded as the $U(1)$ fibration over a six-dimensional Kähler-Einstein base $B_6$ with positive curvature and metric $\dd s^2_T$. In particular, we can write $\contform=\dd\psi+\sigma$, where the Ricci form $\rho$ of $B_6$ is $\rho=4\,\dd\sigma$ and the coordinate $\psi$ is such that the Reeb vector $\xi$ reads $\xi=\partial_{\psi}$. The period of $\xi$ is canonically $2\pi/4$, which coincides with the smallest period required in order to have that the two Killing spinors are globally defined. Smaller periods would imply non-single valued Killing spinors, but larger ones are indeed admissible. In particular, recall that a Sasaki-Einstein is simply connected if and only if $\xi$ has period $2\pi I/4$, where $I\equiv I(B_6)$ is the Fano index of $B_6$. 

We can then consider discrete quotients of Sasaki-Einstein manifolds at the price of losing their simple connectedness; for example we can consider $S^7/\mathbb{Z}_4$ or $S^7/\mathbb{Z}_2$. This discussion can be summarized saying that $\xi$ has periodicity $\frac{2\pi I}{4k}$, where the integer $k$ divides $I$.

In the connection part of the uplift metric in~\eqref{UpliftAnsatz} we see that, since $\xi$ has period $\frac{2\pi I}{4k}$, we must have
\begin{equation} \label{CircleBundleCondition}
	\frac{4k}{2I}\int_{\loci_{2}}\frac{F_{4\mathrm{d}}}{2\pi}=\frac{m}{\labell\, \singm\singp}\,,\quad m\in\mathbb{Z}\,,
\end{equation}   
in order to have a well-defined ${U}(1)$ orbibundle.

Comparing~\eqref{FUpliftquantization} and~\eqref{CircleBundleCondition}, it is immediate to see that
\begin{equation} \label{BoltUpliftCondition}
	k \parflpm\eta\,(\singm\labell+\sigma\singp\labell-\kappa\lens)=m I.
\end{equation} 
When $k=I$, which amounts to requiring canonical period $2\pi/4$ for $\xi$, it is easy to see that the condition~\eqref{BoltUpliftCondition} is always satisfied, and then any spindle Bolt solution can be globally uplifted to a M-theory solution on a regular SE manifold. Examples of internal manifolds with such property are, for example, the homogeneous $S^7/\mathbb{Z}_4$, $V_{5,2}/\mathbb{Z}_3$, $Q^{2,2,2}=Q^{1,1,1}/\mathbb{Z}_2$, $M_{3,2}$, $N_{1,1}$.

However, in general, the condition~\eqref{BoltUpliftCondition} leads to restrictions. For example, consider the case of $M_7=S^7/\mathbb{Z}_2$, which implies $I=2$, $k=1$. By~\eqref{BoltUpliftCondition} this means that $\singm\labell+\sigma\singp\labell-\kappa\lens$ is divisible by 2, which is not satisfied for all values of $(\singm,\singp,\labell,\lens)$; in those cases where this condition is not met, the solution cannot be globally uplifted to a M-theory solution on $\AdS_4 \ltimes S^7/\mathbb{Z}_2$.

To conclude the analysis, let us note that, when $Y_7$ is irregular, the uplift is not possible in general. This is due to the fact that when $Y_7$ is irregular, $\xi$ cannot be periodically identified on $Y_7$. Therefore, since $\eta+\frac{1}{2} A_{4\mathrm{d}}$ defines a global one-form only when $\xi$ is periodically identified in $\eta=\dd\psi+\sigma$, our Bolt solutions cannot be lifted on irregular Sasaki-Einstein manifolds. The caveat here is that the flux of $A_{4\mathrm{d}}$ through the spindle bolt can vanish in some cases: this happens, for example, in the no-twist case in (i) of~\ref{subsubsect: N=0}, or whenever~\eqref{FUpliftquantization} vanishes, namely 
\begin{equation}{\label{no-flux}}
\labell(\singm+\twist\singp)-\kappa t=0.
\end{equation}
Note that, when $\twist=1$ and $\branch=1$, this is precisely the condition~\eqref{fibration-solved-flat}. When~\eqref{no-flux} is met, it is indeed possible to uplift the solution for any choice of $Y_7$.

	\bibliographystyle{JHEP}
	\bibliography{bolts}

\providecommand{\href}[2]{#2}\begingroup\raggedright\begin{thebibliography}{10}

\bibitem{Ferrero:2020laf}
P.~Ferrero, J.P.~Gauntlett, J.M.~P\'erez Ipi\~na, D.~Martelli and J.~Sparks,
  \emph{{D3-Branes Wrapped on a Spindle}},
  \href{https://doi.org/10.1103/PhysRevLett.126.111601}{\emph{Phys. Rev. Lett.}
  {\bfseries 126} (2021) 111601}
  [\href{https://arxiv.org/abs/2011.10579}{{\ttfamily 2011.10579}}].

\bibitem{Ferrero:2021wvk}
P.~Ferrero, J.P.~Gauntlett, D.~Martelli and J.~Sparks, \emph{{M5-branes wrapped
  on a spindle}}, \href{https://doi.org/10.1007/JHEP11(2021)002}{\emph{JHEP}
  {\bfseries 11} (2021) 002}
  [\href{https://arxiv.org/abs/2105.13344}{{\ttfamily 2105.13344}}].

\bibitem{Faedo:2021nub}
F.~Faedo and D.~Martelli, \emph{{D4-branes wrapped on a spindle}},
  \href{https://doi.org/10.1007/JHEP02(2022)101}{\emph{JHEP} {\bfseries 02}
  (2022) 101} [\href{https://arxiv.org/abs/2111.13660}{{\ttfamily
  2111.13660}}].

\bibitem{Ferrero:2021etw}
P.~Ferrero, J.P.~Gauntlett and J.~Sparks, \emph{{Supersymmetric spindles}},
  \href{https://doi.org/10.1007/JHEP01(2022)102}{\emph{JHEP} {\bfseries 01}
  (2022) 102} [\href{https://arxiv.org/abs/2112.01543}{{\ttfamily
  2112.01543}}].

\bibitem{Ferrero:2020twa}
P.~Ferrero, J.P.~Gauntlett, J.M.P.~Ipi\~na, D.~Martelli and J.~Sparks,
  \emph{{Accelerating black holes and spinning spindles}},
  \href{https://doi.org/10.1103/PhysRevD.104.046007}{\emph{Phys. Rev. D}
  {\bfseries 104} (2021) 046007}
  [\href{https://arxiv.org/abs/2012.08530}{{\ttfamily 2012.08530}}].

\bibitem{Cassani:2021dwa}
D.~Cassani, J.P.~Gauntlett, D.~Martelli and J.~Sparks, \emph{{Thermodynamics of
  accelerating and supersymmetric AdS4 black holes}},
  \href{https://doi.org/10.1103/PhysRevD.104.086005}{\emph{Phys. Rev. D}
  {\bfseries 104} (2021) 086005}
  [\href{https://arxiv.org/abs/2106.05571}{{\ttfamily 2106.05571}}].

\bibitem{Inglese:2023tyc}
M.~Inglese, D.~Martelli and A.~Pittelli, \emph{{Supersymmetry and Localization
  on Three-Dimensional Orbifolds}},
  \href{https://arxiv.org/abs/2312.17086}{{\ttfamily 2312.17086}}.

\bibitem{Inglese:2023wky}
M.~Inglese, D.~Martelli and A.~Pittelli, \emph{{The spindle index from
  localization}}, \href{https://doi.org/10.1088/1751-8121/ad2225}{\emph{J.
  Phys. A} {\bfseries 57} (2024) 085401}
  [\href{https://arxiv.org/abs/2303.14199}{{\ttfamily 2303.14199}}].

\bibitem{Colombo:2024mts}
E.~Colombo, S.M.~Hosseini, D.~Martelli, A.~Pittelli and A.~Zaffaroni,
  \emph{{Microstates of Accelerating and Supersymmetric AdS4 Black Holes from
  the Spindle Index}},
  \href{https://doi.org/10.1103/PhysRevLett.133.031603}{\emph{Phys. Rev. Lett.}
  {\bfseries 133} (2024) 031603}
  [\href{https://arxiv.org/abs/2404.07173}{{\ttfamily 2404.07173}}].

\bibitem{Gibbons:1979xm}
G.W.~Gibbons and S.W.~Hawking, \emph{{Classification of Gravitational Instanton
  Symmetries}}, \href{https://doi.org/10.1007/BF01197189}{\emph{Commun. Math.
  Phys.} {\bfseries 66} (1979) 291}.

\bibitem{conference:Dario}
D.~Martelli, \emph{From wrapped branes to spindles},  in
  \emph{\href{https://drive.google.com/file/d/1SiqXSNH7awXqqcrit_LKfQp7JK0NNGSN/view}{Branes,
  black holes and geometry: a meeting in celebration of Jerome Gauntlett's 60th
  birthday}}, 25-26 April 2024, Imperial College, London.

\bibitem{conference:Alessio}
A.~Fontanarossa, \emph{Nuts, bolts, and spindles},  in
  \emph{\href{https://indico.cern.ch/event/1374516/contributions/6043327/}{Eurostrings
  2024}}, 2-6 September 2024, University of Southampton, Southampton.

\bibitem{Martelli:2011fu}
D.~Martelli, A.~Passias and J.~Sparks, \emph{{The gravity dual of
  supersymmetric gauge theories on a squashed three-sphere}},
  \href{https://doi.org/10.1016/j.nuclphysb.2012.07.019}{\emph{Nucl. Phys. B}
  {\bfseries 864} (2012) 840}
  [\href{https://arxiv.org/abs/1110.6400}{{\ttfamily 1110.6400}}].

\bibitem{Martelli:2011fw}
D.~Martelli and J.~Sparks, \emph{{The gravity dual of supersymmetric gauge
  theories on a biaxially squashed three-sphere}},
  \href{https://doi.org/10.1016/j.nuclphysb.2012.08.015}{\emph{Nucl. Phys. B}
  {\bfseries 866} (2013) 72} [\href{https://arxiv.org/abs/1111.6930}{{\ttfamily
  1111.6930}}].

\bibitem{Martelli:2012sz}
D.~Martelli, A.~Passias and J.~Sparks, \emph{{The supersymmetric NUTs and bolts
  of holography}},
  \href{https://doi.org/10.1016/j.nuclphysb.2013.04.026}{\emph{Nucl. Phys. B}
  {\bfseries 876} (2013) 810}
  [\href{https://arxiv.org/abs/1212.4618}{{\ttfamily 1212.4618}}].

\bibitem{Martelli:2013aqa}
D.~Martelli and A.~Passias, \emph{{The gravity dual of supersymmetric gauge
  theories on a two-parameter deformed three-sphere}},
  \href{https://doi.org/10.1016/j.nuclphysb.2013.09.012}{\emph{Nucl. Phys. B}
  {\bfseries 877} (2013) 51} [\href{https://arxiv.org/abs/1306.3893}{{\ttfamily
  1306.3893}}].

\bibitem{Carter:1968ks}
B.~Carter, \emph{{Hamilton-Jacobi and Schrodinger separable solutions of
  Einstein's equations}},
  \href{https://doi.org/10.1007/BF03399503}{\emph{Commun. Math. Phys.}
  {\bfseries 10} (1968) 280}.

\bibitem{PLEBANSKI1975196}
J.F.~Plebañski, \emph{A class of solutions of einstein-maxwell equations},
  \href{https://doi.org/https://doi.org/10.1016/0003-4916(75)90145-1}{\emph{Annals
  of Physics} {\bfseries 90} (1975) 196}.

\bibitem{Alonso-Alberca:2000zeh}
N.~Alonso-Alberca, P.~Meessen and T.~Ortin, \emph{{Supersymmetry of topological
  Kerr-Newman-Taub-NUT-AdS space-times}},
  \href{https://doi.org/10.1088/0264-9381/17/14/312}{\emph{Class. Quant. Grav.}
  {\bfseries 17} (2000) 2783}
  [\href{https://arxiv.org/abs/hep-th/0003071}{{\ttfamily hep-th/0003071}}].

\bibitem{Klemm:2013eca}
D.~Klemm and M.~Nozawa, \emph{{Supersymmetry of the C-metric and the general
  Plebanski-Demianski solution}},
  \href{https://doi.org/10.1007/JHEP05(2013)123}{\emph{JHEP} {\bfseries 05}
  (2013) 123} [\href{https://arxiv.org/abs/1303.3119}{{\ttfamily 1303.3119}}].

\bibitem{PLEBANSKI197698}
J.~Plebanski and M.~Demianski, \emph{Rotating, charged, and uniformly
  accelerating mass in general relativity},
  \href{https://doi.org/https://doi.org/10.1016/0003-4916(76)90240-2}{\emph{Annals
  of Physics} {\bfseries 98} (1976) 98}.

\bibitem{Griffiths:2005qp}
J.B.~Griffiths and J.~Podolsky, \emph{{A New look at the Plebanski-Demianski
  family of solutions}},
  \href{https://doi.org/10.1142/S0218271806007742}{\emph{Int. J. Mod. Phys. D}
  {\bfseries 15} (2006) 335}
  [\href{https://arxiv.org/abs/gr-qc/0511091}{{\ttfamily gr-qc/0511091}}].

\bibitem{Podolsky:2006px}
J.~Podolsky and J.B.~Griffiths, \emph{{Accelerating Kerr-Newman black holes in
  (anti-)de Sitter space-time}},
  \href{https://doi.org/10.1103/PhysRevD.73.044018}{\emph{Phys. Rev. D}
  {\bfseries 73} (2006) 044018}
  [\href{https://arxiv.org/abs/gr-qc/0601130}{{\ttfamily gr-qc/0601130}}].

\bibitem{Podolsky:2022xxd}
J.~Podolsky and A.~Vratny, \emph{{New form of all black holes of type D with a
  cosmological constant}},
  \href{https://doi.org/10.1103/PhysRevD.107.084034}{\emph{Phys. Rev. D}
  {\bfseries 107} (2023) 084034}
  [\href{https://arxiv.org/abs/2212.08865}{{\ttfamily 2212.08865}}].

\bibitem{Astorino:2024bfl}
M.~Astorino, \emph{{Most general Type-D Black Hole and Accelerating
  Reissner-Nordstrom-NUT-(A)dS}},
  \href{https://arxiv.org/abs/2404.06551}{{\ttfamily 2404.06551}}.

\bibitem{Martelli:2005tp}
D.~Martelli, J.~Sparks and S.-T.~Yau, \emph{{The Geometric dual of
  a-maximisation for Toric Sasaki-Einstein manifolds}},
  \href{https://doi.org/10.1007/s00220-006-0087-0}{\emph{Commun. Math. Phys.}
  {\bfseries 268} (2006) 39}
  [\href{https://arxiv.org/abs/hep-th/0503183}{{\ttfamily hep-th/0503183}}].

\bibitem{Martelli:2006yb}
D.~Martelli, J.~Sparks and S.-T.~Yau, \emph{{Sasaki-Einstein manifolds and
  volume minimisation}},
  \href{https://doi.org/10.1007/s00220-008-0479-4}{\emph{Commun. Math. Phys.}
  {\bfseries 280} (2008) 611}
  [\href{https://arxiv.org/abs/hep-th/0603021}{{\ttfamily hep-th/0603021}}].

\bibitem{Futaki:2006cc}
A.~Futaki, H.~Ono and G.~Wang, \emph{{Transverse Kahler geometry of Sasaki
  manifolds and toric Sasaki-Einstein manifolds}}, {\emph{J. Diff. Geom.}
  {\bfseries 83} (2009) 585}
  [\href{https://arxiv.org/abs/math/0607586}{{\ttfamily math/0607586}}].

\bibitem{Farquet:2014kma}
D.~Farquet, J.~Lorenzen, D.~Martelli and J.~Sparks, \emph{{Gravity duals of
  supersymmetric gauge theories on three-manifolds}},
  \href{https://doi.org/10.1007/JHEP08(2016)080}{\emph{JHEP} {\bfseries 08}
  (2016) 080} [\href{https://arxiv.org/abs/1404.0268}{{\ttfamily 1404.0268}}].

\bibitem{BenettiGenolini:2019jdz}
P.~Benetti~Genolini, J.M.~Perez Ipi\~na and J.~Sparks, \emph{{Localization of
  the action in AdS/CFT}},
  \href{https://doi.org/10.1007/JHEP10(2019)252}{\emph{JHEP} {\bfseries 10}
  (2019) 252} [\href{https://arxiv.org/abs/1906.11249}{{\ttfamily
  1906.11249}}].

\bibitem{BenettiGenolini:2023kxp}
P.~Benetti~Genolini, J.P.~Gauntlett and J.~Sparks, \emph{{Equivariant
  Localization in Supergravity}},
  \href{https://doi.org/10.1103/PhysRevLett.131.121602}{\emph{Phys. Rev. Lett.}
  {\bfseries 131} (2023) 121602}
  [\href{https://arxiv.org/abs/2306.03868}{{\ttfamily 2306.03868}}].

\bibitem{BenettiGenolini:2024xeo}
P.~Benetti~Genolini, J.P.~Gauntlett, Y.~Jiao, A.~L\"uscher and J.~Sparks,
  \emph{{Localization of the Free Energy in Supergravity}},
  \href{https://doi.org/10.1103/PhysRevLett.133.141601}{\emph{Phys. Rev. Lett.}
  {\bfseries 133} (2024) 141601}
  [\href{https://arxiv.org/abs/2407.02554}{{\ttfamily 2407.02554}}].

\bibitem{BenettiGenolini:2024hyd}
P.~Benetti~Genolini, J.P.~Gauntlett, Y.~Jiao, A.~L\"uscher and J.~Sparks,
  \emph{{Toric gravitational instantons in gauged supergravity}},
  \href{https://arxiv.org/abs/2410.19036}{{\ttfamily 2410.19036}}.

\bibitem{Faedo:2022rqx}
F.~Faedo, A.~Fontanarossa and D.~Martelli, \emph{{Branes wrapped on orbifolds
  and their gravitational blocks}},
  \href{https://doi.org/10.1007/s11005-023-01671-1}{\emph{Lett. Math. Phys.}
  {\bfseries 113} (2023) 51}
  [\href{https://arxiv.org/abs/2210.16128}{{\ttfamily 2210.16128}}].

\bibitem{Faedo:2024upq}
F.~Faedo, A.~Fontanarossa and D.~Martelli, \emph{{Branes wrapped on
  quadrilaterals}},  \href{https://arxiv.org/abs/2402.08724}{{\ttfamily
  2402.08724}}.

\bibitem{Freedman:1976aw}
D.Z.~Freedman and A.K.~Das, \emph{{Gauge Internal Symmetry in Extended
  Supergravity}},
  \href{https://doi.org/10.1016/0550-3213(77)90041-4}{\emph{Nucl. Phys. B}
  {\bfseries 120} (1977) 221}.

\bibitem{LeBrun:2008kh}
C.~LeBrun, \emph{{The Einstein-Maxwell Equations, Extremal Kahler Metrics, and
  Seiberg-Witten Theory}},  (2008),
  \href{https://doi.org/10.1093/acprof:oso/9780199534920.003.0003}{DOI}
  [\href{https://arxiv.org/abs/0803.3734}{{\ttfamily 0803.3734}}].

\bibitem{Caldarelli:1998hg}
M.M.~Caldarelli and D.~Klemm, \emph{{Supersymmetry of Anti-de Sitter black
  holes}}, \href{https://doi.org/10.1016/S0550-3213(98)00846-3}{\emph{Nucl.
  Phys. B} {\bfseries 545} (1999) 434}
  [\href{https://arxiv.org/abs/hep-th/9808097}{{\ttfamily hep-th/9808097}}].

\bibitem{Gauntlett:2007ma}
J.P.~Gauntlett and O.~Varela, \emph{{Consistent Kaluza-Klein reductions for
  general supersymmetric AdS solutions}},
  \href{https://doi.org/10.1103/PhysRevD.76.126007}{\emph{Phys. Rev. D}
  {\bfseries 76} (2007) 126007}
  [\href{https://arxiv.org/abs/0707.2315}{{\ttfamily 0707.2315}}].

\bibitem{Gauntlett:2009zw}
J.P.~Gauntlett, S.~Kim, O.~Varela and D.~Waldram, \emph{{Consistent
  supersymmetric Kaluza-Klein truncations with massive modes}},
  \href{https://doi.org/10.1088/1126-6708/2009/04/102}{\emph{JHEP} {\bfseries
  04} (2009) 102} [\href{https://arxiv.org/abs/0901.0676}{{\ttfamily
  0901.0676}}].

\bibitem{Cabo-Bizet:2018ehj}
A.~Cabo-Bizet, D.~Cassani, D.~Martelli and S.~Murthy, \emph{{Microscopic origin
  of the Bekenstein-Hawking entropy of supersymmetric AdS$_{5}$ black holes}},
  \href{https://doi.org/10.1007/JHEP10(2019)062}{\emph{JHEP} {\bfseries 10}
  (2019) 062} [\href{https://arxiv.org/abs/1810.11442}{{\ttfamily
  1810.11442}}].

\bibitem{Cassani:2019mms}
D.~Cassani and L.~Papini, \emph{{The BPS limit of rotating AdS black hole
  thermodynamics}}, \href{https://doi.org/10.1007/JHEP09(2019)079}{\emph{JHEP}
  {\bfseries 09} (2019) 079}
  [\href{https://arxiv.org/abs/1906.10148}{{\ttfamily 1906.10148}}].

\bibitem{Astorino:2023ifg}
M.~Astorino, \emph{{Accelerating and charged type I black holes}},
  \href{https://doi.org/10.1103/PhysRevD.108.124025}{\emph{Phys. Rev. D}
  {\bfseries 108} (2023) 124025}
  [\href{https://arxiv.org/abs/2307.10534}{{\ttfamily 2307.10534}}].

\bibitem{Apostolov:2013oza}
V.~Apostolov, D.M.J.~Calderbank and P.~Gauduchon, \emph{{Ambitoric geometry I:
  Einstein metrics and extremal ambik\"ahler structures}},
  \href{https://doi.org/10.1515/crelle-2014-0060}{\emph{J. Reine Angew. Math.}
  {\bfseries 2016} (2016) 109}
  [\href{https://arxiv.org/abs/1302.6975}{{\ttfamily 1302.6975}}].

\bibitem{2013arXiv1302.6979A}
V.~{Apostolov}, D.M.J.~{Calderbank} and P.~{Gauduchon}, \emph{{Ambitoric
  geometry II: Extremal toric surfaces and Einstein 4-orbifolds}},
  \href{https://doi.org/10.48550/arXiv.1302.6979}{\emph{arXiv e-prints} (2013)
  arXiv:1302.6979} [\href{https://arxiv.org/abs/1302.6979}{{\ttfamily
  1302.6979}}].

\bibitem{BenettiGenolini:2016tsn}
P.~Benetti~Genolini, D.~Cassani, D.~Martelli and J.~Sparks, \emph{{Holographic
  renormalization and supersymmetry}},
  \href{https://doi.org/10.1007/JHEP02(2017)132}{\emph{JHEP} {\bfseries 02}
  (2017) 132} [\href{https://arxiv.org/abs/1612.06761}{{\ttfamily
  1612.06761}}].

\bibitem{Toldo:2017qsh}
C.~Toldo and B.~Willett, \emph{{Partition functions on 3d circle bundles and
  their gravity duals}},
  \href{https://doi.org/10.1007/JHEP05(2018)116}{\emph{JHEP} {\bfseries 05}
  (2018) 116} [\href{https://arxiv.org/abs/1712.08861}{{\ttfamily
  1712.08861}}].

\bibitem{Hong:2024uns}
J.~Hong, \emph{{Perturbatively exact supersymmetric partition functions of ABJM
  theory on Seifert manifolds and holography}},
  \href{https://arxiv.org/abs/2411.09006}{{\ttfamily 2411.09006}}.

\bibitem{Martelli:2007pv}
D.~Martelli and J.~Sparks, \emph{{Resolutions of non-regular Ricci-flat Kahler
  cones}}, \href{https://doi.org/10.1016/j.geomphys.2009.06.005}{\emph{J. Geom.
  Phys.} {\bfseries 59} (2009) 1175}
  [\href{https://arxiv.org/abs/0707.1674}{{\ttfamily 0707.1674}}].

\bibitem{Closset:2018ghr}
C.~Closset, H.~Kim and B.~Willett, \emph{{Seifert fibering operators in 3d
  $\mathcal{N}=2$ theories}},
  \href{https://doi.org/10.1007/JHEP11(2018)004}{\emph{JHEP} {\bfseries 11}
  (2018) 004} [\href{https://arxiv.org/abs/1807.02328}{{\ttfamily
  1807.02328}}].

\bibitem{2016arXiv160806844G}
H.~{Geiges} and C.~{Lange}, \emph{{Seifert fibrations of lens spaces}},
  \href{https://doi.org/10.48550/arXiv.1608.06844}{\emph{arXiv e-prints} (2016)
  arXiv:1608.06844} [\href{https://arxiv.org/abs/1608.06844}{{\ttfamily
  1608.06844}}].

\bibitem{Lerman:1995aaa}
E.~Lerman and S.~Tolman, \emph{{Hamiltonian torus actions on symplectic
  orbifolds and toric varieties}},
  \href{https://doi.org/10.1090/S0002-9947-97-01821-7}{\emph{Trans. Amer. Math.
  Soc.} {\bfseries 349} (1997) 4201}
  [\href{https://arxiv.org/abs/dg-ga/9511008}{{\ttfamily dg-ga/9511008}}].

\bibitem{Martelli:2004wu}
D.~Martelli and J.~Sparks, \emph{{Toric geometry, Sasaki-Einstein manifolds and
  a new infinite class of AdS/CFT duals}},
  \href{https://doi.org/10.1007/s00220-005-1425-3}{\emph{Commun. Math. Phys.}
  {\bfseries 262} (2006) 51}
  [\href{https://arxiv.org/abs/hep-th/0411238}{{\ttfamily hep-th/0411238}}].

\bibitem{Martelli:2023oqk}
D.~Martelli and A.~Zaffaroni, \emph{{Equivariant localization and holography}},
  \href{https://doi.org/10.1007/s11005-023-01752-1}{\emph{Lett. Math. Phys.}
  {\bfseries 114} (2024) 15}
  [\href{https://arxiv.org/abs/2306.03891}{{\ttfamily 2306.03891}}].

\bibitem{BenettiGenolini:2023ucp}
P.~Benetti~Genolini and C.~Toldo, \emph{{Magnetic charge and black hole
  supersymmetric quantum statistical relation}},
  \href{https://doi.org/10.1103/PhysRevD.107.L121902}{\emph{Phys. Rev. D}
  {\bfseries 107} (2023) L121902}
  [\href{https://arxiv.org/abs/2304.00605}{{\ttfamily 2304.00605}}].

\bibitem{Emparan:1999pm}
R.~Emparan, C.V.~Johnson and R.C.~Myers, \emph{{Surface terms as counterterms
  in the AdS / CFT correspondence}},
  \href{https://doi.org/10.1103/PhysRevD.60.104001}{\emph{Phys. Rev. D}
  {\bfseries 60} (1999) 104001}
  [\href{https://arxiv.org/abs/hep-th/9903238}{{\ttfamily hep-th/9903238}}].

\bibitem{Skenderis:2002wp}
K.~Skenderis, \emph{{Lecture notes on holographic renormalization}},
  \href{https://doi.org/10.1088/0264-9381/19/22/306}{\emph{Class. Quant. Grav.}
  {\bfseries 19} (2002) 5849}
  [\href{https://arxiv.org/abs/hep-th/0209067}{{\ttfamily hep-th/0209067}}].

\bibitem{Duistermaat:1982vw}
J.J.~Duistermaat and G.J.~Heckman, \emph{{On the Variation in the cohomology of
  the symplectic form of the reduced phase space}},
  \href{https://doi.org/10.1007/BF01399506}{\emph{Invent. Math.} {\bfseries 69}
  (1982) 259}.

\bibitem{berline1982classes}
N.~Berline and M.~Vergne, \emph{Classes caract{\'e}ristiques {\'e}quivariantes.
  formule de localisation en cohomologie {\'e}quivariante}, {\emph{CR Acad.
  Sci. Paris} {\bfseries 295} (1982) 539}.

\bibitem{Atiyah:1984px}
M.F.~Atiyah and R.~Bott, \emph{{The Moment map and equivariant cohomology}},
  \href{https://doi.org/10.1016/0040-9383(84)90021-1}{\emph{Topology}
  {\bfseries 23} (1984) 1}.

\bibitem{BenettiGenolini:2023ndb}
P.~Benetti~Genolini, J.P.~Gauntlett and J.~Sparks, \emph{{Equivariant
  localization for AdS/CFT}},
  \href{https://doi.org/10.1007/JHEP02(2024)015}{\emph{JHEP} {\bfseries 02}
  (2024) 015} [\href{https://arxiv.org/abs/2308.11701}{{\ttfamily
  2308.11701}}].

\bibitem{Boido:2022iye}
A.~Boido, J.P.~Gauntlett, D.~Martelli and J.~Sparks, \emph{{Entropy Functions
  For Accelerating Black Holes}},
  \href{https://doi.org/10.1103/PhysRevLett.130.091603}{\emph{Phys. Rev. Lett.}
  {\bfseries 130} (2023) 091603}
  [\href{https://arxiv.org/abs/2210.16069}{{\ttfamily 2210.16069}}].

\bibitem{Boido:2022mbe}
A.~Boido, J.P.~Gauntlett, D.~Martelli and J.~Sparks, \emph{{Gravitational
  Blocks, Spindles and GK Geometry}},
  \href{https://doi.org/10.1007/s00220-023-04812-8}{\emph{Commun. Math. Phys.}
  {\bfseries 403} (2023) 917}
  [\href{https://arxiv.org/abs/2211.02662}{{\ttfamily 2211.02662}}].

\bibitem{BenettiGenolini:2024kyy}
P.~Benetti~Genolini, J.P.~Gauntlett, Y.~Jiao, A.~L\"uscher and J.~Sparks,
  \emph{{Localization and attraction}},
  \href{https://doi.org/10.1007/JHEP05(2024)152}{\emph{JHEP} {\bfseries 05}
  (2024) 152} [\href{https://arxiv.org/abs/2401.10977}{{\ttfamily
  2401.10977}}].

\bibitem{Hristov:2024cgj}
K.~Hristov, \emph{{Equivariant localization and gluing rules in 4d
  $\mathcal{N}=2$ higher derivative supergravity}},  6, 2024
  [\href{https://arxiv.org/abs/2406.18648}{{\ttfamily 2406.18648}}].

\bibitem{Ferrero:2021ovq}
P.~Ferrero, M.~Inglese, D.~Martelli and J.~Sparks, \emph{{Multicharge
  accelerating black holes and spinning spindles}},
  \href{https://doi.org/10.1103/PhysRevD.105.126001}{\emph{Phys. Rev. D}
  {\bfseries 105} (2022) 126001}
  [\href{https://arxiv.org/abs/2109.14625}{{\ttfamily 2109.14625}}].

\bibitem{Couzens:2021cpk}
C.~Couzens, \emph{{A tale of (M)2 twists}},
  \href{https://doi.org/10.1007/JHEP03(2022)078}{\emph{JHEP} {\bfseries 03}
  (2022) 078} [\href{https://arxiv.org/abs/2112.04462}{{\ttfamily
  2112.04462}}].

\bibitem{Benini:2015noa}
F.~Benini and A.~Zaffaroni, \emph{{A topologically twisted index for
  three-dimensional supersymmetric theories}},
  \href{https://doi.org/10.1007/JHEP07(2015)127}{\emph{JHEP} {\bfseries 07}
  (2015) 127} [\href{https://arxiv.org/abs/1504.03698}{{\ttfamily
  1504.03698}}].

\bibitem{Papadimitriou:2005ii}
I.~Papadimitriou and K.~Skenderis, \emph{{Thermodynamics of asymptotically
  locally AdS spacetimes}},
  \href{https://doi.org/10.1088/1126-6708/2005/08/004}{\emph{JHEP} {\bfseries
  08} (2005) 004} [\href{https://arxiv.org/abs/hep-th/0505190}{{\ttfamily
  hep-th/0505190}}].

\bibitem{Anabalon:2018qfv}
A.~Anabal\'on, F.~Gray, R.~Gregory, D.~Kubiz\v{n}\'ak and R.B.~Mann,
  \emph{{Thermodynamics of Charged, Rotating, and Accelerating Black Holes}},
  \href{https://doi.org/10.1007/JHEP04(2019)096}{\emph{JHEP} {\bfseries 04}
  (2019) 096} [\href{https://arxiv.org/abs/1811.04936}{{\ttfamily
  1811.04936}}].

\bibitem{AST_1985__S131__95_0}
C.~Fefferman and C.R.~Graham, \emph{Conformal invariants},  in \emph{\'Elie
  Cartan et les math\'ematiques d'aujourd'hui - Lyon, 25-29 juin 1984},
  no.~S131 in Ast\'erisque, pp.~95--116, Soci\'et\'e math\'ematique de France
  (1985).

\bibitem{Cvetic:2005ft}
M.~Cvetic, H.~Lu, D.N.~Page and C.N.~Pope, \emph{{New Einstein-Sasaki spaces in
  five and higher dimensions}},
  \href{https://doi.org/10.1103/PhysRevLett.95.071101}{\emph{Phys. Rev. Lett.}
  {\bfseries 95} (2005) 071101}
  [\href{https://arxiv.org/abs/hep-th/0504225}{{\ttfamily hep-th/0504225}}].

\bibitem{Martelli:2005wy}
D.~Martelli and J.~Sparks, \emph{{Toric Sasaki-Einstein metrics on S**2 x
  S**3}}, \href{https://doi.org/10.1016/j.physletb.2005.06.059}{\emph{Phys.
  Lett. B} {\bfseries 621} (2005) 208}
  [\href{https://arxiv.org/abs/hep-th/0505027}{{\ttfamily hep-th/0505027}}].

\bibitem{Cacciatori:2009iz}
S.L.~Cacciatori and D.~Klemm, \emph{{Supersymmetric AdS(4) black holes and
  attractors}}, \href{https://doi.org/10.1007/JHEP01(2010)085}{\emph{JHEP}
  {\bfseries 01} (2010) 085} [\href{https://arxiv.org/abs/0911.4926}{{\ttfamily
  0911.4926}}].

\bibitem{Ovcharenko:2024yyu}
H.~Ovcharenko, J.~Podolsky and M.~Astorino, \emph{{Black holes of type D
  revisited: relating their various metric forms}},
  \href{https://arxiv.org/abs/2409.02308}{{\ttfamily 2409.02308}}.

\bibitem{Chong:2005hr}
Z.W.~Chong, M.~Cvetic, H.~Lu and C.N.~Pope, \emph{{General non-extremal
  rotating black holes in minimal five-dimensional gauged supergravity}},
  \href{https://doi.org/10.1103/PhysRevLett.95.161301}{\emph{Phys. Rev. Lett.}
  {\bfseries 95} (2005) 161301}
  [\href{https://arxiv.org/abs/hep-th/0506029}{{\ttfamily hep-th/0506029}}].

\bibitem{Cassani:2015upa}
D.~Cassani, J.~Lorenzen and D.~Martelli, \emph{{Comments on supersymmetric
  solutions of minimal gauged supergravity in five dimensions}},
  \href{https://doi.org/10.1088/0264-9381/33/11/115013}{\emph{Class. Quant.
  Grav.} {\bfseries 33} (2016) 115013}
  [\href{https://arxiv.org/abs/1510.01380}{{\ttfamily 1510.01380}}].

\bibitem{Cvetic:2005zi}
M.~Cvetic, G.W.~Gibbons, H.~Lu and C.N.~Pope, \emph{{Rotating black holes in
  gauged supergravities: Thermodynamics, supersymmetric limits, topological
  solitons and time machines}},
  \href{https://arxiv.org/abs/hep-th/0504080}{{\ttfamily hep-th/0504080}}.

\bibitem{2015arXiv151206391A}
V.~{Apostolov} and G.~{Maschler}, \emph{{Conformally K{\"a}hler,
  Einstein-Maxwell Geometry}},
  \href{https://doi.org/10.48550/arXiv.1512.06391}{\emph{arXiv e-prints} (2015)
  arXiv:1512.06391} [\href{https://arxiv.org/abs/1512.06391}{{\ttfamily
  1512.06391}}].

\bibitem{LeBrun_2015}
C.~LeBrun, \emph{The einstein–maxwell equations, kähler metrics, and
  hermitian geometry},
  \href{https://doi.org/10.1016/j.geomphys.2015.01.009}{\emph{Journal of
  Geometry and Physics} {\bfseries 91} (2015) 163–171}.

\bibitem{LeBrun_2016}
C.~LeBrun, \emph{The einstein–maxwell equations and conformally kähler
  geometry},
  \href{https://doi.org/10.1007/s00220-015-2568-5}{\emph{Communications in
  Mathematical Physics} {\bfseries 344} (2016) 621–653}.

\bibitem{2015arXiv151106805K}
C.~{Koca} and C.W.~{T{\o}nnesen-Friedman}, \emph{{Strongly Hermitian
  Einstein-Maxwell Solutions on Ruled Surfaces}},
  \href{https://doi.org/10.48550/arXiv.1511.06805}{\emph{arXiv e-prints} (2015)
  arXiv:1511.06805} [\href{https://arxiv.org/abs/1511.06805}{{\ttfamily
  1511.06805}}].

\bibitem{2017arXiv170801958F}
A.~{Futaki} and H.~{Ono}, \emph{{Conformally Einstein-Maxwell K{\"a}hler
  metrics and structure of the automorphism group}},
  \href{https://doi.org/10.48550/arXiv.1708.01958}{\emph{arXiv e-prints} (2017)
  arXiv:1708.01958} [\href{https://arxiv.org/abs/1708.01958}{{\ttfamily
  1708.01958}}].

\bibitem{2017arXiv170607953F}
A.~{Futaki} and H.~{Ono}, \emph{{Volume minimization and Conformally
  K{\"a}hler, Einstein-Maxwell geometry}},
  \href{https://doi.org/10.48550/arXiv.1706.07953}{\emph{arXiv e-prints} (2017)
  arXiv:1706.07953} [\href{https://arxiv.org/abs/1706.07953}{{\ttfamily
  1706.07953}}].

\bibitem{Couzens:2018wnk}
C.~Couzens, J.P.~Gauntlett, D.~Martelli and J.~Sparks, \emph{{A geometric dual
  of $c$-extremization}},
  \href{https://doi.org/10.1007/JHEP01(2019)212}{\emph{JHEP} {\bfseries 01}
  (2019) 212} [\href{https://arxiv.org/abs/1810.11026}{{\ttfamily
  1810.11026}}].

\bibitem{Apostolov2001TheGO}
V.~Apostolov, D.M.J.~Calderbank and P.~Gauduchon, \emph{The geometry of weakly
  selfdual kahler surfaces}, {\emph{arXiv: Differential Geometry} (2001) }.

\bibitem{10.4310/jdg/1146169934}
V.~Apostolov, D.M.~Calderbank and P.~Gauduchon, \emph{{Hamiltonian 2-Forms in
  Kähler Geometry, I General Theory}},
  \href{https://doi.org/10.4310/jdg/1146169934}{\emph{Journal of Differential
  Geometry} {\bfseries 73} (2006) 359 }.

\bibitem{Ferrero:2024vmz}
P.~Ferrero, \emph{{D6 branes wrapped on a spindle and Y$^{p,q}$ manifolds}},
  \href{https://doi.org/10.1007/JHEP05(2024)182}{\emph{JHEP} {\bfseries 05}
  (2024) 182} [\href{https://arxiv.org/abs/2403.03988}{{\ttfamily
  2403.03988}}].

\bibitem{Gibbons:1978tef}
G.W.~Gibbons and S.W.~Hawking, \emph{{Gravitational Multi - Instantons}},
  \href{https://doi.org/10.1016/0370-2693(78)90478-1}{\emph{Phys. Lett. B}
  {\bfseries 78} (1978) 430}.

\end{thebibliography}\endgroup

\end{document}